# The State of AI Ethics
## January 2021

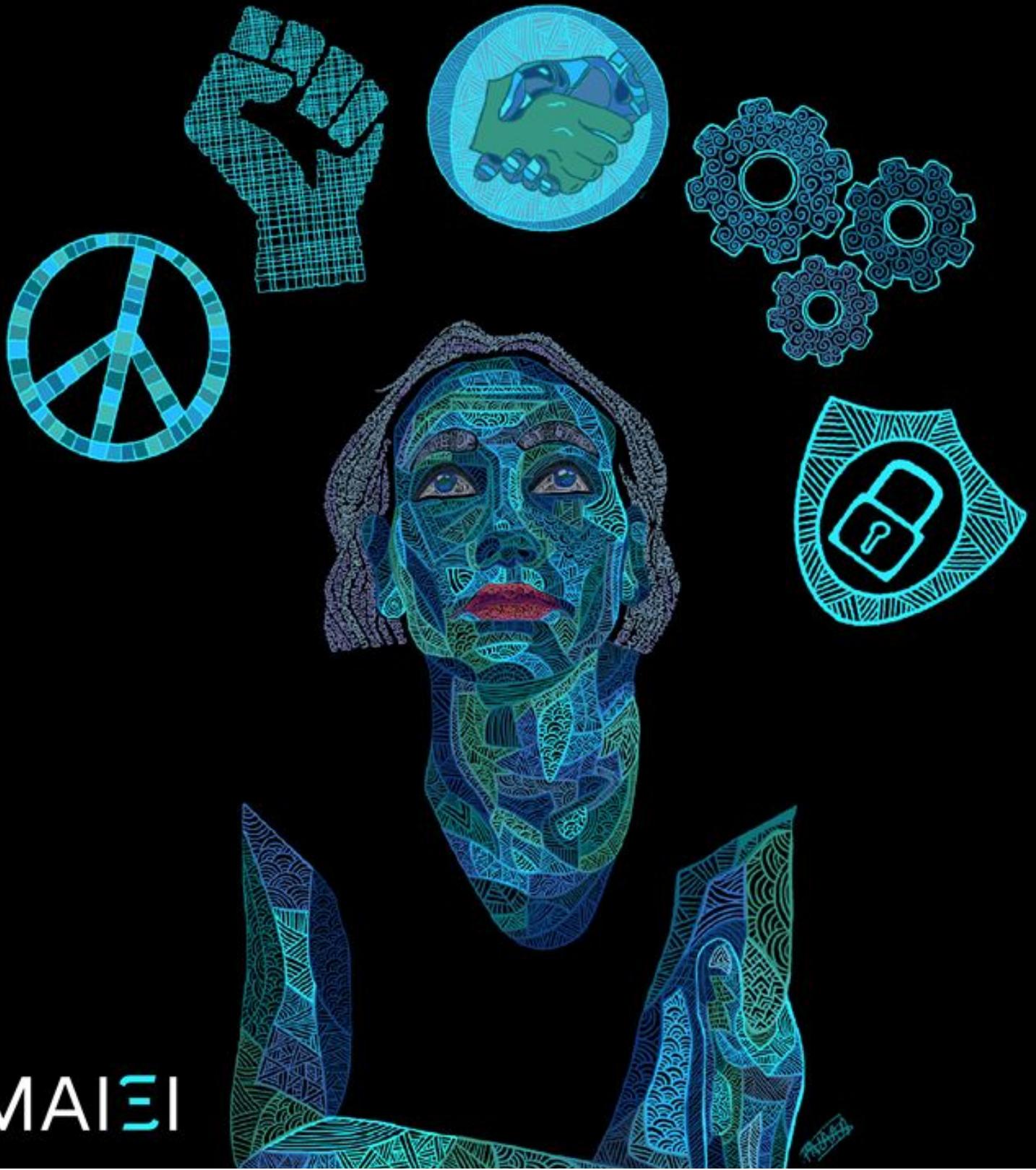

MAIEI

This report was prepared by the **Montreal AI Ethics Institute (MAIEI)** — an international non-profit organization democratizing AI ethics literacy. **Learn more on our website or subscribe to our weekly newsletter The AI Ethics Brief**.

This work is licensed open-access under a **Creative Commons Attribution 4.0 International License**.

Primary contact for the report: **Abhishek Gupta (abhishek@montrealethics.ai)**

**Full team behind the report:**

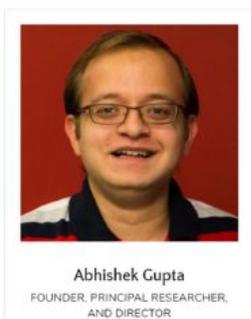 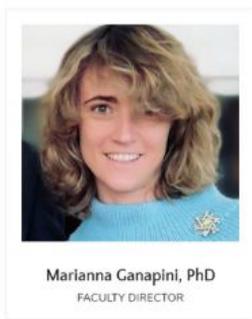 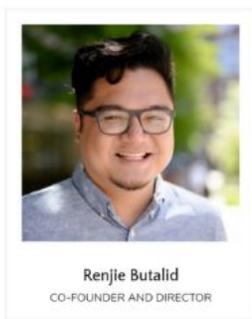 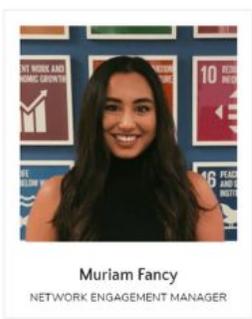 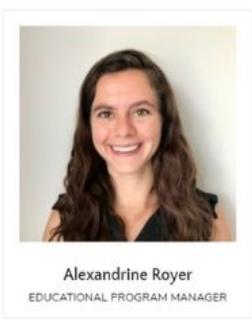 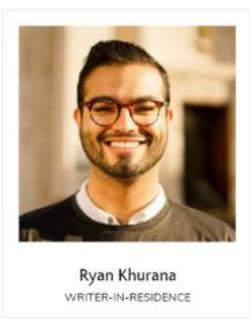

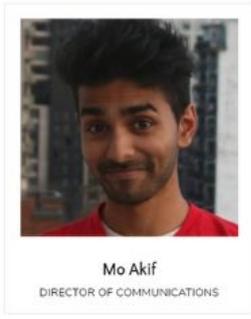 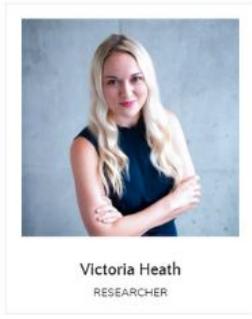 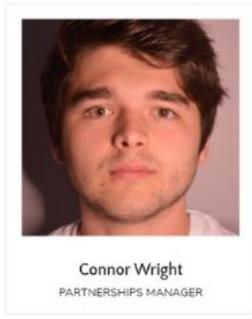 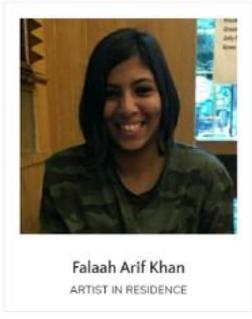 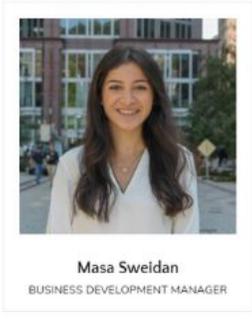 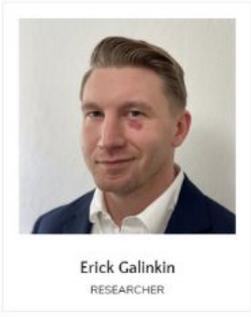

Special thanks to external research summary contributors:
**Anne Boily, Andrew Buzzell, Khoa Lam** and **Nga Than**

Cover Artwork '*Seer*' by **Falaah Arif Khan** (@FalaahArifKhan), *Artist in Residence*
Graphic Line Art (2021)

> The problems at the forefront of AI Ethics today – injustice, discrimination and retaliation – are battles that marginalized communities have been fighting for decades. It only took us millions of dollars and immense public interest into our darling technology to notice. Algorithms manifest and further exacerbate the structural inequalities in our society. We're finally starting to see 'bias' – algorithmic or otherwise – for what it really is: a fundamentally human problem.

# The State of AI Ethics Report (January 2021) is supported by

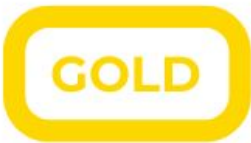 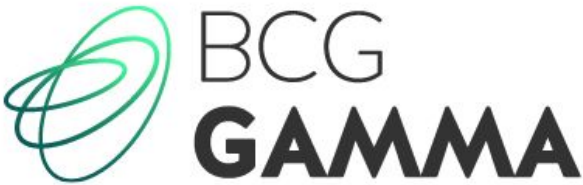

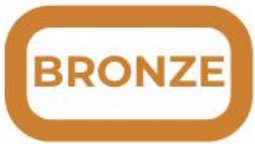 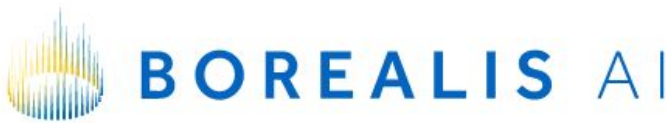

# Table of Contents

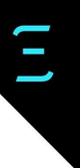

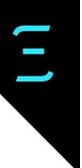

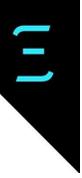

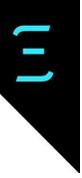

**\*Note:** The original sources are linked under the title of each piece. The work in the following pages combines summaries of the original material supplemented with insights from MAIEI research staff, unless otherwise indicated.

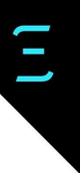

# Foreword by Steven Mills (Managing Director & Partner, Chief AI Ethics Officer, Boston Consulting Group)

As we reflect on this latest report on the current state of AI ethics, we are struck by the ways in which AI continues to reach deeper and deeper into our lives—often in unexpected ways that challenge the very foundation of our collective notions of society. Algorithms now inform decisions ranging from the seemingly inconsequential to those that have a profound, direct effect on our lives. Only a few years ago, we looked to AI to help us decide which song or movie to enjoy. Today, it dictates whether entire groups of people will qualify for mortgages, get job interviews, or receive affordable insurance policies. The content here reflects this evolving reality. And as the depth and breadth of AI applications themselves expand, so too does the research into the ethical issues associated with them. We are grateful to MAIEI for presenting a concise overview of this research, and are excited to be part of this latest edition.

After instilling artificial intelligence with so much power, it is incumbent upon us to make sure this power is wielded responsibly. It is encouraging to see this body of work re-affirm the role of human beings at the center of AI. After years of focus on automating human decision processes, there is a strong movement to implement what we refer to as "Human + AI," which involves embracing and even elevating the role of humans in decision processes. We have already seen how humans and AI working together produce better business outcomes than either one operating in isolation. And, as the research in this report demonstrates, humans have a critically important role to play in minimizing the unintended consequences of AI systems.

While we all understand the long-reaching implications of the decisions we make today regarding AI development, the report brings this reality into stark focus. We were particularly struck by the research herein that explores the impact these systems are having on the youngest, most vulnerable members of society. Our children find themselves navigating a world with technology proliferation, algorithmic discrimination, widespread misinformation, and the racialization of AI. As the report observes, children need protection from the subtle influence of AI, perhaps through the early introduction of AI literacy programs. This line of research reminds us of the difference this community of researchers can make and amplifies the feeling that we are moving in the right direction.

The 2021 report is particularly meaningful given the backdrop of the tumultuous year we have just collectively experienced—one of many unprecedented challenges including a global pandemic, an unfinished reckoning on racial justice, natural disasters intensified by the climate crisis, and growing global political unrest and division. It does not escape us that this report highlights how AI can sit on both sides of these issues. AI can amplify the spread of fake news, but it can also help humans identify and filter it; algorithms can perpetuate systemic societal biases, but they can also reveal unfair decision processes; training complex models can have a significant carbon footprint, but AI can optimize energy production and data center operations. The challenge for us as a community is to minimize the unintended consequences and shape a future in which AI will be used to right wrongs, deliver positive solutions, and increase optimism and hope. The work summarized here, and the critical conversations it triggers, are important tools to help us achieve that goal. Indeed, the report provides valuable guideposts by which AI professionals can assess their progress and look over the horizon to help navigate the future.

Many of the ethical issues brought into focus by AI are not new. Many of them will be difficult to solve. But in the face of so much change, it is crucial to remember that, ultimately, humans—not the machines we construct—control our fate. We build the algorithms, and if we are not satisfied with the results, we can work together to improve them. If our datasets reflect societal biases, we can continue to identify and mitigate those biases as we push to make change in the world. If we fear unethical applications of AI, we can redouble our efforts to bring ethics to the very center of the development process. As this report reminds us, we must not lose our sense of hope. We are far from helpless in making sure AI is used for the benefit of society.

There is, and should be, a sense of urgency to get AI right. Organizations like MAIEI and the work featured in this report continue to empower our community and advance our understanding of why we must quickly address these ethical issues. AI holds tremendous promise to help so many lives and lead us towards a much brighter future. Working together, it is well within our power to create a future in which all people—in every corner of the world—share equally in the benefits of AI.

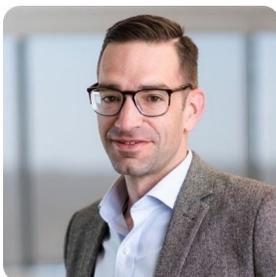

**Steven Mills (@stevndmills)**
Managing Director & Partner, Chief AI Ethics Officer
Boston Consulting Group

Steven Mills is a Managing Director & Partner at Boston Consulting Group (BCG) where he serves as the Chief AI Ethics Officer and the global lead for the Artificial Intelligence topic in the firm's Public Sector practice area. He is responsible for developing BCG's internal Responsible AI program as well as guiding clients as they implement their own Responsible AI programs. Steve serves private and public sector clients across defense, health, finance, aerospace, environment, and social impact on topics including implementing complex AI/machine learning use cases, developing artificial intelligence and analytic strategies, and helping clients build internal analytics capability.

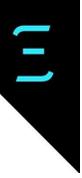

# Founder's Note

We've all heard many formulations of how 2020 was unprecedented in many respects and certainly, it was a year that brought together the best and worst of humanity. There are far too many things that have transpired in the past year feeling as if we had compressed a decade's worth of life events into a single year. Each year presents us with the opportunity to reflect critically on what took place in the previous year and reset our aspirations for the coming year so that we hopefully inch towards achieving our goals in a way that is sustainable and respectful of those around us.

We've been humbled by all the support that we have received from the AI ethics community in 2020 which has helped us move closer to our own aspiration of *Democratizing AI Ethics Literacy.* The edition that you hold in your hands (digitally) is a reflection of that trust that *you,* the AI ethics community, have placed in us to bring forth the most relevant research and reporting in the domain in a concise manner. We are also particularly humbled to have received support from **[BCG Gamma](#)** and **[Borealis AI](#)** that helps us continue to do the work that we do to ensure open-source and open-access for everyone, while being able to provide fair compensation to those who contribute to creating the report and run the activities at MAIEI.

We are also delighted to bring to you some special inclusions in this report that go beyond the research and reporting, including a marquee piece from researchers at MIT titled, *The Abuse and Misogynoir Playbook*; highlights from community-nominated researchers that showcase diverse initiatives from around the world; a special section on the future of AI ethics; and remarks from luminaries in the field reflecting on the content of each of the chapters helping to further contextualize the content.

I sincerely hope that you enjoy this January 2021 edition of the *State of AI Ethics R*eport and that it helps you gain a broad-based view of the very vast landscape of AI ethics.

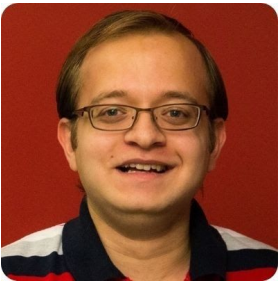

**Abhishek Gupta ([@atg_abhishek](#))**
Founder, Director, & Principal Researcher
Montreal AI Ethics Institute

Abhishek Gupta is the founder, director, and principal researcher at the Montreal AI Ethics Institute. He is also a machine learning engineer at Microsoft, where he serves on the CSE Responsible AI Board. His book '[Actionable AI Ethics](#)' will be published by Manning in 2021.

KATLYN TURNER, DANIELLE WOOD, CATHERINE D'IGNAZIO

# THE ABUSE AND MISOGYNOIR* PLAYBOOK

**1 A BLACK WOMAN MAKES A CONTRIBUTION**
- Uses tools of society
- Advances knowledge
- Reveals inequity, truth
- Changes status quo
- Is unacceptable in some way to dominant class & stakeholders

**2 DOMINANT CLASS & STAKEHOLDERS RESPOND WITH DISBELIEF**
- "How could you have done/made this?"
- "How can this be true?"
- "This can't be right."

**3 CONTINUED DISBELIEF**

**A. DISMISSAL**
- "This isn't important."
- "Moving on…"
- Business as usual
- Ignores

**BELIEF**

**B. GASLIGHTING**
- "You must be mistaken."
- "This isn't actually what happened."
- Makes contributor seem crazy and isolated in views
- Uses pity, sympathy as weapon

**BELIEF**

**C. DISCREDITING**
- "This isn't as big of a deal as they claim."
- "This isn't as important as they claim."

**4 ERASURE**
- Moves on in news cycle, company statements
- No longer mentions contributor
- Gatekeeps contributor from further contributions
- Silences contributor and those who agree

**5 REVISIONISM**
- New narrative
- Negative reaction to reminders of contributions
- Loss of original knowledge over generations
- Leads to genuine unawareness over time of original events

\* *Misogynoir*, a term coined by Dr. Moya Bailey in 2010, describes the particular form of anti-Black sexism faced by Black women.

DESIGN: MELISSA 青 TENG

*The Abuse and Misogynoir Playbook diagram by Katlyn Turner, Danielle Wood, and Catherine D'Ignazio. Design by [melissa 青 teng](#).*

# 1. The Abuse and Misogynoir Playbook

**By Katlyn Turner, Danielle Wood, Catherine D'Ignazio**

**Disbelieving, devaluing, and discrediting the contributions of Black women has been the historical norm. Let's write a new playbook for AI Ethics.**

"...come celebrate
with me that everyday
something has tried to kill me
and has failed."

- Lucille Clifton

In the past decade, Black women have been producing leading scholarship that challenges the dominant narratives of the AI and Tech industry: namely that technology is ahistorical, "evolved", "neutral" and "rational" beyond the human quibbles of issues like gender, class, and race. Safiya Noble demonstrates how search algorithms routinely work to dehumanize Black women and girls (Noble 2018). Ruha Benjamin challenges what she calls the "imagined objectivity" of software and explains how Big Tech has collaborated with unjust systems to produce "the New Jim Code", software products that work to reproduce racial inequality (Benjamin 2019). Joy Buolamwini and Timnit Gebru definitively expose racial and gender bias in facial analysis libraries and training datasets (Buolamwini & Gebru 2018). Meredith Broussard challenges the "technochauvinism" embedded in AI and machine learning products (Broussard 2018). Rediet Abebe calls for us to confront the limitations of the concept of fairness and center our analysis on power (Kasy & Abebe 2020). Simone Browne teaches us that today's cutting-edge technologies are part of a long history of surveillance of Black bodies in public spaces (Browne 2015).

These scholars, along with many others, are sounding the alarm that tech is neither neutral nor ahistorical. Rather, how *evolved* it is reflects how quickly it can reproduce and entrench our historical biases. How *rational* it is indicates our collective desire to forget and erase the ugliness of racism, sexism, classism, xenophobia — and assign it instead to an opaque algorithm and output. And these instincts are far from *ahistorical*, rather they are part of a centuries-old playbook employed swiftly and authoritatively over the years to silence, erase, and revise contributions and contributors that question the status quo of innovation, policy, and social theory.

*The Abuse and Misogynoir Playbook,* as we name it here, has been used successfully by individuals and institutions to silence, shame, and erase Black women and their contributions for centuries. *Misogynoir* is a term introduced [by Dr. Moya Bailey in 2010](#) (Bailey and Trudy, 2018; Bailey 2021) that describes the unique racialized and gendered oppression that Black women systemically face. We see the Playbook in operation in the recent [well-publicized](#) and swift firing of Google's Ethical AI Co-Lead, Dr. Timnit Gebru. We see the Playbook in operation in the case of poet Phillis Wheatley in the 1700s. The Playbook's tactics, described in the accompanying diagram, are **disbelief**, **dismissal**, **gaslighting,** **discrediting**, **revisionism**, and **erasure of Black women and their contributions**.

## Google Fires, Discredits, Dismisses, Gaslights and Attempts to Silence Internationally Recognized AI & Ethics Researcher Dr. Timnit Gebru

Until December 2020, [Dr. Timnit Gebru](#) was the Staff Research Scientist and Co-Lead of Ethical Artificial Intelligence (AI) team at Google. Dr. Gebru has been long respected both in and beyond the field of AI Ethics for her insightful contributions to the space — often focusing on the uncanny details of how we live, how we choose to present ourselves, what we don't say — and what it says about us in the bigger picture of society. Her [groundbreaking paper](#) *(Gebru, Krause, Wang, Chen, Deng, Lieberman Aiden, and Fei-Fei 2017)* on the political preferences of Americans and its link to the cars we choose to drive was not just a sharp use of Google Earth and satellite data, but a subtle commentary on how the choices we make as consumers say more about us than whether we prefer to drive a pickup truck or a hybrid. Dr. Gebru is well-known for her work on bias in algorithms in machine learning, has authored numerous scholarly papers, and collaborated with leading experts and institutions. She is also a high-profile champion of diversity, equity and inclusion in the AI and machine learning communities. Together with Rediet Abebe, Dr. Gebru founded [Black in AI](#), a group for Black people to build community and collaborations while working on artificial intelligence research. Following her graduation from Stanford University and the publication of such groundbreaking work, Dr. Gebru was employed as the Technical Co-Lead for Google's Ethical AI team, a position that led many — both in the field and outsiders — to give the team and the company credibility over its stated commitment to ethics and justice in AI.

Dr. Gebru was fired in December 2020 over a paper that she co-wrote titled, ["On the Dangers of Stochastic Parrots: Can Language Models Be Too Big?"](#) The paper outlines risks posed by

The State of AI Ethics, January 2021                                                                  16

large-scale language models, including potential ecological harms, misuse by bad actors, and lack of transparency. While initially approved through Google's internal review process, leadership later reversed its decision and demanded that Dr. Gebru and co-authors retract it from being published. When Dr. Gebru wrote to request answers about this censorship and demand accountability, she was "[resignated](#)", a term coined by her team to describe her firing that Google subsequently and fallaciously described as a "[resignation](#)."

Dr. Timnit Gebru's firing is a use of the Playbook that we can't look away from. It produced collective outrage in the form of thousands of signatures to a [petition](#), a [Congressional letter](#) to Google, and the firm being dropped by [HBCU recruiters](#). Yet many of us are unaware that the tactics used to silence and erase Dr. Gebru and her contributions are part of a centuries old pattern. Until we recognize these actions for what they are, and name them, we will be stuck in a pattern that precludes all progress towards true ethics, equity, and justice.

**The Timeless Tactics of the Abuse and Misogynoir Playbook**

Throughout the history of the United States, the march toward true justice and equity has not been a linear, steady, or constant thread. Rather, much like water boiling in a kettle, small bubbles questioning everyday injustice, inequity, and the status quo rise up — insignificant at first, until overwhelmingly a rolling boil demanding change whistles out, and the nation responds to this chorus: culturally and judicially. The nation was founded as a British colony upholding settler colonialism, *de facto* and *de jure* segregationist racism, and patriarchal misogyny (Collins 1990; Kendi 2016). In the early days of the republic, the only members endowed with the full rights of citizenship and participation in the governance of the budding country were landowning white men — with ideas about self-governance and participation drawing from models of antiquity like the ancient Romans. Over time, many fought passionately and bitterly to form a "more perfect Union" that systemically included and recognized the humanity, rights, and citizenship of "all": Black people, Indigenous Peoples, women, immigrants and people of non-white ethnicities, LGBTQ people, and those who practice religions other than Christianity. Though mainstream attitudes around these issues have changed over the country's history, all of these fights continue today (Ortiz 2018).

The fight to end enslavement and gain citizenship for Black people cost trillions of dollars, a forever-changed economy, and over 600,000 lives including that of President Abraham Lincoln (Mullen and Darity 2020). The cause for women's suffrage gained momentum over eighty years and thousands of voices to produce the 19th amendment (Crawford 2001). The call for equality, protection, and freedom under the law to love independent of gender demanded centuries of action, both clandestine and in the open, and cost countless literal and figurative lives — until a seemingly innocuous and humble court case for a loving widower to have the dignity of legal recognition of his deceased husband reached the United States Supreme Court and made marriage equality the law of the land. The issues surrounding equity and justice in STEM, Big Tech, and AI will undoubtedly follow a similar bubbling path: as voices gather, actions mount, knowledge is learned, sacrifices — willing and unwilling — are made, and the people demand sweeping change. We can take heart that the path our field is on now is part of a long tradition of questioning, perfecting, and ensuring that reality reflects our timelessly stated ideals: justice for all. And as part of that process, we can name and root out dynamics that have long been employed to preclude, erase, and silence progress.

From the United States' beginnings, Black women have been uniquely situated at the bottom of a gendered, racial, and class-based hierarchy that systematically dehumanized them and aimed to strip them of dignity (Collins 1990; Crenshaw 2017; Taylor 2017). Racialized capitalism (Kelley 2017; Ortiz 2018) and segregationist ideas (Kendi 2016) about the supposed inferiority of Africans and the "Negro race" ensured the legal status of Black people in the United States for hundreds of years was relegated to chattel slavery as a means to a permanent supply of cheap labor (Mullen and Darity 2020). This economic racism robbed individuals of their rights to autonomy, self-determination, citizenship, protection, and freedom. The legal regimes upholding this caste included measures to criminalize actions such as religious gatherings and literacy, and stripped Black people of their rights: to bear arms, to protect themselves, to speak freely, to secure property, and to advocate for their presents and futures — individually or as a collective. Similarly, women have faced systemic struggles targeting their autonomy, independence, and protections. Women's right to vote was legalized only a hundred years ago (Crawford 2001), the right to financial independence only legalized in the 1960s, and measures such as the Violence Against Women Act only became law in the 1990s.

As groups, Black people and women have faced similar types of limitations on their rights, as well as a sustained struggle for equal rights and protection under the law. However, a crucial difference between these two groups is related to their perceived value and roles in society, and understanding this difference is key to understanding *misogynoir* — the unique racialized and gendered oppression that Black women systemically face[1]. As a group, Black people have been cast as laborers: untrustworthy and irresponsible with their own lives and futures, but strong capable workers when under organized (white) direction. Their racial status as cheap laborers dehumanized them, relegated them to the status of property, and did not afford them protection even as they built and engineered the country, raised the children of the white elite, and were considered the ideal domestic servant. In contrast, women, particularly white women, have been seen as a protected class: to be honored, to be revered, to be treated kindly, and to be guarded from both perceived and actual harm (Friedan 1964). For (white and white-adjacent) women, this protection comes at a price: adhering to the arbitrary and persnickety standards of femininity, which has included giving up one's autonomy and any desire for intellectual pursuits, suppressing any proclivities not deemed "ladylike", and ideally focusing entirely on being a proper object — someone's wife, mother, or daughter — rather than a nuanced individual subject. Despite gains from the advent of mainstream awareness and advocacy around feminism, women are still punished societally for activities not deemed feminine — being too assertive, too competent in areas unrelated to child-rearing and caregiving, being insufficiently feminine in their outward appearance (Mullany 2010), or even running for President (Wilz 2016).

---

[1] For further reading on these issues, see: Bailey, Moya. Misogynoir Transformed: Black Women's Digital Resistance. New York University Press, 2021. hooks, bell. Ain't I A Woman: Black women and feminism. New York, NY: Routledge, 2014.; Davis, Angela Y. Women, race, & class. Vintage, 2011.; Taylor, Keeanga-Yamahtta, ed. How we get free: Black feminism and the Combahee River Collective. Haymarket Books, 2017.

Black women were not afforded the protections typically afforded to white women in exchange for the "price" of femininity; nor were they expected or allowed to be traditionally feminine (Cottom 2018). Rather, enslaved Black women in the United States were expected to be dutiful, loyal, and selfless laborers in chief. Everything from agricultural duties to domestic labor to sexual exploitation was their "responsibility" to endure. Black women were placed in a role as a thankless workhorse: to birth more Black children into slavery, to raise and nurse children who were not their own, to competently manage a household, toil on plantation crops, and to comply when desired by any white man. For this brutal role, the price Black women were expected to pay was silence and complicity: an expectation that they would work tirelessly in all facets of the plantation system when demanded, and that they would neither complain nor fight back against this system even when it brutally and violently dehumanized them and their loved ones. This unique brand of oppression is misogynoir, and far from a historical relic, it is alive and well today (Cabrera 2016). Though Black women, like white women, face penalties and are policed on their femininity and their assertiveness, the ultimate societal taboo and danger for Black women is not being too feminine or too competent, but being too loud: to use one's voice to speak truth, or to advocate for oneself and one's community — rather than to obey and serve. While white women using their voice in a similar way may be perceived as annoyances (at worst), or as morally righteous (at best), Black women who do so often face retribution in the way of the abusive tactics we describe herein, or worse.

Despite the abuse Black women have suffered at the hands of misogynoir, many throughout history have made vital contributions to lasting justice and equity, often at their own erasure or even peril. Black women's contributions have set precedents across disciplines and society: from Harriet Tubman's [raid on the Combahee River](#) freeing hundreds of enslaved African Americans, to Tarana Burke's creation of the now viral and mainstream [#MeToo movement](#). However, society has not often reacted warmly to the contributions of Black women, particularly when those contributions challenge the status quo that keeps them in a uniquely vulnerable position.

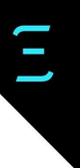

In fact, the reactions from society to contributions from Black women are so much of a pattern, that we describe them here as a Playbook. The Abuse and Misogynoir Playbook functions in the short term to use abusive tactics such as **gaslighting**, **dismissal**, and **discrediting** to **erase** and invalidate the **contributions** of Black women that challenge the status quo and aim to advance justice. These tactics often end up harming the women themselves, which may serve as a convenient deterrent for potential future truth tellers. In the long term, the impacts of this Playbook are even more devastating: the erasure of valuable contributions by Black women, supplanted by a more whitewashed narrative of events, that over time the public accepts as truth. As a result of the tactics used in this timeless Playbook entangling abuse and misogynoir, contributions by Black women who challenge the status quo, no matter how sound, brilliant, and timely — are often erased, the women themselves discredited and forgotten, and their voices deplatformed, such that no organized movement can build from their thoughts, actions, and contributions.

Dr. Timnit Gebru's firing at Google, when viewed in light of these tactics and their historical context within misogynoir, is not an isolated event nor an erratic incident of misunderstanding between Dr. Gebru and her former supervisors. It is a pattern that continues to unfold now that has a history, a context, and hopefully now, a name. What is novel is that we know about it as it's happening. What is novel is that Dr. Gebru is already respected and well-known in her field, as she should be. What is noteworthy is that it's happening in 2020 and 2021, where thousands of supporters can lend their support with a retweet or a signature to an online petition. But unfortunately, as Black women of diverse walks of life know — being dehumanized, humiliated, stripped of dignity, gaslit, silenced, manipulated, and abused at the hands of power is par for the course. If anything, Dr. Gebru joins a long lineage of Black women who, over the course of history, have dared to dream, create, sound the alarm, push for change; to be scientists, artists, creatives, engineers, women, and human — and have faced the wall of silence, pushback, threats, and danger. Black women have pushed and insisted on progress at a great cost — often their careers, their reputations, their dignities, their health, their legacies, and their lives.

If AI and Tech want to claim to care about ethics — we must recognize this pattern. We must own that the mechanisms within STEM and Tech used to silence, discredit, gaslight, and dismiss Black women are part of a historic pattern. We must grapple with the truth that Tech is not ahistoric, evolved, or "beyond" these issues. Rather, it is as much as part of them as any other field, and its treatment of Black women like Dr. Timnit Gebru is as much a part of a legacy of entrenched racism, anti-Blackness, and misogynoir. If AI and Tech want to claim a better, more ethical future — we must break out of this pattern and Playbook, and **stop abusing Black**

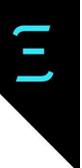

**women**.

## The Playbook from History to Present

The tactics used by Google against Dr. Gebru have been used from the colonial era of the US to the present. These tools have been used to uphold the status quo, limit progress towards true equity, and keep marginalized groups like Black women "in their place" for centuries[2]. By examining notable examples of Black women over history, we can see that what happened to Dr. Gebru is part of a continuous thread, rather than an aberration, or an isolated interpersonal incident.

**Taboo Contributions & the Master's Tools**

The Abuse and Misogynoir Playbook starts with a **contribution**: a Black woman using her voice. Using the tools of societal currency and competency — be it data science and academic publications, or simply literacy and the written word — these unwanted **contributions** expose or reveal the truths of an unjust facet of society *(see step 1 in the diagram)*. The use of social currency in this way by Black women has historically incurred a backlash; this is due to the taboo Black women are crossing by using "the master's tools" in such a subversive way. Historically, Black women were supposed to do all the work without question, to observe and competently take note of all details, but say nothing.

Black women have long realized that oftentimes it is necessary to use the so-called "neutral" and "legitimate" power of the written word — whether by publishing books, speeches, or journal articles — to make uncomfortable truths known. This is rooted in systemic oppression as the power of literacy was systematically withheld from Black people for centuries. Black people using what Audre Lorde calls the ["master's tools"](#) to bring light to uncomfortable truths has always been fraught, especially for people who had enslaved and subjugated them (Lorde 1984). Historically, the dominant white class placed literacy, such as the ability to read, write and create, as gifts and privileges. While in rare exceptions, they sometimes taught and allowed Black people to use these skills, the implicit end of that bargain is that Black people should not "abuse" this "gift" by using it to upend a power structure that advantages whiteness. Words, publications, literacy, data science, statistics, maps and more have long been treated with contempt when coming from Black voices. And while non-Black voices are often heralded as visionary, genius, or ahead of their time when writing and publishing uncomfortable truths — this has not been the case for Black people.

---

[2] For more reading on the strategies that have been used to undermine the liberatory contributions of Black women, see the following: Lorde, Audre. Sister outsider: Essays and speeches. Crossing Press, 1984; Carruthers, Charlene. *Unapologetic: A Black, queer, and feminist mandate for radical movements*. Beacon Press, 2018. Morrison, Toni. *The source of self-regard: Selected essays, speeches, and meditations*. Vintage, 2020. Smith, Barbara. *The Combahee River collective statement: Black feminist organizing in the seventies and eighties*. Vol. 1. Kitchen Table: Women of Color Press, 1986.

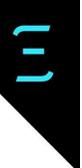

At Google, Dr. Gebru's team was known for groundbreaking research that challenged and brought light to the inequities and disparities that the software and AI industry sustained and often exacerbated. Through her and her team's dedicated work, they repeatedly described and quantified bias in AI systems as well as bias and exclusion inside of large corporations — and the paper that led to her firing was no exception. By working within the system, at Google itself, and using the currency of the scientific method, research, publications, and academic platforms to share this work, it enabled her work and the truths it uncovered to circulate to high-profile mainstream audiences who were not already inclined to see this truth. However, this currency of research also became a double-edged sword. By aiming for Dr. Gebru's team to retract their paper, Google attempted to silence this truth, and threaten Dr. Gebru's safety and security (*via* her job) as a bargaining chip. This is not new — Black women have dealt with this for centuries.

A historic example of particularly violent whitelash to controversial **contributions** by Black women is found in the life of Ida B. Wells. In 1893, Ida B. Wells[3] gave a controversial speech entitled "LYNCH LAW IN ALL ITS PHASES" at Tremont Temple Baptist Church, which still stands in Boston, only a few miles from where the authors work. This speech aimed to raise Northern awareness of lynching in the South, and to seek Northern support for antilynching policies. Wells was motivated to speak because she hoped that if her Northern audience understood the evils of lynching, they would take action, stating, "*I cannot believe that the apathy and indifference which so largely obtains regarding mob rule is other than the result of ignorance of the true situation.*" In the speech, Wells described her position, stating "*three years ago last June, I became editor and part owner of the Memphis Free Speech...I set out to make a race newspaper pay a thing which older and wiser heads said could not be done.*"

Wells saw her newspaper as a tool to support self-driven liberation among her Black community. In this role, Wells started to focus on lynching after the death of several of her friends. She noted in her speech that she did not expect lynchings in Memphis, however, "On the morning of March 9, the bodies of three of our best young men were found in an old field horribly shot to pieces." Wells had used her position in the Memphis Free Speech to advance knowledge on the fact that lynchings did indeed occur in the area and the authorities did not prosecute the deaths. Wells used the Memphis Free Speech to encourage Black residents to move to states further west, since the police were not protecting Black lives from mob violence.

---

[3] Read more of Ida B. Wells analysis of the lynching of Black people in her book, Wells-Barnett, I. B. (1892). *Southern horrors: Lynch law in all its phases*. Good Press.

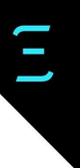

After publishing an antilynching editorial, she received numerous death threats, did not feel safe to stay in Memphis, and instead traveled to cities such as Boston to seek support for the antilynching movement. During her 1893 speech, Wells detailed her experience with death threats and harm that came to her due to reporting data and writing newspaper stories about the truth of the pervasive evil of lynching. She used tools such as data tables alongside her written articles to make her point, which has made her a hero to present-day data journalists and led to the founding of the [Ida B. Wells Society for Investigative Reporting](#).

The fact that Ida B. Wells was so empowered in her position at the newspaper, and that she used this position — one not normally afforded to Black people or to women — to advance and publish stories that upended the status quo of white supremacy and domestic racial terrorism — was intolerable. This was different than merely rumors or gossip of lynching; Wells had used the power of data science and the written word to reveal a depraved truth. For a nation decades away from even beginning to legalize racial equity, this **contribution** was not just unwanted, it was unbearable for those in power, and dangerous for Wells.

In a similar way, Dr. Gebru's team made a **contribution**, using her rightful position as the lead of Google's AI Ethics team, in order to advance knowledge of inequity in relation to large-scale language models. Perhaps, in a similar way — she and her team's **contribution** were years too early for an industry still so dominated by elite white men. But it does not change the fact that she paid for this **contribution** with her job and her security, and it does not change the fact that rather than working with her, her employers and many of her peers in Tech have employed the tactics of the Abuse and Misogynoir Playbook in attempt to revise and reshape the narrative.

**Disbelief and the Devaluing of Black Women's Creations**

Dr. Gebru's team's findings in the pivotal work that led to her firing were significant: they held implications not just for Google as a company, but for the entire tech industry and for the fields of AI and natural language processing specifically. The paper outlined four risks of large-scale language models including ecological harms, lack of transparency, research opportunity costs, and potential for misuse**.**  In response, Dr. Gebru's employers showed signs of **disbelief** *(see step 2 in the diagram)*: they argued that these **contributions** simply couldn't be true, and they attempted to undermine her credibility, academic honesty, and her psychological well-being in their remarks for her to retract the paper that came after the corporation had previously approved its publication. This **disbelief** that a Black woman could contribute to a field's discourse in such a groundbreaking way hearkens back to the life of Phillis Wheatley[4], the first

---

[4] Historical information about the life of Phillis Wheatly is drawn from the following: Poetry Foundation, "Phillis Wheatley," https://www.poetryfoundation.org/poets/phillis-wheatley, Accessed January 2021; Kendi Ibram, Stamped from the Beginning, Nation Books 2016.

African American author of a published book of poetry in the 1770s, whose statue adorns the high-end Commonwealth Avenue park in Boston.

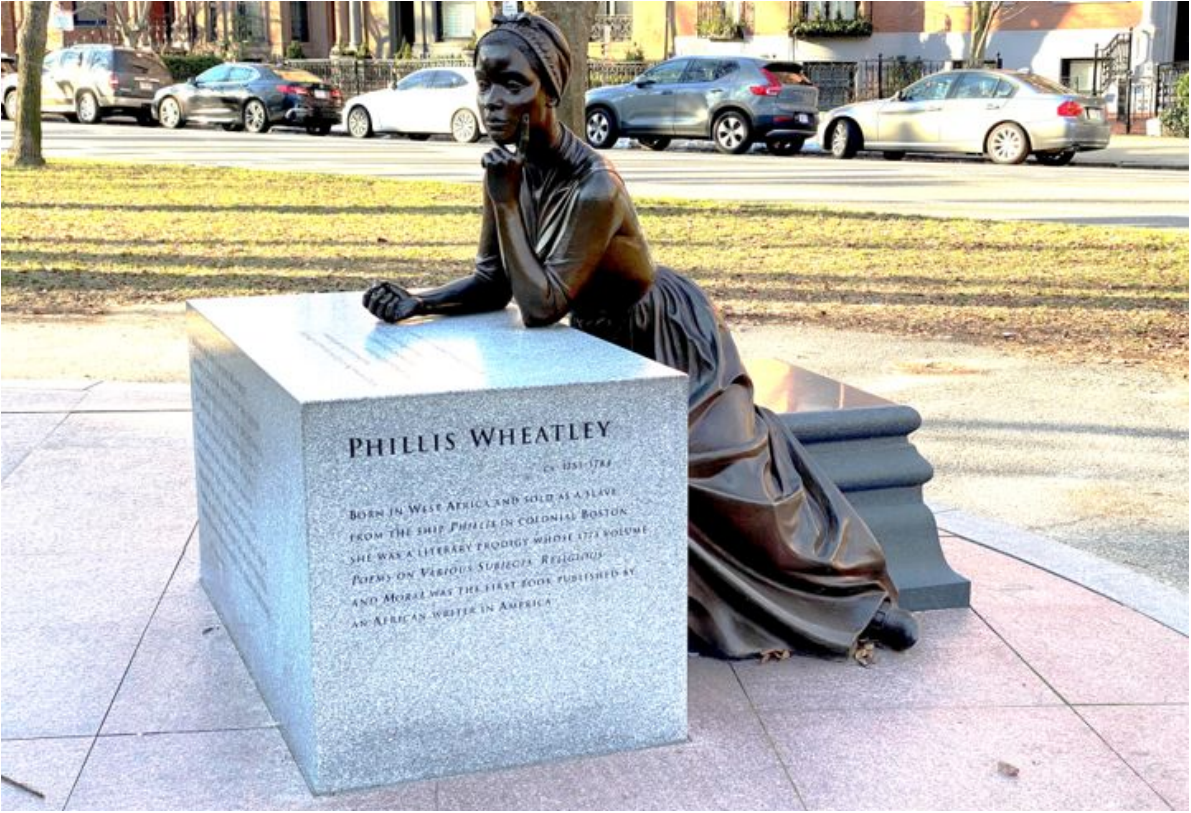

Statue of Phillis Wheatley on Commonwealth Avenue Park in Boston, Massachusetts
(Credit: Danielle Wood)

As a young girl, Phillis Wheatley was enslaved somewhere near modern-day Senegal and Republic of the Gambia and brought to Boston when she was around 7 to 9 years old. The family that enslaved her treated her as a servant but also provided her the opportunity to learn to read and write. Initially, even as young as 13, Phillis wrote individual poems that were shared at major events — such as elegies for a funeral — or were published in newspapers. Around the age of 18 (1772), Phillis had written enough poems for a book called *Poems on Various Subjects, Religious and Moral*, and she sought "subscribers'' in Boston to support the publication cost. However, financial support was not enough for publication in the United States. The opening pages of the book *Poems on Various Subjects* reveals that her owner, John Wheatley expected that white colonists would not believe Phillis was the true author of the poems. John Wheatley arranged for a panel of distinguished Bostonian leaders — all male — to interview Phillis in order to confirm that she was the real author. Ultimately, the panel all agreed and signed a document confirming this. Their names are included in a preface to the book, including the Governor of Massachusetts and John Hancock, a well-known leader during the American

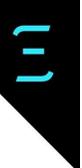

Revolution. Still, no publisher in the United States would take on the work.

Phillis Wheatley's status as an enslaved Black woman certainly contributed to the difficulties she faced — in publishing her work at all, in the public correctly attributing her authorship, and in earning fair compensation for the contributions she made. In part, the white power-holding class simply could not believe that Phillis could have created the poems because they were regarded as *good*. When it became clear that Wheatley was the author of the book of poems — the public attempted to devalue her and her work, deplatform her, and make invisible her legacy and groundbreaking contributions to literature as an enslaved African American woman. Phillis was later able to publish *Poems on Various Subjects* via a publisher in London. She received some amount of notoriety and income for her poetry during the lifetime of the Wheatley family. After their death, Phillis married a free Black man, and her later life was economically very difficult. Although she continued to publish poetry, she still faced difficulty in publishing her books in the United States. In 1779, she tried again to gain subscriber support to publish a book of poetry in the United States and found support lacking. Ironically, however, in the year of her death, 1784, she published an influential poem called *Liberty and Peace* celebrating the American Revolution, but she and her family were poor and debt-ridden. Phillis Wheatley died in a state of poverty and ill health while her second book of poetry was finally published two years later in the United States, but this was too late to provide her any financial relief.

Although centuries separate their stories, the wondrous **disbelief** and subsequent devaluing of Dr. Gebru and her work are steeped in vestiges of the same phenomenon that happened to Phillis Wheatley. Although legalized segregationist racism no longer exists in the United States today, the impacts of colonial attitudes and ideas about the inferiority and limitations of Black women leave a lasting and noticeable legacy that is active today. Even Dr. Gebru's non-Black colleagues and former coworkers remarked that the way Google treated her case was [different from that of other employee activists](#) — noting an inequity even amongst how the institution had treated those it had fired in the past. An insidious thread that connects the stories of Phillis Wheatley and Dr. Timnit Gebru is the thread of misogynoir — in this case, manifesting as **disbelief** that a person of this class could produce such an authoritative contribution.

**Silencing the Truth and the Teller: Discrediting and Gaslighting**

In cases when a Black woman's status-quo-upending contribution is believed — at least by the dominant class and stakeholders — but not welcomed, stakeholders have sent signals through the mechanisms of **discrediting** and **gaslighting** to silence the truths revealed, as well as their authors *(see steps 3b and 3c in the diagram)*. These mechanisms work well when a stakeholder aims to fight back against a narrative revealed by a particular contribution, because they serve to make the author of the contribution seem irrational, crazy, isolated, and mistaken — rather than attacking the narrative itself. By using **discrediting** and **gaslighting** tactics, such as weaponizing pity and sympathy, as signers of the petition [accuse Jeff Dean of doing](#) in his initial responses to Dr. Gebru's firing, the public response to such backlash is managed. Rather than accept the truth that an institution may be perpetuating harm, which is difficult to swallow, we can accept one that says that an individual like Dr. Gebru is mistaken or unwell — and that is what led to this errant "contribution." The public is especially primed to accept the **gaslighting** and **discrediting** of people from marginalized groups — like Black women — who as noted, were historically expected to play a role of silent and accepting laborer, rather than activist or academe.

**Discrediting** and **gaslighting** have two subjects, though — the contributor in question, as well as the community and public at large. When the community and public accept this gaslight, it lures them into further complicity, and takes away their means or drive towards self-advocacy. Dr. Gebru's paper was controversial within Google because it outlined the risks of employing large-scale language models, models that Google has made strategic financial investments in. Allowing truth like this to be published does not bode well for those who'd prefer that the status quo – big data and big money, for some – not change. The way Dr. Gebru was fired, **discredited**, and **gaslit** in the response to her work, is part of this timeless pattern of Abuse and Misogynoir.

Another historic example of misogynoir and **gaslighting** can be found within the life of Harriet Jacobs (1861), author of the autobiographical "[Incidents in the Life of a Slave Girl,](#)" whose grave is located in [Mount Auburn Cemetery](#) close to the home of the authors. Jacobs experienced cruelty and danger in her life as an enslaved woman. She was consistently tormented emotionally and threatened when she did not succumb to her owner's sexual desires. Jacobs went into hiding for several years to avoid further mistreatment, which required her to confine herself to a crawl space in which she could not stand. After her escape to the North and freedom, Jacobs realized that Northerners were being **gaslit** by a narrative that slavery in the American South was not that bad. She wondered if the inaction and indifference to slavery by many Northerners was because they simply did not understand the reality of slavery, which often included abuse, assault, family separation, neglect, torture, and murder. Her autobiographical work discussed how the entire society was being **gaslit** in order to maintain slavery and white supremacy, and addresses particularly the indifference she observed among white women in the North. She describes slaves being told that runaways in the North lived in "deplorable conditions" in order to decrease desire to flee. Additionally, she discussed the reality of marital and sexual relations in the South in order to attempt to dissuade support for the Fugitive Slave Act. Jacobs described how men in the South took sexual gratification from whomever they wished — female slaves in particular — and how easily marriage vows were cast aside. Furthermore, she detailed the evils that slavery did to families, for example, how separation of mothers and children were a common occurrence. Her book was seen as groundbreaking for its exposure of the ways that myths and narratives about slavery **gaslit** people on all sides into complicity.

Black women have often been **gaslit** as a way of **discrediting** not only their work, but taking away their means to advocate for themselves, and their greater communities. This has been effective at slowing progress towards justice and equity because society wants to believe that things just aren't that bad. Just as society wanted to believe that slavery flat out wasn't evil, it also wants to disregard the societal harms from practices pursued by Big Tech and AI. **Gaslighting** contributors like Dr. Gebru as well as the greater public, therefore, is in service of maintaining this status quo.

**Dismissal, Erasure, and Revisionism**

When Black women's contributions are **dismissed** and **discredited**, and the public is **gaslit**, as described above — a crucial next step concerns what to replace the contribution and subsequent scandal with. How can an individual or institution move forward from such revelations, and how can it ensure that unwanted contributions and contributors are no longer empowered to gather respect and clout? By practicing **erasure** and **revisionism**, the final steps in the Abuse and Misogynoir Playbook *(see steps 4 & 5 in the diagram)*, perpetrators erase and silence the contributions and legacy of the contributors by supplanting their contribution with a new or revised narrative, which all but ensures that only their desired narratives will be remembered, and everything else forgotten.

**Erasure** aims to delete the original contribution, silence the contributor, and stigmatize her so that she does not gain a platform or respect elsewhere. In Dr. Gebru's case, she and others at Google were alarmed by how quickly those at the company aimed to [pivot its messaging](#), move forward, and erase the fact that Dr. Gebru had ever been the co-lead — not just a member, but a leader — of its Ethical AI program. **Revisionism** is a natural next step: by providing a new narrative of what happened, confused employees and the public at large are given an opportunity to move away from the discomfort and distress of grappling with injustice, and to move on with their lives.

A final example of the impact of **erasure** and **revisionism** can be shown by the life and work of Zora Neale Hurston (1942), a literary giant who grew up in Eatonville, Florida, in the first [incorporated Black municipality in the United States](#), just a few miles from where co-author Danielle Wood spent her childhood. Zora Neale Hurston was posthumously ['rediscovered'](#) by Alice Walker as a literary genius and is now lauded as a talented [anthropologist, playwright,](#)

The State of AI Ethics, January 2021    27

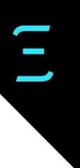

folklorist, novelist and poet. One of her books, "Barracoon", was published in 2018, many years after her death in 1960 (Hurston 2018). Zora tells the story in her own letters and autobiography of the long journey to the publication of the book (Hurston 2006). She spent several years collecting interviews from a man named Kossola who had survived capture and enslavement. Zora describes in the book all the times she visited Kossola and her methods for interviewing him. She was not able to find a publisher after compiling the interviews into a manuscript. In the preface to the book, Deborah Plant, the editor of the newly published edition cites Zora's own letter to her sponsor stating that one publisher told her they would not publish the book unless Zora changed the quotes from Kossola from "dialect" to what they called "language" or the mainstream white form of English used more commonly in written publications. Zora refused to make that change. Although Hurston was able to publish multiple books during her lifetime, she died in poverty and was only recognized much later for her strong literary contributions.

The cost of **erasure** and **revisionism** is a loss of knowledge: by erasing the original contributor and her **contributions**, it ensures that knowledge is not passed down and shared intergenerationally. Additionally, it serves to cast doubt on the contributor in the event that she is "rediscovered" — as in the case of Hurston.

**Erasure** and **revisionism** can impact individuals, such as Dr. Gebru and Zora Neale Hurston, and they can also impact entire events and swaths of history. These mechanisms are why, for example, the US does not teach that Abraham Lincoln tried for years to buy out Southern slaveholders with compensated and phased emancipation in order to avoid a Civil War (Mullen and Darity 2020). Or similarly, why the story of the Tulsa race massacre — by historical accounts an organized domestic terrorist attack against well-resourced African Americans and their neighborhood — has only recently become well-known in the mainstream, despite being such a significant event (Johnson 2020). Erasure and revisionism have the insidious effect of making victims of injustice seem alone, deranged, or partisan in their struggles, rather than part of a centuries-long struggle for liberty and self-determination. These are themselves forms of **gaslighting**. **Erasure** and **revisionism** is why children in many states are taught that the Civil War is the War of Northern Aggression (Anderson 2013), and why some Americans believe today that unarmed Black individuals like Trayvon Martin or Breonna Taylor deserved to be murdered.

The ability to form a new narrative is powerful; especially when the truth reveals a systemic and metastatic rot.

**Breaking Out of the Playbook**

As should be clear by now, this cycle of abuse of Black women is not new. Dr. Gebru's story is not so much personal as it is *Playbook* – a classic example of how those in power cannot make space for truth bearers when the truth threatens the very basis of their power. The playbook is plucked right from history, as we have shown by the stories of Ida B. Wells, Phillis Wheatley, Harriet Jacobs, and Zora Neale Hurston. The playbook continues to be wielded today: during her run for governor of Georgia Stacey Abrams had to face down multiple smear campaigns, including [attempts to paint her as a gerrymanderer](#) or as an extremist for [burning a Confederate flag](#) while in college (to which we say "good work, young Abrams"). These attempts at **discrediting** Abrams didn't work, and she was very close to winning, so her opponent resorted to **erasure** through [overt, but legal, voter suppression](#).

And then there is the case of the Pulitzer prize-winning project *1619* by journalist Hannah Nikole Jones which centers American history around the development of slavery. Princeton historian Sean Wilentz wrote [a public letter](#) with other senior historians against the project. Their letter, ostensibly about "facts", but in fact about scholarly minutiae, uses the playbook tactics of **discrediting and dismissal** to undermine the work of *1619*. To his credit, New York Times executive editor Jake Silverstein roundly [dismisses the dismissal](#), but doesn't actually identify what is happening in this letter. It is a public letter from prominent white historians about the work led by a Black woman and centering Black experience. It is evidence that elite white people in power cannot tolerate the idea that the world as they know it – professionally and personally, historically and presently – is riddled with white supremacy and that they themselves may in fact be the chief conduits. The thought is intolerable. Thus, they devolve to the Playbook.

In this sense, there is also nothing special about the domain of AI and tech, other than the fact that it is an industry exceptionally dominated by elite, white, Anglo, Christian, heterosexual, cisgender men from the Global North. This is a notably infinitesimal slice of the global population which possesses an outsized proportion of the world's wealth and power. As D'Ignazio & Klein describe in their book *Data Feminism*, tech and AI have a particularly acute *privilege hazard* problem (D'Ignazio & Klein 2020). What this means is that the Abuse and Misogynoir Playbook gets pulled out sooner and more often and with more impunity than in other industries. We see this in Google's brazen firing of Dr. Gebru as well as their surprise at the angry and widespread reaction from their own employees and the broader AI community. More than 1100 Google employees and 4300 researchers signed [a statement demanding accountability from Google](#). Recruiting firm HBCU 20x20 [dropped Google](#) as a client, stating "We cannot morally move forward with a company who will blatantly disregard the concerns of the people. Black people." Nine members of Congress wrote [a letter to Google demanding](#)

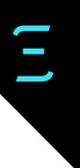

answers for the firing.

We see the Abuse and Misogynoir Playbook in action in Amazon's 2019 campaign to **discredit** Joy Buolamwini and Deborah Raji's research into their Rekognition product used by law enforcement (Raji and Buolamwini 2019; Raji, Gebru, Mitchell, Buolamwini, Lee and Denton, 2020). In their audit of facial technologies described in the New York Times, Buolamwini and Raji demonstrated that Rekognition exhibits significant gender and racial bias. Amazon had a senior level executive respond with false assertions that Buolamwini and Raji had refused to share their training data (**discrediting**), confused facial analysis and facial recognition (**dismissal**) and thus their research was flawed (**devaluing**). Alvaro Bedoya, founding director of Georgetown's Center on Privacy & Technology called these tactics "trademark ignore/deny/attack." In response, Buolamwini had to Black-woman-splain back to Amazon in a first statement and then a second statement how error-ridden gender classification in face detection might spill over into error-ridden gender classification in face recognition and what the effects might be for Black people and women. Moreover, more than seventy artificial intelligence researchers, including a Turing Award winner, came forward to author a statement in support of Buolamwini and Raji's research.

There is a pattern here. When brilliant Black women expose misogynoir and speak truth to power, why does it take a great assembly of multi-racial professionals to defend their claims? Let us think of the many Black women who have been silenced for speaking their truths because they did not have access to accredited professionals, to white institutions, to the written word. We do not know their stories because they have been silenced by the Playbook. As Dr. Gebru tweeted in Dec 2020, "It's about a pattern that many people from specific backgrounds who do not have my platform, visibility or support experience in silence and have experienced at Google. So it's much bigger than the personal harm to me — which is why I noted that a lot of Black women are speaking up."

We call on the AI ethics community to take responsibility for rooting out white supremacy and sexism in our community, as well as to eradicate their downstream effects in data products. Without this baseline in place, all other calls for AI ethics ring hollow and smack of DEI-tokenism. This work begins by recognizing and interrupting the Playbook – along with the institutional apparatus – that works to **disbelieve**, **dismiss**, **gaslight**, **discredit**, **silence** and **erase** the leadership of Black women. This work continues by transforming our labs and institutions from cultures of white supremacy towards cultures of racial justice. This work is hard when we work in white-dominant institutions and yet it is still possible. As co-authors, we draw inspiration from the liberatory work and brilliance of many people who counter the Abuse and Misogynoir Playbook, such as Keeanga-Yamahtta Taylor, Hannah Nikole Jones, Tressie McMillan Cottom, Cathy O'Neil, Safiya Noble, Ruha Benjamin, Mikki Kendall, Yeshimabeit Milner, Virginia

Eubanks, Meredith Broussard, and Tawana Petty. Within our MIT community, we are thankful to strive alongside colleagues such as Charlotte Braithewaite, Erica Caple James, Helen Elaine Lee, Dayna Cunningham & Colab, Taina McField, Ceasar McDowell, Ekene Ijeoma, D. Fox Harrell, Michel deGraff, Craig Steven Wilder, Melissa Nobles, Amah Edoh, Sasha Costanza-Chock, Delia Wendel, Karilyn Crockett, Devin Michelle Bunten, Eric Huntley, Sarah Williams, Mariana Arcaya, Dasjon Jordan, Géraud Bablon, Kevin Lee, the urban planning students who organized the "Black DUSP Thesis," Alexis Hope, Randi Williams, Joy Buolamwini and the Algorithmic Justice League, all the instructors and staff in Women and Gender Studies, African and African Diaspora Studies as well as the Consortium for Graduate Studies in Gender, Culture, Women, and Sexuality. We know there are many more colleagues working towards justice whom we do not know personally yet and we uplift your work here as well.

At our own research labs at MIT we do not hand-wring about the "pipeline" and wait for our chief diversity officers to deal with racism and sexism. We lead research labs that have made racial justice a central commitment. Within the Space Enabled Research Group at the MIT's Media Lab, Professor Wood and Dr. Turner pursue a mission to "Advance Justice in Earth's Complex Systems Using Designs Enabled by Space." Research projects within Space Enabled include activities to apply current space technology in support of societal needs — such as coastal resilience in Indonesia, Brazil and Benin — and efforts to design next-generation space systems. Dr. Turner and Prof. Wood also collaborate to develop theories and methods that technologists can use to infuse Intersectional Antiracism (Kendi 2019) into the inputs and outputs of complex sociotechnical systems (DeWeck 2011), such as nuclear and space (Wood 2019) infrastructure. Drawing from this research, Professor Wood teaches courses for undergraduates and graduate students to learn how to apply Critical Race Theory, Feminism (Taylor 2017) and Anticolonial Thinking (Wood 2020) in their work as engineers, designers, architects and artists. Professor D'Ignazio's Data + Feminism Lab at MIT designs technology with the explicit purpose of creating racial and gender equity in data-driven systems as well as physical spaces and places. In the project Data Against Feminicide, she and collaborators are building AI systems to support civil society efforts to monitor and challenge gender-based, racialized violence. The D+F Lab prioritizes the creation of a welcoming community for BIPOC, queer and women students, who have recently been working on auditing monuments, streets and place names for equity. D'Ignazio helped support the DUSP Racial Justice Teach-in last fall and is working together with faculty to integrate a racial justice lens into the Geographic Information Systems (GIS) spatial analysis class, a core requirement in the urban planning curriculum.

In the midst of Dr. Gebru's firing, a person tweeted to her that she'd be able get a job "just about anywhere" and to "put your resume in." Dr. Gebru pushed back. "For me this is not just about me getting a job but changing a toxic field that has been toxic to others like me for too long." It has been far too long that the Abuse and Misogynoir Playbook has been used to silence Black women speaking truth to power. It's time to write a new playbook in AI Ethics built on respecting, acknowledging and supporting the liberatory work of Black women.

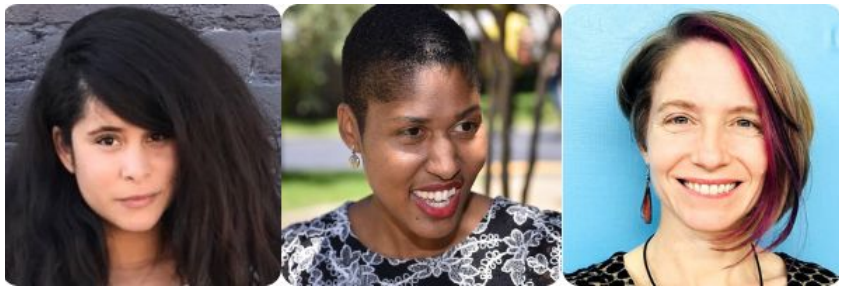

*(from left to right)*

**Katlyn Turner** (@KatlynMTurner) is a Research Scientist within the Space Enabled research group. In that role, her primary research includes work on inclusive innovation practices, and on principles of anti-racist technology design. She additionally mentors students, works on proposal writing efforts, and helps to communicate the team's work. Dr. Turner earned her PhD in Geological Sciences from Stanford University, where she researched novel nuclear waste forms. From 2017-2019, Katlyn was a postdoctoral fellow at the Project on Managing the Atom & the International Security Program at Harvard Kennedy School's Belfer Center for Science & International Affairs, where she researched environmental and socio-political impacts of nuclear energy. Dr. Turner additionally holds an M.S. in Earth & Environmental Sciences from the University of Michigan, and a B.S. in Chemical & Biomolecular Engineering from the University of Notre Dame. Dr. Turner is passionate about issues of diversity, justice, inclusion, and accessibility within society — particularly in higher education and within STEM employment sectors.

**Danielle Wood** (@space_enabled) serves as an Assistant Professor in the Program in Media Arts & Sciences and holds a joint appointment in the Department of Aeronautics & Astronautics at the Massachusetts Institute of Technology. Within the Media Lab, Prof. Wood leads the Space Enabled Research Group which seeks to advance justice in Earth's complex systems using designs enabled by space. Prof. Wood is a scholar of societal development with a background that includes satellite design, earth science applications, systems engineering, and technology policy. In her research, Prof. Wood applies these skills to design innovative systems that harness space technology to address development challenges around the world. Prior to serving as faculty at MIT, Professor Wood held positions at NASA Headquarters, NASA Goddard Space Flight Center, Aerospace Corporation, Johns Hopkins University, and the United Nations Office of Outer Space Affairs. Prof. Wood studied at the Massachusetts Institute of Technology, where she earned a PhD in engineering systems, SM in aeronautics and astronautics, SM in technology policy, and SB in aerospace engineering.

**Catherine D'Ignazio** (@kanarinka) is an Assistant Professor of Urban Science and Planning in the Department of Urban Studies and Planning at MIT. She is also Director of the Data + Feminism Lab which uses data and computational methods to work towards gender and racial equity. D'Ignazio is a scholar, artist/designer and hacker mama who focuses on feminist technology, data literacy and civic engagement. She has run reproductive justice hackathons, designed global news recommendation systems, created talking and tweeting water quality sculptures, and led walking data visualizations to envision the future of sea level rise. With Rahul Bhargava, she built the platform Databasic.io, a suite of tools and activities to introduce newcomers to data science. Her 2020 book from MIT Press, Data Feminism, co-authored with Lauren Klein, charts a course for more ethical data science.

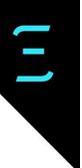

# 2. Algorithmic Injustice

**Opening Remarks** by Blakeley H. Payne (Independent researcher)

For me, and I imagine for many of you reading this, the issue of algorithmic bias was the doorway through which I entered the AI ethics space. In 2016, Joy Buolamwini's TED Talk, *Code4Rights Code4All,* was the first time I learned the phrase "algorithmic bias," and reading Cathy O'Neil's book *Weapons of Math Destruction* inspired me to apply and attend graduate school. When working on my graduate thesis, an AI + Ethics curriculum for middle school students, I knew it would not be complete without at least one lesson on algorithmic bias.

In a world full of techno-optimism, I find that sharing stories of algorithmic bias can be a powerful rhetorical tool in the sometimes-uphill battle of illustrating that technical decisions have social consequences. Highlighting how technical decisions - such as the curation of training datasets or benchmarks, or choosing different definitions of fairness to optimize for - impact a person's ability to get a job, access to education, welfare benefits, or even due process, make it clear that technology is inherently political. Highlighting the impact of technical decisions can also act as an invitation for technologists to consider how they might use their skills and power to bend technology more toward justice. However, what often gets lost in this opening gambit is that technical solutions are not the only tool in the technologist's toolbox.

In the same way that examples of algorithmic bias were my entry point to the AI ethics space, it feels familiar to me that the publications highlighted in this first chapter are also each concerned with algorithmic bias. But what I love about this set of articles is that they use algorithmic bias as an invitation to reflect on infrastructure and processes - both within our community and beyond - which enable algorithmic injustice.

As the research publications featured in this section discuss, those of us in the AI Ethics community are not only charged as a community to carefully consider the technology we are building, but also to consider the culture and norms we're building together. How do we ensure that conversations about mathematical parity don't distract from advancing true social justice? How do we construct data collection practices that resist hegemony? Beyond our community, how do we seek out and respect domain experts with different ways of knowing than us?

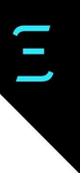

The news articles summarized in this chapter also highlight something important: that no amount of work we do to improve the AI pipeline, algorithms embedded in oppressive systems will continue to be oppressive. There is no hope for "fairer" facial recognition technologies while the systemically racist institution of policing and imprisonment continues to exist. It is thus our job as a community to make space for systemic analyses, communicate it to those who might champion techno-solutionism, and stand in solidarity with community organizers who seek to abolish these oppressive systems.

At the end of this chapter, I must admit that I do not expect criticisms of how algorithmic bias is often framed in light of techno-solutionism to go away soon. As long as we are still doing work to convince others that technology is inherently political, I believe we will continue to see algorithmic bias framed as a problem impacted and "solved" first and foremost by technical decisions, and we will have to repeat the importance of structural, social analysis.

Looking forward, I do not expect the conversation around how we should frame algorithmic bias to go away soon either. As a community, we will have to repeat the importance of social analysis and encourage each other to radically reimagine anti-oppressive, or even liberatory, social structures.

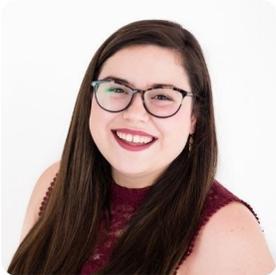

**Blakeley H. Payne (@blakeleyhpayne)**
Independent researcher

Blakeley H. Payne (she/her) is an independent researcher, writer, and consultant interested in how we might promote justice in an increasingly algorithmic world. She holds a master's degree from the MIT Media Lab, where she developed the first K-12 curriculum on the ethics of artificial intelligence. Previously, Blakeley worked at Adobe, Inc. as a machine learning researcher, and earned her Bachelor of Science in Computer Science and Mathematics at the University of South Carolina in 2017. She tweets, perhaps too often, at @blakeleyhpayne.

# Go Deep: Research Summaries

### Bring the People Back In: Contesting Benchmark Machine Learning
([Original paper](#) by Emily Denton, Alex Hanna, Razvan Amironesi, Andrew Smart, Hilary Nicole, Morgan Klaus Scheuerman)
(Research summary by Alexandrine Royer)

In recent years, industry and non-industry members have decried the prevalence of biased datasets against people of colour, women, LGBTQ+ communities, people with disabilities, and the working class within AI algorithms and machine learning systems. Due to societal backlash, data scientists have concentrated on adjusting the outputs of these systems. Fine-tuning algorithms to achieve "fairer results" have prevented, according to Denton et al., data scientists from questioning the data infrastructure itself, especially when it comes to benchmark datasets.

The authors point to how new forms of algorithmic fairness interventions generally center on the parity of representation between different demographic groups within the training datasets. They argue that such interventions fail to consider the issues present within data collection, which can involve exploitative mechanisms. Academics and industry members alike tend to disregard the question of why such datasets are created. Factors such as what and whose values are determining the type of data collected, in what conditions are the collection being done, and whether standard data collection norms are appropriate often escape data scientists. For Denton et al., data scientists and data practitioners ought to work to "denaturalize" the data infrastructure, meaning to uncover the assumptions and values that underlie prominent ML datasets.

Taking inspiration from French philosopher Michel Foucault, the authors offer the first step in what they termed the "genealogy" of machine learning. For a start, data and social scientists should trace the histories of prominent datasets, the modes of power as well as the unspoken labour that went into its creation. Labelling within datasets is organized through a particular categorical schema, but it is seen as widely applicable, even for models with different success metrics. Benchmarking datasets are treated as gold standards for machine learning evaluation and comparison, leading them to take on an authoritative status. Indeed, as summarized by the authors, "once a dataset is released and established enough to seamlessly support research and development, their contingent conditions of creation tend to be lost or taken for granted."

The State of AI Ethics, January 2021     37

Once datasets achieve this naturalized status, they are perceived as natural and scientific objects and, therefore, can be used within multiple institutions or organizations. Publicly available research datasets, constructed in an academic context, often provide the methodological backbone (i.e. infrastructure) for several industry-oriented AI tools. Despite the disparities in the amount of data collected, industry machine learners will still rely on these datasets to undergird the material research in commercial AI. Technology companies treat these shifts as merely changes in scale and rarely in kind.

To reverse the taken-for-granted status of benchmark datasets, the authors offer 4 guiding research questions:

1. How do datasets developers in machine learning research describe and motivate the decisions that go into their creation?

2. What are the histories and contingent conditions of the creation of benchmark datasets in machine learning? As an example, the authors offer the case of Henrietta Lacks, an Afro-American woman whose cervical cancer cells were removed from her body without her consent before her death.

3. How do benchmark datasets become authoritative, and how does this impact research practice?

4. What are the current work practices, norms, and routines that structure data collection, curation, and annotation of data in machine learning?

The research questions offered by Denton et al. are a good start in encouraging machine learners to think critically as to whether their dataset is aligned with ethical principles and values. Any investigation into the history of science will quickly reveal how data-gathering operations are often part of predatory and exploitative behaviours, especially towards minority groups who have little recourse to contest these practices. Data science should not be treated as an exception to this long-standing historical trend. The creators of data collection should merit as much ethical consideration as the subjects that form this data. By critically investigating the work practices of technical experts, we can begin to demand greater accountability and contestability in the development of benchmark datasets.

## Algorithmic Bias: On the Implicit Biases of Social Technology

([Original paper](#) by Gabbrielle M Johnson)
(Research summary by Abhishek Gupta)

The paper investigates how there are similarities between human cognitive biases and algorithmic biases and how this might provide some clues in helping us combat these ethical issues in the systems. One of the things highlighted in the paper is the proxy problem, where there are other attributes in the dataset that strongly correlate with sensitive or protected information and even upon removing them, the biases reflected in the proxies are still present and lead to unjust outcomes.

The paper makes the argument that biases exist everywhere around us where there is induction putting them into the following categories: "biases can be cognitive, algorithmic, social, non-social, epistemically reliable, epistemically unreliable, morally reprehensible, or morally innocuous." The way our thoughts are captured and represented in online datasets form the basis for the biases that get represented in the systems, for example, people searching for things like "Is my son gifted?" more often than "Is my daughter gifted?" gets captured in language models that use training data from such searches.

Statistical regularities when paired with the utilization of even neutral technologies can manifest biases since ultimately they are a way of surfacing both implicit and explicit patterns from the data. The paper utilizes a toy use-case with the k-nearest neighbor algorithm to show how even innocuous conceptions of a use-case can quickly create biases that are wholly unintended by the creators of the system.

One of the places where the paper shines is in giving quite accessible examples that illustrate the problems when people make inferential jumps, be they intended or unintended, that lead to the codification of biases in algorithmic systems. Take the case for the common societal perceptions that the elderly are bad with technology; while it may be true that it is the case for some subset of that demographic, generalizing that assumption to others without verification and other information can lead to patronizing behaviour towards them that is unwarranted. Many other such instances are manifested through the errors that might seep in from those who are labelling the data when they might believe in some of stereotypes and allocate a label that the elderly are bad with technology even when there isn't evidence to support that. Perhaps, this then raises the question on whether there are unproblematic datasets and if not then by the adage of garbage-in, garbage-out, we're bound to a fate where such systems will continue to produce problematic outputs. When thinking about implicit bias, in the case of the algorithmic system, one can argue that this is a case of codification, whether through errors or intentionally, and hence the system behaves quite in the same way as humans do with their own cognitive biases.

As mentioned in this paper, given that biases are inevitable when considering induction, achieving a normatively neutral notion of bias where we regard bias as problematic would then make induction impossible and limit the usefulness of some of the more advanced techniques that we employ in predictive systems. The paper provides an interesting example on proxy variables that eschews the common race proxy codified in the zip codes in the US. Specifically, the author mentions a case of how watchers of Fox News are somehow perceived to be bad with technology, building on the previous example, not realizing that even though the person doesn't have any biases against those who watch Fox News, or the elderly, because the viewership of Fox News skews towards the elderly, there is a risk of transferring and encoding that in the judgements made about them in terms of their ability to use technology.

Another example demonstrates how biases can arise even when the sensitive or protected attributes are stripped from consideration; looking at historical presidential candidates in the US and mapping them onto a 2D space with skin tone on one axis and the propensity to dress in a feminine manner on another, even without codifying the sensitive attributes into the training, someone who doesn't fit the past patterns will be deemed to be "unfit" to become President of the United States. To counteract this problem, the author suggests finding counterexamples that beat the expressed stereotypes and embed them into the dataset allowing for a more holistic representation within the data which can then be picked up by the system. While there are large systemic issues that need to be addressed when trying to find such counterexamples, it at the very least bolsters the case for why representation matters.

Yet another example covered by the author talks about hiring in academia and how one might rely on publication records as an indication of a candidate who is worthy. In the field of philosophy, what has been found is that women represent a much lower percentage of published authors than their actual presence in the field. Now, one might argue to remove this attribute from consideration given that it has a negative impact on women in the hiring process but then we run into the issue of being able to find another attribute that can help us make decisions when hiring faculty and it is not clear if there is something else that can substitute for it.

Ultimately, the paper concludes that the study of biases in both humans and machines will lead to a better understanding in both domains and at least make us realize that purely algorithmic interventions to address machine bias will be futile efforts.

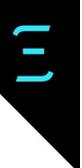

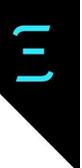

### Language (Technology) is Power: A Critical Survey of "Bias" in NLP
([Original paper](#) by Su Lin Blodgett, Solon Barocas, Hal Daumé III, Hanna Wallach)
(Research summary by Falaah Arif Khan)

The authors ground their analysis in the recognition that social hierarchies and power dynamics deeply influence language. With this in mind, they make the following recommendations for future scholarship on Bias in Natural Language Processing (NLP). They implore researchers to engage with relevant literature outside of the technical NLP community, in order to better motivate a deeper, richer formalization of "bias" - it's sources, why it is harmful, in what ways and to whom. They also underline the importance of engaging with communities who are most affected by NLP systems and to take into account their lived experiences.

Their critical survey on recent scholarship demonstrates that perspectives that reconcile language and social dynamics are currently lacking. They find that most papers contain poorly motivated studies that leave unstated what algorithmic discrimination even entails or how it contributes to social injustice. This is further exacerbated by papers that omit normative reasoning and instead focus entirely on system performance. When motivations are enumerated in papers, they often remain brief and overlook an exposition on what type of model behaviors are deemed as harmful or 'biased', in what ways do these behaviors cause harm and to whom do they inflict harm. In the absence of a strong, well-articulated motivation for studying bias in NLP, papers on the same task end up operating with different notions of "bias" and hence take different approaches to mitigating this "bias".

With opposing notions of "bias", scholars tend to treat "bias" that is inherently representational (the model represents certain social groups less favorably than others) as allocational (discriminatory allocation of resources to different groups) and so authors tend to incorrectly treat representational norms as problematic only due to the fact that they can affect downstream applications that result in allocations.

In terms of shortcomings of techniques used to study "bias" in NLP, the paper identifies a lack of engagement with relevant literature outside of NLP, a mismatch between motivation and technique, and a narrow focus on the sources of bias.

With these limitations of existing scholarship in mind, the authors propose a fundamental reorientation of scholarship on analysing 'bias' in NLP towards the question: How are social hierarchies, language ideologies and NLP systems co-produced? Language is a tool for wielding power and language technologies play a critical role in maintaining power dynamics and enforcing social hierarchies. These dynamics influence every stage of the technological lifecycle and hence scholarship focused only on algorithmic interventions will prove to be inadequate.

The authors also validate their recommendations using a case study on African-American English (AAE). They explain how models such as toxicity detectors that perform extremely poorly on AAE perpetuate social stigmatization of AAE speakers. The case study drives home the authors' point that analysis of 'bias' in such a context cannot be limited to merely algorithmic analyses, without taking into account the underlying systemic and structural inequalities.

The authors conclude with an open call to the scientific community, reiterating the need to unite scholarship on language with scholarship on social and power hierarchies.

## Fairness Definitions Explained

([Original paper](#) by Sahil Verma, Julia Rubin)
(Research summary by Abhishek Gupta)

**Statistical definitions**
- Basic
    - **True Positive (TP):** when the predicted outcome and the actual outcome are both in the positive class
    - **True Negative (TN):** when the predicted outcome and the actual outcome are both in the negative class
    - **False Positive (FP):** when the predicted outcome is positive but the actual outcome is negative.
    - **False Negative (FN):** when the predicted outcome is negative but the actual outcome is positive.

- Intermediate
    - <u>Definitions looking at predicted outcomes:</u>
        - **Positive Predictive Value (PPV):** Divide the correct positive predictions that you made by all the cases where you predicted positive outcomes. This is also called precision (a term that you will encounter frequently in the field)
        - **Negative Predictive Value (NPV):** Similar to the one above, divide the correct negative predictions that you made by all the cases where you predicted negative outcomes. *You can choose to just remember the previous definition and you'll know that this is just the case for the negative classes.*

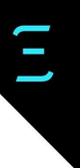

- **False Discovery Rate (FDR):** This is the probability of false (or wrong) acceptance.
- **False Omission Rate (FOR):** This is the probability of wrongly rejecting someone. *Again, here you can choose to just remember the previous one and know that this is the opposite case.*

○ Definitions looking at actual outcomes
- **True Positive Rate (TPR):** This is the probability of identifying a truly positive subject as such. This is also called sensitivity or recall.
- **True Negative Rate (TNR):** This is the probability of identifying a truly negative subject as such. Again, here you can choose to just remember the previous one and know that this is the opposite case.
- **False Positive Rate (FPR):** This is the probability of false alarms, that is falsely accepting a negative case.
- **False Negative Rate (FNR):** This is the probability of getting a negative result for actually positive cases.

**Definitions based on predicted outcomes**

Let's give a quick definition for protected attributes: these are attributes which are sensitive and are the subject of potential discrimination. Someone who belongs to the protected group is the one who has positive values for this attribute, and the unprotected group is the one that has negative values for this attribute.

- **Group Fairness:** Members from both the protected and unprotected group have the same probability of being assigned to the positive predicted class. In the case of the credit example, this means that regardless of gender, you have an equal probability of being assigned a good credit score. *Also called equal acceptance rate, benchmarking, and statistical parity (not to be confused with the following definition).*
- **Conditional Statistical Parity:** Extending the previous definitions, this definition allows for the inclusion of a group of legitimate factors that won't be considered discriminatory when included in the decision making process. So, controlling for these factors, the groups should have equal probabilities of landing with positive predicted values.

**Definitions based on predicted and actual outcomes**
- **Predictive Parity:** The fraction of correctly predicted positive cases should be the same across the protected and the unprotected group. *Also called outcome test.*

- **False positive error rate balance:** Getting similar results for both groups when they actually belong to the negative class. *Also called predictive equality.*
- **False negative error rate balance:** The probability of a subject in a positive class to have a negative predictive value should be the same across the groups. *Also called equal opportunity.*
- **Equalized odds:** Combining the previous two definitions, this definition tells us that, across the two groups, we should get a similar classification for those with actual positive outcomes and actual negative outcomes. *Also called disparate mistreatment and conditional procedure accuracy equality.*
- **Conditional use accuracy equality:** This definition gives equivalent accuracy, across both groups, for both positive and negative predicted classes.
- **Overall accuracy equality:** Across both groups, this definition gives equal prediction accuracy for those who are in the positive and negative classes to be correctly assigned to those classes. A thing to note here is that we are assuming that we value equally true positives and true negatives which might not always be the case.
- **Treatment equality:** Across both groups, this definition looks that the ratio of errors made is the same (i.e. the ratio of false positives to false negatives).

**Definitions based on predicted probabilities and actual outcomes**
- **Test fairness:** Given the probability score from the classifier, across both groups, this definition checks if the subjects have equal probability of truly belonging to the positive class. *Also called matching conditional frequencies and calibration.*
- **Well-calibration:** This extends the previous definition by saying that not only should the probability of truly belonging to the positive class is the same across the groups but that it is also equal to the predicted probability score. The reason to do so is that we are trying to *calibrate* the classifier by checking if it indeed meets the base rate prevalence for truly belonging to the positive class as it predicts.
- **Balance for positive class:** The subjects constituting the positive class from both groups have equal average predicted probability scores.
- **Balance for negative class:** This is the flipped version of the previous definition.

**Similarity-based measures**
- **Causal discrimination:** Satisfying this definition means that if all the attributes except the *protected attribute* are the same, then we should get the same predicted outcomes across both groups.
- **Fairness through unawareness:** This definition is essentially the same as the previous one, except that we don't include the *protected attributes* in making the decisions.

- **Fairness through awareness:** Expanding the previous two definitions, similar subjects should have similar classification.

**What are causal graphs?**

They are basically a way to map causal (cause-effect) reasoning and expressing them as nodes and edges helps to understand how one thing influences another.

**What is a proxy attribute?**

In the context of a causal graph, it is an attribute that can be used to derive the values in another attribute.

- **Counterfactual fairness:** If the predicted outcome in the graph is not dependent on any of the descendants of the protected attribute.

**What is a resolving attribute?**

An attribute that is influenced by the protected attribute but in a non-discriminatory way.

- **No unresolved discrimination:** If the only paths from the protected attributes to the predicted outcome in the causal graph are through a resolving attribute, then we satisfy this definition.
- **No proxy discrimination:** We satisfy this definition if there is no path from the protected attribute to the predicted outcome that is blocked by a proxy attribute.
- **Fair inference:** We classify the different paths from the protected attribute to the predicted outcome as legitimate or illegitimate. Some attributes even if they are proxies might be considered legitimate in making a prediction and hence the path could be classified as legitimate. We satisfy this definition then if there are no illegitimate paths in the causal graph.

*For the above 3 definitions, we strongly recommend looking at the graphic in the original paper and then working through the definition description in the paper as they are slightly complicated and require a couple of pass-throughs to grasp.*

Conclusion
- Although statistical definitions are easy to measure, we need to have verified, measured outcomes. If those are not available, there is a problem. That is, even though we have a certain distribution for the training data, there is no guarantee that it holds for the real-world data on which the system will operate.

- The causal definitions and similarity-based metrics require expert input and are hence subject to expert bias.
- For some definitions like *fairness through unawareness* it requires that we are able to find individuals who differ in some attributes but are identical in others. This is not always (in fact quite rarely!) possible in the real-world and hence limits the viability of these approaches.

# Go Wide: Article Summaries (summarized by Abhishek Gupta)

### 'F*ck The Algorithm'? 5 Ways Algorithms Already Rule Our Lives
([Original *Politico* article](#) by Vincent Manancourt)

With the cancellation of the A-level examinations in the UK, automated systems were used to compute grades for students. Turns out that these grades boosted the scores for those who came from advantaged backgrounds and depressed the scores for those who were from disadvantaged backgrounds. In turn, protests were held that led to the rescinding of the scores allocated by the system in favor of those assigned by the teachers of those students. This article shows that this is not the only time where such egregious faults have come to light, there are many more instances of algorithmic faults within the European countries.

In particular, algorithmic systems have been used to automate things like getting loans where criteria outside of traditional evaluation metrics are used in determining eligibility. In making hiring decisions, some companies claim to utilize external social media data to get a better grip on whether the candidate would be a good fit, unfairly penalizing some in favor of others. Criminal investigations and arrests made based on inputs from algorithmic systems are subject to wide coverage and would be more than familiar to readers of the previous reports. Getting access to welfare and crossing borders is something that isn't as well discussed yet happens with alarming frequency in some of the Scandinavian and other European nations as highlighted in the article. As public awareness rises concerning where these systems might fail, we must continue to demand higher levels of transparency in the operations of algorithmic systems that are involved in making important decisions about our lives.

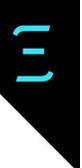

## An A.I. Training Tool Has Been Passing Its Bias to Algorithms for Almost Two Decades

([Original *OneZero* article](#) by Hayden Field)

Co-NLL 2003 is a staple in the world of Natural Language Processing (NLP). It is used to benchmark performance on Named Entity Recognition (NER) and is used as fodder to train a lot of machine learning systems that are utilized for tasks like creating knowledge panels, identifying contacts, and much more. The creators behind the Co-NLL 2003 dataset weren't thinking too much about bias 17 years ago when they worked hard to create the dataset by parsing through newswires and annotated the different entities in the text from them. Only upon deeper analysis did researchers find that there were severe problems with representation in the dataset. Specifically, male names were more frequent than female names and there was only limited consideration given to how some names are gender-neutral.

As some researchers interviewed in the article pointed out, just enhancing the representation within the dataset would also not fix the problem. There are even more issues when trying to infer gender from names since gender is something that people determine for themselves, and shoehorning them into predetermined categories from an undersampled and biased dataset exacerbates the problem. What we at MAIEI believe is that communal resources should be invested to create more representative datasets including the involvement of people from diverse backgrounds in order to embody the principle of "nothing about us, without us."

## Councils Scrapping Use of Algorithms in Benefit and Welfare Decisions

([Original *The Guardian* article](#) by Sarah Marsh)

The A-levels grades fiasco in the UK highlighted how automated systems can go wrong in a very public way. Chants on the street led to the government having to make a U-turn and get rid of the decisions that were offered by the system. This article highlights how this has triggered, and builds on a trend of, abandonment of the use of automated systems in things like welfare decisions, visa processing, and more. They point to examples in the UK where councils have now given up, for example, the use of such systems to determine if certain claims for welfare benefits are fraudulent. This led to delays for some people without their knowledge, when processing officers were looped in to do an extra review of the application. There was also the case recently when visa applications were denied disproportionately for people who were not white.

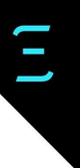

In some cases, the organizations using these systems have also realized that they don't get too many benefits in terms of speed or efficiency in using these systems. They also realized that some of the outcomes are antithetical to the very reason for their existence which has led to abandonment. But, as the article points out, there are still companies that are pitching their products and services to government agencies since these are lucrative contracts. The article argues that having greater transparency in this process along with the push for including public consultation as a part of this process might aid in alleviating some of these concerns. This idea was explored in more detail in this article titled "[Prudent Public Sector Procurement of AI Products](#)".

### Algorithms Are Automating Fascism. Here's How We Fight Back
([Original *Vice* article](#) by Janus Rose)

Facial recognition technology is unfortunately utilized in subversive ways, often without transparency, by law enforcement which is one of the drivers for the recent backlash against this technology. While there are glaring issues with bias, and other concerns, the article highlights a recent case where someone faced a siege from police officers after facial recognition was applied to their social media accounts to identify them. Yet, technology is only partly to blame here; systemic problems of racism and brutality underscore the importance of looking at these systems with a sociotechnical lens to ensure that we don't fall into the situation of analyzing problems in a unidimensional manner. These systems merely amplify the existing biases of the people and institutions that build the systems.

Pointing to an example of a proposed rule a few months ago that instructed federally run shelters to eject transwomen based on features like facial hair, Adam's apple, etc. This is particularly problematic because these features are well within the capabilities of existing machine vision systems and hence prone to automation. A project highlighted in the article called the White Collar Crime Risk Zones flipped the script on over policing in ethnic and minority neighborhoods via predictive policing by spotlighting zones where financial crimes are likely to take place to show how predictive policing leads to amplification of systemic faults.

A proposed mechanism for rebelling against these injustices is, for example, to use anti-surveillance makeup patterns or wear masks to combat facial recognition systems. The problem is that this a cat-and-mouse game where the surveillance systems learn how to circumvent such protection measures with more training data and hence don't provide a lasting solution.

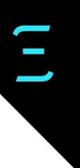

The current demand for seemingly magical solutions, based on flawed methods, will always be there which is something that the larger public and those with decision-making powers need to wake up to and realize the flaws. As the article concludes, we need to attack these problems through all vectors possible in the hopes of preventing them from realizing a dystopic future.

### Google's Autocomplete Ban on Politics Has Some Glitches
(Original *Wired* article by Tom Simonite)

Google, the default search engine for billions of people across the world, is taking steps to ensure that falsehoods and bias are as far removed as possible from search results. But, how successful that is going to be remains to be seen. For starters, one of the things that we need to consider seriously is the role that auto-complete plays in how people search for and navigate issues. The helpful dropdowns while useful when you're searching for innocuous things like patio furniture have much more serious implications when they guide and nudge you towards donating towards one campaign vs. another or to surface more negative coverage about one candidate vs. another.

While reports have been filed with Google to not skew in any one direction, as reported by Wired in this article, it is clear that for some things actions have been taken by Google but for others, that doesn't seem to be the case based on what their perception of the issue is. For example, when talking about Black Lives Matter, some factions might consider this to be a partisan issue and ask for auto-complete to be altered for it while others will talk about it in a non-partisan manner making waters murky. For now, surfacing places where there are problems and flagging them is our best approach though the experiences of everyone are not consistent because of how search results are personalized based on your own search history and interactions on the web.

### Health Care AI Systems Are Biased
(Original *Scientific American* article by Amit Kaushal, Russ Altman, Curt Langlotz)

This article is about the highly fragmented nature of the data ecosystem in the healthcare system. There are concerns about privacy given the sensitive nature of the information that is stored in these databases. A side effect of this is that each institution trying to use techniques that require large amounts of data struggle because they are unable to compile together enough data that can form representative samples and hence open up the possibility of bias.

The use of techniques like differential privacy can potentially help to mitigate some of these concerns, but there are vested interests by some healthcare providers who want to keep the data sequestered in their silos for the fear of losing business. A study mentioned in the article talks about how healthcare institutions that are more open with their data see higher rates of patient attrition compared to those that don't.

Perhaps, there is also risk mitigation and aversion that comes into play whereby administrators don't want to take on potential problems where data sharing can lead to adverse outcomes from a legal and compliance perspective. National repositories and data commons structured through trusts can be a way around some of these problems, but they require concerted efforts on the part of multiple parties before they become a reality. Most importantly, we believe that awareness on the part of patients so that they don't just accept the status quo but see data as an extension of their self can also help to push this agenda and ensure that there is more openness in sharing data and hence better chances at mitigating bias in AI systems used in healthcare.

# 3. Discrimination

**Opening Remarks** by Kathryn Hume (Interim Head, Borealis AI)

**Discrimination and bias: Building fairer models**

That artificial intelligence is capable of discrimination is clear. Equally clear is that AI can be used to reinforce discrimination and racial bias. The big question is what the AI community, business leaders, and regulators can do about it.

As the articles selected for this section of the publication clearly illustrate, discrimination in artificial intelligence (AI) is about much more than just biased data sets and models. Abebe Birhane's article, and the paper by Stephen Cave and Kanta Dihal both provide ample evidence of how historical narratives surrounding AI celebrate a kind of intelligence prized by those who have traditionally led the field: North American white males. Many of the other authors demonstrate how AI can be (intentionally or unintentionally) manipulated, influenced, or created to institutionalize bias and discrimination.

The good news is that, over the past few years, we have started to see significant discussion and debate on the topic. Awareness – amongst the developer community, business leaders, citizens, and regulators – has started to increase. This publication by the Montreal AI Ethics Institute is just one example of the force of academic and policy-led research now going into the field of AI Ethics. The more the issues around discrimination and bias in AI are discussed and debated, the better the quality of the dialogue.

**The urgency of action**

While we are seeing encouraging action in defining the challenge, the unfortunate reality is that progress on defining a set of actions that can help reduce the potential for, and impact of, discrimination and bias on AI has been slow. The gap between talk and action is growing. Yet the need for action is becoming more urgent by the day.

In part, that is because AI is becoming embedded into decisions that can have a significant influence on our lives. Consider, for example, how a credit decision made by an algorithm could impact an individual's ability to buy a new house, start a new business or pay their rent. Or how algorithms embedded in apps like YouTube Kids could influence how toddlers develop (both intellectually and socially). The more infused AI becomes in our daily lives, the more urgent the need to reduce or remove existing biases.

For organizations that operate in highly-regulated industries like healthcare and banking, the stakes are particularly high. Not only are many of the decisions made in these sectors incredibly impactful on peoples' lives; they can also be complex and better informed by data, so great problems for AI to solve.

**Making it practical**

At Borealis AI, the machine learning research institute for the Royal Bank of Canada, our mission goes beyond applying machine learning to shape the future of banking. It is also to use our privilege in the AI community to assess and address the ethical implications of AI. In our view, the two must go hand-in-hand.

Our mandate has allowed us to think more practically about how we (and other businesses) can help to progress the agenda on AI Ethics while simultaneously taking steps to reduce the potential for, and impact of, discrimination in our algorithms. And while our research has yet to result in a concrete set of universal actions for eliminating discrimination and bias in AI (if that is even possible), it has led to a number of valuable insights that could drive positive action for business leaders and the AI community in general.

1. *Understand the context:* The values that shape how we build tools in one context rarely translate perfectly into another context. As such, organizations need to take great care to understand the unique contexts in which their tools and AI models will be used and adjust accordingly. Anthropological sensitivity – a capability not often taught to AI developers – will be needed.

2. *Enhance model validation:* The best way to uncover bias in your models is to test them continuously, thoroughly, against a wide range of scenarios, using a broad set of test data, and leveraging a wide range of diverse perspectives. Model validation and governance is perhaps not the most exciting part of AI development and management, but it is likely the most important step in ensuring models are free from bias and discrimination.

3. *Focus on diversity:* As many of the articles selected for this publication point out, the AI community remains largely white and male-dominated. And that (among other factors) has led to a general bias towards white males. Businesses and governments will want to redouble their efforts to encourage greater diversity within the AI community. More diverse voices at the table will help fuel the creation of technologies that work for everyone.

4. ***Get people talking:*** Developers, business leaders, policy-makers, and – perhaps most importantly – citizens need to be more aware of the risks and positive implications of AI so they can be better prepared to ask the tough questions and manage the right risks. Developers, in particular, must be encouraged to challenge their own biases and those of the people around them. The best place to catalyze change is from within.

There is no doubt that AI has the potential to create, reinforce and exacerbate discrimination and bias. Yet it also has the potential to illuminate latent biases in society, and catalyze changes that transcend the limits of technology. But only if it is developed and overseen with eyes wide open to the risk of discrimination, and the courage to design algorithms that promote the future society we want. That is the only way we can build better models – and a better tomorrow for everyone.

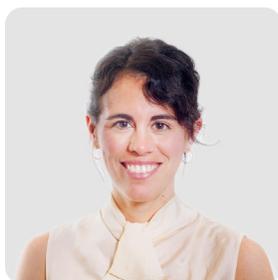

**Kathryn Hume, PhD (@HumeKathryn)**
Interim Head
Borealis AI

Kathryn Hume is passionate about building companies and products that unlock the commercial value of emerging technologies. Prior to joining Borealis AI, Kathryn held leadership positions at integrate.ai and Fast Forward Labs (Cloudera), where she helped large Fortune 500 companies apply machine learning to increase revenue and operational efficiency. A widely respected speaker and writer on AI, Kathryn holds a PhD in Comparative Literature from Stanford University. She speaks seven languages and has taught courses on enterprise adoption of AI, law, and ethics at Harvard Business School, MIT Media Lab, Stanford, and the University of Calgary Faculty of Law.

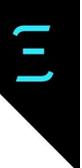

# Go Deep: Research Summaries

## Algorithmic Colonization of Africa
([Original paper](#) by Abebe Birhane)
(Research summary by Falaah Arif Khan)

Taking Facebook's bid to map out the population density of Africa as a representative example, Birhane presents evidence as to how historic patterns of conquest have begun to re-emerge. The first step is to come up with a justification for invasion and Big Tech does this under the umbrella of "Tech solutions for the developing world". They then mobilize to assert territorial claims and large volumes of data are indiscriminately "mined" from these demographics. The very notion of "Data Mining" is reminiscent of the colonial mindset that resources are for the taking. In the 21st century, people (and their private data) are the resources and are thereby treated as passive objects for manipulation by colonizers. And so, moral, social questions start to be dictated by corporate interest.

The last step is to institutionalize this domination, which Big Tech easily does by enforcing Western ideals and notions of what is "correct" and "good", implicitly through Algorithmic Decision-making Systems. Technology is increasingly invasive and encoded values start to influence the social, political, and cultural landscape of its consumers.

Birhane makes the distinction between the potential of algorithms and the effect of monopolistic interventions using algorithms. The former has the capacity to alleviate some of the most pressing problems of the African continent. Examples include: bringing about grassroots change by improving awareness about existing disparities and issues in sectors such as healthcare, agriculture, etc, and facilitating the local development of technology. Conversely, Western algorithmic invasion hinders local products and leaves the continent dependent on external software and infrastructure. Another example is micro-financing in Africa. FinTech was envisioned to be the holy grail for the poor, but it ended up being another case of the rich (stakeholders) getting richer, while the poor found themselves in perpetual debt.

Birhane explains how the effectiveness of a solution is inexplicably linked with the context in and for which it was developed. This is why an Africa-specific technological revolution is essential. As value systems change from culture to culture, so does the notion of a 'pressing' problem. Hence, solutions created in one setting for one demographic, with the value system of one culture may not transfer well into a different setting. This is evidenced in the attempted use of mammograms for early breast cancer detection in sub-Saharan Africa, which were trained on and worked well for women in the West, but did not scale to the African settings due to the differences in age profiles and access to treatment for women.

Another concern is that algorithms tend to perpetuate historical bias, discrimination, and injustice. The most vulnerable are already left out of the data and so any subsequent policy that is built on digital identities would further preclude them. Moreover, a utilitarian approach to algorithmic decision-making fails to consider solutions that help minorities. It is absolutely essential to engage with and consult these demographics while building systems but is seldom a consideration for profit-maximizing corporations.

Birhane makes a strong case for the African youth to be the ones to step in and effect change. They can be the most effective in delivering technological solutions for their continent, since they uphold the same community values, understand the problems better than anyone else, and care most deeply about effecting positive changes in their communities, something that is completely lacking in the Western approach.

## The Whiteness of AI
([Original paper](#) by Stephen Cave, Kanta Dihal)
(Research summary by Victoria Heath)

**Introduction**

Robots and artificially intelligent machines are often depicted as white. *Type in "robot" in Google Images and you'll notice it immediately*. This often unnoticed phenomenon is problematic and an example of how technology can be racialized. Authors Stephen Cave and Kanta Dihal from the University of Cambridge explain, "intelligent machines are predominately conceived and portrayed as White. We argue that this Whiteness both illuminates particularities of what (Anglophone Western) society hopes for and fears from these machines, and situates these affects within long-standing ideological structures that relate race and technology."

In this paper, the authors examine "how the ideology of race shapes connections and portrayals of artificial intelligence (AI)," contributing to an increasing number of studies and books looking at the connections between race and technology (e.g. Safiya Noble's *Algorithms of Oppression*). Their discussion is informed and framed by the philosophy of race and critical race theory, relying on Black feminist theories and work in Whiteness studies. More specifically, they utilize the "white racial frame" developed by Joe R. Feagin (2006) to examine representations of AI (e.g. white robots). By drawing attention to the "the operation of Whiteness" in technology, which is often normalized and made invisible, we can expose the "myth of colour-blindness" that is prevalent in tech culture that prevents "serious interrogation of racial framing" in technology.

The authors offer three interpretations of the Whiteness of AI: 1) normalization of Whiteness in the Anglophone West, 2) the White racial frame primarily ascribes intelligence to White people, and 3) it allows for "a full erasure of people of color from the White utopian imaginary."

**Seeing the Whiteness of AI**

Machines Can Be Racialized

The idea of a "racialized" machine or technology may seem counterintuitive (and make designers uncomfortable), but research shows that when humans are asked to define the race of a robot presented to them, they do so easily. According to Cave and Dihal, this is primarily due to the fact that machines (including robots) are often anthropomorphised. In our society, appearing more "human-like" means to have a race. This is not only achieved through the color of a robot, but also the name it's given, the voice, dialect, and more.

This racialization has significant consequences. Research has shown that people may feel more comfortable with a machine if it has the same racial identity as them. Unfortunately, this has a flip-side: people often project prejudice onto machines too. In one study, after participants watched three videos depicting a female-gendered android that appeared Black, White, and East Asian, they were asked to provide commentary. The researchers found that "the valence of commentary was significantly more negative towards the Black robot than towards the White or Asian ones and that both the Asian and Black robots were subject to over twice as many dehumanizing comments as the White robot." Further studies have shown that the patterns of bias, discrimination, and prejudice in human-to-human interactions also show up in human-to-robot interactions.

Whiteness as the Norm for Intelligent Machines

To illustrate how White is the norm for intelligent machines, the researchers look at four categories: real humanoid robots, virtual personal assistants, stock images of AI, and portrayals of AI in film and television.

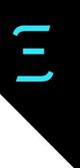

One of the most obvious (and infamous) examples of a White humanoid robot is Sophia from Hanson Robotics, but there are other examples such as Kismet, Cindy Smart Doll, My Friend Cayla, and more. When it comes to chatbots and virtual assistants, it's often the sociolinguistic markers that indicate if they are racialized as White. Mark Marino wrote in 2014 about the natural language processing program ELIZA (created in 1966), "it was "performing standard white middle-class English." Today's virtual assistants are an example of how designers often make conscious decisions to ascribe certain sociolinguistic characteristics to their technology depending on their target market, which historically (and perhaps still) skewed wealthy and White.

In regards to stock images of AI, particularly anthropomorphized robots, the dominance of white (and White) depictions is stark. This can be particularly problematic, as these images are utilized widely to illustrate this technology in the media which influences how society perceives it. Closely connected is the predominance of Whiteness of AI in film and television, examples include robots or voice assistants from the *Terminator, RoboCop, Blade Runner, I, Robot, Ex Machina, Her, A Space Odyssey*, and more. It's only recently that we've begun to see more "diverse" racial depictions, such as in Westworld and Humans.

**Understanding the Whiteness of AI**

Whiteness Reproducing Whiteness

"In European and North American societies, Whiteness is normalized to an extent that renders it largely invisible," write Cave and Dihal, and it "confers power and privilege." This has led to other colors being racialized while white is considered "an absence of color" and in technology, the "default." This has also led to what was referred to earlier as "color blindness," which further normalizes Whiteness, marginalizes other racialized groups, and dismisses or ignores the real world impacts of technology on different groups of people.

Some explanations for the Whiteness of AI have rested on the idea that the normalization of Whiteness leads designers to almost unconsciously design a machine as White. However, that doesn't explain why not all created entities are portrayed as White. Cave and Dihal point out that often in science fiction writing, for example, extraterrestrials are racialized as non-White, more specifically, East Asian. Petty-criminals are often portrayed as Afro-Caribbean. To the authors, this suggests that the racialization of AI is more a choice than an unconscious decision due to the aforementioned White racial frame.

AI and the Attributes of Whiteness

Building on the previous section, Cave and Dihal argue, "AI is predominantly racialized as White because it is deemed to possess attributes that this frame [White racial frame] imputes to White people." Those attributes are: intelligence, professionalism, and power. "Throughout the history of Western thought," they write, "intelligence has been associated with some humans more than others." This strain of thought is what bolstered colonial and imperialist campaigns under the guise of "civilizing" other nations, ethnic groups, and races (summarized by Rudyard Kipling as the "the white man's burden") and attempts to "measure" the intellectual capabilities of different racial groups (e.g. IQ test). Thus, Cave and Dihal argue, "it is to be expected that when this culture [White, European] is asked to imagine an *intelligent* machine, it imagines a White machine." Another important aspect is generality. Being generally intelligent is often attributed to White people, and seeing that "general artificial intelligence" is the primary aim of the AI field, it is not difficult to link these together. "To imagine a truly intelligent machine," they write, "one with general intelligence is therefore to imagine a White machine."

Historically, professional work (e.g. medicine, business, etc.) has also been deemed unsuitable for most people. People of color were (and still are) systematically excluded from entering these fields. In White European and North American mainstream culture, this has led to a warped perspective of *who* is a professional. When one imagines a doctor, it's more often than not a White male. Thus, as AI systems are believed to be capable of *professional* work in the future, "to imagine a machine in a white-collar job is therefore to imagine a White machine." Here, Cave and Dihal mention the issue of power as well. "Alongside the narrative that robots will make humans redundant," they write," an equally well-known narrative is that they will rise up and conquer us altogether." Interestingly, those narratives, especially in science fiction, portray the rising robots as hyper-masculine, White machines. Why? They argue, "it is unimaginable to a White audience that they will be surpassed by machines that are Black." Thus, for a White designer (or science fiction writer) to imagine these machines overpowering and supplanting humanity, they "imagine a White machine."

White Utopia

Cave and Dihal's final hypothesis regarding the Whiteness of AI is that this racialization "allows the White utopian imagination to fully exclude people of color." AI is believed to eventually lead to a "life of ease" for most people. Historically, leisure was restricted to the wealthier class of society and often came at the expense of working-class women of color who conducted the necessary labour that allowed for that leisure—often remaining "invisible" themselves due to the "dirty" nature associated with their work. Thus, in the White utopian imaginary, write Cave and Dihal, people of color would be removed altogether, "even in the form of servants." Further, this omission may even extend to all women. "Seen as a form of offspring, artificial intelligence offers a way for the White men to perpetuate his existence in a rationally optimal manner," they write, "without the involvement of those he deems inferior."

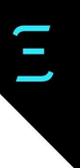

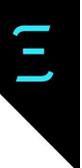

**Conclusion and Implications**

The racialization of AI as White can cause several representational harms. Cave and Dihal point out three in particular:

1. Amplify existing racial prejudices by sustaining and mirroring discriminatory perceptions that the attributes discussed in the section above (i.e. intelligence, professionalism, and power) are only ascribed to White people.
2. Exacerbate injustices by situating these machines "in a power hierarchy above currently marginalized groups, such as people of color." This may exacerbate automation bias, in which the machines (racialized White) are trusted above that of the people oppressed by them.
3. Distort society's perception of the risks and benefits of AI. For instance, the conversation around the labour impacts of AI could be framed primarily about the risks/benefits posed to White middle-class men or white collar professionals, while ignoring other groups. This could lead to policy focusing more on protecting one group while ignoring more marginalized groups.

Cave and Dihal position this paper in a larger conversation and process called *decolonizing AI*, in which scholars aim to break down "the systems of oppression that arose with colonialism and have led to present injustices that AI threatens to perpetuate and exacerbate." They hope by drawing attention to the Whiteness of AI, it can be denormalized, noticed, and raised for discussion.

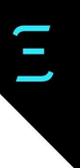

# Go Wide: Article Summaries (summarized by Abhishek Gupta)

## How the Racism Baked Into Technology Hurts Teens
([Original *The Atlantic* article](#) by Avriel Epps-Darling)

For the most part, when we talk about intersectionality in the bias discussion, we focus on the ideas of ethnic origin, geography, gender, etc. but there are fewer discussions on the intersection of race and age. This article makes that distinction and backs it up with some pretty solid arguments as to why this is an important issue.

Sustained and pervasive exposure to negative stereotypes has a noticeable influence on how we perceive ourselves. In an age where we spend a significant amount of time online, it is not surprising that racism online will play a big role in our psychological development. Especially children who are young and whose minds are still malleable, they are quite susceptible to the racism that they have to endure online. The effects are manyfold: negative impacts on sleep, self-respect, academic performance, relationships, and more.

Using the phrase *technology microaggressions*, the author illustrates that the frequency with which they occur online is much higher than that experienced by people in-person. Some of the protection measures offline like parents being able to intervene are much more limited online partially because of a lack of understanding on the parents' part in terms of how the platform algorithms work. There is no "stranger danger" with recommendation systems. Ultimately, the trauma that teens will experience in their adolescent years is going to have lasting impacts on them and hence should be addressed proactively.

## In French Daycare, Algorithms Attempt to Fight Cronyism
([Original *Algorithm Watch* article](#) by Alexandre Léchenet)

When they say that algorithms have penetrated all aspects of our lives, this isn't far from the truth. Daycare allocations, something highly prized by parents in places like Paris, France being replaced by algorithmic systems have the potential to crack open obfuscated selection criteria that are opaque and lead to cronyism. Transparency in public processes is always something that we should strive for, especially when it comes to how decisions are made about different aspects of people's lives that have the potential to affect them significantly, for example, the care of their children.

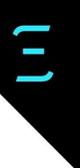

Codifying existing criteria that are used and acted upon by people who previously sat in committees to make allocations are now done in a more transparent way by the machine and utilizing something called the student optimal fair matching algorithm which ensures that no student who prefers a school to her outcome will be rejected while another student with lower priority is matched to the school.

A complaint raised by the people who previously held the positions that made this allocation on paper by hand is that the algorithm might not be able to grasp the nuances of each family that is seeking this service. But, if we can combat cronyism and offer higher levels of transparency, then the outcomes, both quantitatively and qualitatively, from the perspective of the parents should be the final driver in deciding how successful the program has been.

## How to Make a Chatbot That Isn't Racist or Sexist
(Original *MIT Tech Review* article by Will Douglas Heaven)

GPT-3 has shown tremendous capabilities, to the point that people think that it exhibits human-like intelligence which can be disorienting to understand when it comes to its limitations. But, if you've read our previous reports, you will not be surprised to learn that when you utilize such large-scale models (GPT-3 is the largest language model ever built), it comes with "internet-scale" biases, as the creators of the system like to say.

This article points out some of the problematic outputs from the system which we warn readers to take a look at since we don't want to reproduce that vile here. Given that we are solutions-oriented, the mere identification of the problem is just the first step in solving the problem. The article mentions a safe conversational AI workshop that was held recently that made some recommendations on how we can address these problems.

Specifically, they point to the bolting on of filters that can bleep out the content that is harmful (though identification of that content is in itself a challenge). Another approach is to avoid utilizing data from contentious domains like politics and religion. But, this has the risk of throwing out good training data with bad training data.

Ultimately, legislative approaches combined with technical interventions will be the ideal combination of strategies that can help mitigate the harmful effects of machine learning models gone awry. Most of all, sandboxing and testing the systems prior to release is going to be essential, a recommendation made by the participants of the workshop that tried to "adversarially" get the model to spew out harmful content, something that trolls will inflict on the system when it is released into the wild.

## The Man Who Helped Turn 4chan Into the Internet's Racist Engine
([Original *Vice* article](#) by Rob Arthur)

A horrifying story about one of the darkest corners of the internet, this article takes a deep dive into how the 4chan platform went from being a community for discussions on everyday subjects to a cesspool of hate, bigotry, and everything wrong with humanity. It documents the rise of a single moderator who now controls what passes muster on the platform and has allowed it to spiral into a forum that creates and disseminates content that actively harms people's well-being and the integrity of our democratic processes.

While other moderators on the platform have tried to rise up against him, for the most part, it has been ineffective because of the way the policies are specified and how they are enforced. For example, the rule against racist content on the platform is interpreted by this moderator as to only look at the intent of the poster and not the content. The intent is always guesswork and an easy way to exculpate people spreading hateful content. A consequence of this was that more people piled into this phenomenon and started flooding other boards on the platform with this content creating "gateways" for other users to this type of content.

This is a warning for other platforms that rely on human moderation to watch for signs of power accumulation and skewing of the moderation and review process on the platform. Small actions can have a cumulative effect taking a platform down the path from where it is impossible to bring back. 4chan now represents the worst that humanity has to offer and it has essentially been weaponized into an instrument for pushing particular political agendas by channeling the content from this unfortunately fertile ground to other platforms used more widely. The lesson to be learned from this is that technology is a powerful lever and people who are an intricate part of it do have choices that should be rightly exercised to steer the development of technology in a direction that is good for us all.

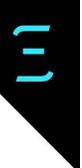

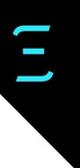

### Does Facebook Still Sell Discriminatory Ads?
**(Original *The Markup* article by Jeremy B. Merrill)**

Discrimination is a constant struggle that automated systems encounter, especially when trying to regulate content that is posted on online platforms at web-scale. As fewer humans are involved in the operations of how decisions are made with regards to what is accepted for publishing on the platform, opportunities emerge for transferring over discriminatory practices from the analog to the digital world. In this article, The Markup unearths how discriminatory ads are still run on the Facebook platform, especially as they relate to housing ads that target specific people. This is illegal as per regulations in the US owing to how landlords used this historically to exclude people from access to housing based on their race and ethnicity. The difference compared to print ads is that while those were geared to appear next to related content, we can now target people directly.

Civil rights activists argue that what Facebook does is not show users what they want, but what it thinks that they want based on the stereotypes of their demographics. As there are very strong correlations between seemingly innocuous actions and race, gender, and other demographic information that Facebook might be interpreting circumstantial evidence as expressed preferences. What has also frustrated activists is that Facebook has mentioned efforts to combat this sort of problem, but it hasn't been transparent in what it is doing exactly and what is the efficacy of those adopted measures. Finally, relying on Section 230(c) partially absolves them of their responsibility in ensuring that the content posted meets certain normative guidelines but many argue that the antiquated law needs to be adapted for the Internet Age to avoid these problems.

### The U.S. Needs a National Data Service
**(Original *Scientific American* article by Julia Lane)**

The article provides 3 primary recommendations on how a National Data Service might be created and potential benefits associated with its establishment. But before diving into that, it's important to consider why we even need such a service in place. As readers of our reports, you're no stranger to the requirement of large corpora of data to train machine learning systems. It is without a doubt an invaluable resource that nations are trying their best to structure and make available to glean insights to inform policy. There have been numerous efforts in the US for example where they have attempted to put in place requirements for policymaking to become more data-driven.

With the US Census that took place in 2020, the need for data collection has never been more front and center. The pandemic has made in-person collection hard, yet the census data collection process can't be halted because the insights from this data is used to inform all sorts of downstream decisions including representation in government. So, why do we need a national data service then? (And this applies to not just the US, but other countries too!) The article makes the case that it would help to harmonize, measure, and validate the data that is collected for example in the census to ensure its accuracy; "Trust, but verify."

The recommendations made by the article are: empowering local governments to utilize existing data in their decision making, ensure replicability and comparability across the nation, and finally ensuring privacy and confidentiality of the data collected to engender trust from people. These are crisp and provide a great starting point for countries to consider as a part of their national AI strategies, since without data, it is really hard (at least in the current paradigm of AI) to build meaningful systems.

# 4. Ethical AI

**Opening Remarks** by Mona Sloane, PhD (Fellow, NYU Institute for Public Knowledge)

2020 firmly exposed big tech's lip service to equity and justice — peaking right at the end of the year with the firing of Dr. Timnit Gebru from Google's research team. After an early enthusiasm about AI ethics, we are now, at long last, in the midst of an "AI ethics-lash": a shifting discourse that, thanks to many years of activism and scholarship, questions the often oppressive power structures that are being upheld by the broad-strokes deployment of "ethics" in AI research, practice, and policy. The pieces in this chapter show this nicely and center views from the "Global South" and "East", as well as indigenous knowledges, make "ethical concerns" specific by examining what they can mean concretely in ML research, focus on corporate power by comparing big tech to big oil, and hone in on the notion of "fairness" in AI.

As we move forward with expanding our notions of how AI affects the social, and vice versa, and how this relates to "ethics", there are a couple of important things to keep in mind.

First, "ethics'' as often promoted in AI research is grounded in moral philosophy, a Western ontology that makes universal claims and that itself is characterized by vast underrepresentations that bear striking resemblance to many STEM disciplines. We cannot talk about "ethics" without calling out the structural inequities that characterize much of philosophy (with some notable exceptions, of course, such as the work by Shannon Vallor, or new scholarship emerging in feminist philosophy) and that dominate the nomenclature of "ethical AI".

Second, there are some important terms that have found their way into the "ethical AI" discourse, but that tend to not be deployed to their full potential. The most important one is "intersectionality", a term coined by Kimberlé Crenshaw that provides a framework for understanding how intersecting identities combine discrimination and privilege. Importantly, intersectionality prioritizes lived experience and self-identification over classification. Taking intersectionality seriously in the context of "ethical AI", then, means to critically examine existing data classifications and put them into a historic context (which for some AI applications firmly points towards eugenics, such as facial recognition technology) and take as a new point of departure the expertise and self-identification of individuals and communities that are most adversely affected by AI.

And third, we should bear in mind that "ethics", first and foremost, is a social practice, it is something that we *do* — not something that *is*. Along those lines, it may be fruitful to think about "ethical AI" more along the lines of research ethics, grounded in the legacies of the Helsinki Declaration and the Belmont Report.

In 2021, I hope and expect that we can move beyond the polarization of tech-dystopia and tech-utopia, or the pronounced split between the idea that social problems can best be fixed through tech fixes, and the idea that the critique of this approach is all but productive. Moving forward we will have to develop more nuanced understandings of the intersection of AI and society, and engage in genuine interdisciplinary discourse and practice that brings together critique with careful technical approaches that can help address the adverse impacts of AI. I also expect more regulation to kick in, especially on the local level as cities and local decision makers take leadership in gaps left in federal regulation - a trend Stefaan Verhulst and I have called "[AI Localism](#)". Another hope that I have for 2021 is that we can center issues and questions of environmental justice in the AI and tech discourse and mobilize collective action to address the climate emergency.

The next hope that I have is that we can integrate the idea of generational justice into "ethical AI" and provide the next generation with the conceptual tools, and the power, they need to create the long-lasting social change they clearly long for. I would also like to see the academe more decidedly involved in the public discourse around AI and society, and take seriously the obligation scholars have to provide expertise to important questions around technology, policy and equality.

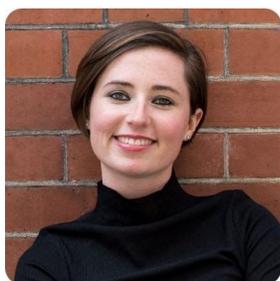

**Mona Sloane, PhD ([@mona_sloane](#))**
Fellow
NYU Institute for Public Knowledge

Mona Sloane is a sociologist working on inequality in the context of AI design and policy. She frequently publishes and speaks about AI, ethics, equitability and policy in a global context. Mona is a Fellow with NYU's Institute for Public Knowledge (IPK), where she convenes the ['Co-Opting AI' series](#) and co-curates the ['The Shift' series](#). She also is an Adjunct Professor at NYU's Tandon School of Engineering, an Affiliate of the Center for Responsible AI, and is part of the inaugural cohort of the Future Imagination Collaboratory (FIC) Fellows at NYU's Tisch School of the Arts. Mona is also affiliated with The GovLab in New York and works with Public Books as the editor of the [Technology section](#). Her most recent project is '[Terra Incognita: Mapping NYC's New Digital Public Spaces in the COVID-19 Outbreak](#)' which she leads as principal investigator. Mona is also affiliated with the Tübingen AI Center in Germany where she leads research on the operationalization of ethics in German AI startups. She holds a PhD from the London School of Economics and Political Science and has completed fellowships at the University of California, Berkeley, and at the University of Cape Town.

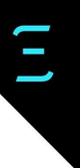

# **Go Deep: Research Summaries**

### **Corporate Logics, Silicon Valley, and the Institutionalization of Ethics**
([Original paper](#) by Jacob Metcalf, Emmanuel Moss, danah boyd)
(Research summary by Nga Than)

Google's recent firing of Timnit Gebru, a prominent AI ethics researcher, has called into question two main issues in the tech industry: its lack of diversity, and its faulty relationship with ethical considerations of technological development. Dr. Gebru represents the growing "ethics owners class of tech workers" who champion ethical causes, ethical designs, development, and deployment of technology from within the tech industry. This research article provides a conceptual framework to understand this emerging occupation and the various day-to-day struggles that these ethics professionals like Dr. Gebru are facing within the industry.

The authors employed a mixed qualitative methods approach by gathering ethnographic, textual and interviewing data. They interviewed 17 ethics professionals from different well-known companies whose formal roles have been to address ethics within their companies or within the industry. This particular role has become institutionalized after a series of public scandals such as Cambridge Analytica in the 2016 US presidential election, racial biases in facial recognition technology broke out.

Ethics owners' daily activities are to examine the social consequences of technology products. Their jobs are similar, but not the same as the business ethics, legal ethics, and sometimes PR teams. They respond to external pressures to tech companies within corporate boundaries. Sometimes they are seasoned engineers, MBA holders, trained in social sciences, or humanities. This class of tech workers are trying to change the tech industry from within by "fulfilling fundamental ethics commitments." The article defines the process of institutionalization of ethics "as a set of roles, and responsibilities," and "operationalized as a set of practices and procedures." They argue that ethics owners "operate inside a fraught dynamic." While "attempting to resolve critical external normative claims about the core logics of the tech industry," they have to do so within the corporate structure, and being embedded within corporate logics. This might lead to structural, cultural and social pitfalls.

The authors use the ethnographic approach to ethics or the "ordinary ethics approach." Instead of thinking about ethics as a set of abstract concepts, and principles, they examine "how ethics and morality structure social life," and "how everyday practices reveal the moral commitments embedded in actions." They found that "ethics as everyday practice" meets with challenges because tech workers, managers, and other stakeholders are not necessarily aware of ethics. Inside these companies, ethics owners are in charge of developing "strategies to align everyday practices with corporate logics" while navigating the everydayness of corporate life. They actively define, and help their companies to locate where ethics responsibility lies within the organizational hierarchy.

The role of ethics owners within the tech industry is ambivalent. Ethics owners operating within the industry are up against corporate logics that might prevent them from implementing their work or from achieving intended results. The three main corporate and industry logics that the authors examine are meritocracy, technological solutionism, and market fundamentalism.

**Meritocracy** is an ideological framework that legitimizes unequal distributions of wealth and power as arising from differences in individual abilities. This has defined the modern subject: as autonomous and responsible for perpetual self-improvement. The tech industry was founded on the myth that it is a meritocratic segment where talents should be rewarded handsomely. This meritocratic belief manifests in the idea that engineers are best at solving ethical issues that their products might create. Similarly, meritocratic logics place a strong emphasis on individual ethics rather than regulation and legislation. Companies and teams try to come up with their own codes of ethics to drive off legislation. The authors conclude that despite their best efforts, ethics owners' perspectives on larger societal problems are partial, as are their roles within the industry.

**Technological solutionism** is the belief that technology can solve social problems, which are then reinforced by the financial rewards that the industry has gained for producing technology that they believe solve the problems. Critics have pointed out that many so-called "solutions" can actually cause problems such as rising income and housing inequalities. The tech industry often responds by proposing even more technical solutions. Similarly, ethical problems are also framed as could be solved by technological solutions. This logic leads to creation of checklists, procedures or evaluative metrics to ensure the design and implementation of ethical products. The authors however point out that this approach is limited, and problematic because it centers ethics in the practices of technologists, and not in the social worlds wherein technical systems are created.

The State of AI Ethics, January 2021    68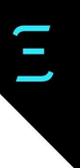

**Market Fundamentalism**, or market logics, refers to the idea that companies are there to make money, and if ethics initiatives are cut into the bottom line, companies should not do it. Besides, there is a belief that ethical initiatives are often costly, and antithetical to corporate profits. Furthermore across the industry, if other companies do not implement similar ethical considerations on their products, one should not do it. In the context of the absence of a legal framework, implementing ethics initiatives might be a business problem rather than a solution. In other words, the works of ethics owners in practice are constrained by what the market can allow.

These three different corporate logics reinforce each other and create a dynamic in which ethics owners have to navigate. Operating within these different logics can create scenarios which the authors termed "normalizing ethical mishaps," and "blinkered isomorphism." Normalizing ethical mishaps refers to situations when tech companies create structures that "normalize ethical transgressions;" while "blinkered isomorphism" refer to the process whereby tech companies converge to one structure by learning from each other's extreme cases while overlooking everyday ethical failings.

The article sheds light to the current events around the departure of Dr. Gebru from Google. The research shows ethics owners such as Dr. Gebru have to negotiate different competing logics between corporate interests, personal, and professional commitments. The recent events seem to suggest that when corporate logics appear to be more important, ethics owners regardless of how prominent they are could be let go. In other words, "ethics owners" occupy both ambivalent and precarious positions within the tech industry hierarchy.

### A Snapshot of the Frontiers of Fairness in Machine Learning
([Original paper](#) by Alexandra Chouldechova, Aaron Roth)
(Research summary by Falaah Arif Khan)

The motivation behind the paper is to highlight the key research directions in Fair ML that provide a scientific foundation for understanding algorithmic bias. These broadly include- identifying bias encoded in data without access to outcomes (for example we have access to data about who was arrested and not who committed the crime), the utilitarian approach to optimization and how it caters purely to the majority without taking into account minority groups and the ethics of exploration. The role of exploration is a key one since in order to validate our predictions we must have data that enumerates how the outcome in fact played out. This brings up several important questions such as: Is the impact of exploration overwhelmingly felt by one subgroup? If we deem the risks of exploration too high, by how much does a lack of exploration slow learning? Is it ethical to sacrifice the well-being of current populations for the perceived well-being of future populations?

The next important research direction is one that seeks to formalize the definition of Fairness. There are several proposed definitions, the most popular one being the statistical definition of Fairness. Such a formulation enforces parity in some chosen statistical measure across all groups in the data. The simplicity, assumption-free nature, and the ease with which a statically fair allocation can be verified makes this definition popular. However, a major shortcoming is the proven impossibility of simultaneously equalizing multiple desirable statistical measures. A statistical definition of fairness can also be computationally expensive to model.

The second popular notion is that of Individual Fairness, which enforces that, for a given task, the algorithm treats individuals who are similar, similarly. While this is richer, semantically, it makes strong assumptions that are difficult to realize practically.

Chouldechova and Roth then go on to present questions around Intersectional Fairness, namely: how different algorithmic biases compound for individuals who fall at the intersection of multiple protected groups. They also question the feasibility of a 'good' metric of fairness and whether such a notion will be accessible while making predictions, and the existence of an 'agnostic' notion of Fairness that does not rely on any one measure, but instead takes human feedback to correct for bias.

Another important consideration is the dynamics of Fairness. Models are seldom deployed in one-shot settings and are usually used in conjunction with several other predictors. In such a setting, how does compositionality affect algorithmic fairness? i.e. do individual components that satisfy conditions of 'fairness', continue to adhere to the same degree of fairness when composed together to decide a single outcome?

Another source of dynamism is the impact that algorithmic decision-making systems have on the environment. Models that determine outcomes also influence the incentives of those who interact with them and hence it becomes imperative to consider long-term dynamics when designing 'fair' algorithms. We also need to reconcile the individual motives of the different actors in the system and incentivize them to behave ethically.

Lastly, Chouldechova and Roth enumerate open questions in modeling and correcting for bias in data, namely: How does bias arise in data? How do we correct for it? How do we take into account feedback loops, where biased predictions further lead to biased training data in future epochs? Enforcing any notion of fairness on biased data would see a drop in model accuracy and this begs the question of how we go about validating our 'fair' predictions.

The State of AI Ethics, January 2021                                                                70

## Perspectives and Approaches in AI Ethics: East Asia

([Original paper](#) by Danit Gal)
(Research summary by Victoria Heath)

**Introduction**

Although artificial intelligence (AI) and robots are *tools*, "their perception is increasingly that of *partners*," writes author Danit Gal. In this research, Gal places the perceptions of AI and robots in South Korea, China, and Japan along the spectrum ranging from "tool to partner" by exploring a) policies and ethical principles, b) academic thought and local practices, and c) popular culture. Further, Gal examines the relationships between these perceptions and local approaches to AI ethics. identifying three AI and robotics-related ethical issues that arise: 1) female objectification, 2) the Anthropomorphized Tools Paradox, and 3) "antisocial" development.

**South Korea**

According to Gal, South Korea is placed in "the tool range" of the spectrum "due to its establishment of a clear human-over-machine hierarchy" and "demonstrates a clear preference for functional AI application and robots." There are several policies and ethical principles that enshrine this idea, such as the Robots Ethics Charter (revised in 2016), from which the South Korean National Information Society Agency (NIA) built its Ethics Guidelines for the Intelligent Information Society (April 2018). These Guidelines outline four positions: 1) responsibility on users to regulate use, 2) responsibility for assessing AI and robots' negative social impact on providers, 3) responsibility on developers for eliminating bias and discriminatory characteristics in AI, and 4) calls for developing AI and robots that do not have "antisocial" characteristics." Broadly, most policies and ethical principles in South Korea place an emphasis on balancing protecting "human dignity" and "the common good," as well as reaffirm the idea that these are "tools meant to protect human dignity and promote the common social good."

In academic thought and local practices, Gal highlights the Korea Advanced Institute of Science and Technology (KAIST) Code of Ethics for Artificial Intelligence (2018) released in response to protests regarding its involvement in developing lethal autonomous weapons systems. The third principle found in this Code of Ethics is most unique, stating: "AI shall follow both explicit and implicit human intention," with a note that the "AI should follow the person with the highest priority or closet relationship" if multiple people are involved. Gal critically examines the conflict between this idea, which could reinforce existing power structures and discriminatory practices, with the "developer's mandate to act as eliminators of social bias and discrimination under NIA's Ethical Guidelines."

The aforementioned hierarchical structure has been challenged by South Korean popular culture. Korean dramas, writes Gal, often place AI and robots as "family members, friends, and love interests." This leads some to wonder if the more reliant on, or normalized, to social robots people become, the more people will lose basic ethical values and devalue human relationships. Hence the emphasis by the NIA on "avoiding the antisocial development of AI and robots." For now, however, it appears South Koreans are more comfortable with "functional robots" because they retain more "control" over it, unlike a more "biologically inspired" robot.

**China**

China's overall perception of AI and robots, especially on the government and corporate-level, is similar to South Korea, sitting more so on the "tool range" of Gal's spectrum. The Chinese Association for Artificial Intelligence (CAAI), led by Professor Xiaoping Chen, is responsible for creating ethical guidelines for the development of AI and robots in China. There is evidence that the CAAI is looking at the ethical challenges related to creating technology that can be used as an "intelligent tool but designed with the characteristics of a desirable partner." More often than not, these "tools" are designed with "feminine" characteristics, touching on the Anthropomorphized Tools paradox and female objectification issues aforementioned by Gal.

Robin Li Yanhong, the CEO of Baidu, emphasized in 2019 at a government-run event the importance of sharing "Chinese wisdom" globally to inform international AI Ethics discourse. This includes the "integration of the Chinese government's twelve 'core socialist values,'" which are divided into three groups: 1) national values, 2) individual values, and 3) social values. Within the national Engineering Ethics textbook, four unique Chinese characteristics are highlighted: "responsibility precedes freedom, obligation precedes rights, the group precedes the individual, and harmony precedes conflict."

Among academic thought, local practices, and popular culture, however, there is "strong interest in imbuing AI and robots with partner-like capabilities to help them realize their full positive potential." The Harmonious Artificial Intelligence Principles (HAIP) led by Professor Yi Zeng promotes ideas that aim to achieve harmony between humans and AI and robots through mutual respect, empathy, and altruism. For example, one idea is that AI should have privacy. Another is that humans shouldn't show bias towards the machine. There is even an idea, proposed by Hanniman Huang, that AI and robots should be considered a new species and should be eventually considered a part of human society. This idea aligns with ideas in Chinese Buddhism that "everything can be cultivated toward enlightenment and become the Buddha.

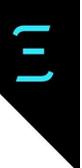

An example of how this can manifest itself in popular culture is the intelligent robot monk, Xian'er, which has over one million followers on social media and engages with Buddhist scriptures.

AI and robots have been depicted as love interests since the 1990s in China, with movies like Funny Robot Talk (1996), Robot Boyfriend (2017), and Robot Maid from Heaven (2017). Raising the female objectification issue (plus other issues) is the social chatbot XiaoIce, created by Microsoft, which is modeled after a female teenager and has over 660 million users who, as Gal writes, "often perceive it as a friend or love interest." There are even AI and robots replicating famous entertainers and music groups, like May Wei VIV, or acting as news anchors on state television. Even though these are developed as "tools" for entertainment purposes, people often engage with them as "friends and partners," similar to human entertainers.

**Japan**

Gal places Japan on the "partner range" of the spectrum "due to its exceptionally strong mix of pro human-AI-robots partnership academic thought, local practices, and popular culture." Interestingly, while the policy approach to AI in the country appears to be moving more toward the "tools" range of the spectrum in order to stay in line with international discourse, "the extent of its societal vision for coexistence and coevolution with AI and robots is distinct." The 5th Science and Technology Basic Plan (2016) introduces Society 5.0: a future in which AI and robots enable a more "convenient and diverse" society by responding to the needs of humans and even anticipating those needs—potentially creating a "push rather than pull culture" (similar to the current world of online advertising). The Cabinet Office Council on the Social Principles of Human-centric AI has warned of the "overdependence on AI and robots," emphasizing "the need to maintain human dignity" while still calling for an "AI-based human living environment." This basically outlines a future where Japan's social systems and "individual character" may need to be redesigned to accommodate the use of AI and robots as social tools.

In academic thought, local practices, and popular culture, there seems to be a divergence from the idea of AI and robots as merely tools for enabling social progress. This may be explained by Japan's history of "robot-friendly" media, as well as the perception that robots can help solve many of the social problems Japan is facing, including an aging population (robots can offer elder care) and a slowing economy (AI offers automation). In Japanese popular culture, two cartoons have made a significant impact on perceptions regarding AI and robots, as well as inspired would-be developers: Astro Boy, first introduced in 1952, and Doraemon, first introduced in 1969. Softbank's robot Pepper, which is a conversational humanoid robot, has also had profound influence. It functions as everything from an assistant to a Buddhist priest, and has been marketed by the company as a "friend, sibling, potential love interest, entertainer, and caretaker." There's also Aibo, Sony's pet robot dog, which in some instances has been given religious rites when it breaks down. "This derives from the concept of animism," writes Gal, which is found in the Shinto belief that the "spirits of otherworldly beings can dwell in animate and inanimate objects," and in the Buddhist belief that "both animate and inanimate objects are a part of the natural world and possess the character of the Buddha."

Finally, Japanese media has many stories of AI and robots acting as partners, especially love interests, such as Absolute Boyfriend (2008), Cyborg She (2008), or Ando Lloyd—A.I. Knows Love? (2013). This exists outside of television and movies, however, there is also Vinclu Inc.'s Gatebox AI lab's holographic virtual wife and home assistant which is "modeled after a young female character named Hikari Azumu." Although its popularity is a byproduct of the loneliness epidemic in Japan, it also constitutes a "rare edge case of intentional tool anthropomorphizing and female objectification, where a functional home assistant is specifically designed to act as a meaningful romantic partner."

**Conclusion**

Although these three countries are at different places on the "tool to partner" spectrum, they are all shifting their place due to changes in international and local discourse regarding AI and robots, as well as emerging tensions between the social benefits and harms of these technologies become clearer. In particular, the three AI and robotics-related ethical issues that Gal discusses: 1) female objectification, 2) the Anthropomorphized Tools Paradox, and 3) "antisocial" development, will increasingly become tension points—not only for the countries studied here, but everywhere.

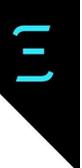

## Go Wide: Article Summaries (summarized by Abhishek Gupta)

### Big Oil Faded. Will Big Tech?
(Original *NY Times* article by Shira Ovide)

It was only 10 years ago that ExxonMobil was the most valuable company in the world and our world was dominated by large energy companies. A key vulnerability of these energy companies that has led to their decline is that they weren't nimble enough to adapt to the changing energy landscape in the move away from fossil fuels. They are also hostage to the demand forces for the product that are out of their control. It seems that the technology companies don't face that same struggle. With the technology companies now possessing more market capitalization than the entire European stock market, it calls for a moment to pause and reflect on the magnitude of power that these companies have.

They are able to shape policy, dictate what constitutes anti-trust, our notions of privacy, and to some extent how we live our lives since they are the conduits through which we manage our digital existence. This is exacerbated when governments don't step up in their role to regulate these players and they run amok reshaping the landscape to suit their monetary goals. Companies like Apple are pushing for changes to how users' personal data is protected but as the article points out, it takes a lot of effort on the part of the users to check off the right boxes that enable them to reap the benefits of enhanced privacy protections. We are all presented with an interesting conundrum in the sense that we have these large companies who might shape policies to benefit end-users, especially in the absence of government efforts, yet the fact that they have this power in the first place to so significantly impact our lives is in itself quite dangerous.

### Google Offers to Help Others With the Tricky Ethics of AI
(Original *Wired* article by Tom Simonite)

An interesting moniker, Ethics-as-a-service (EaaS), is used in this article to aptly sum up what Google is trying to do with this latest announcement. While there is a fantastic team of folks doing very meaningful research at PAIR (People + AI Research) at Google, when it comes to outsourcing ethics to a for-profit organization, many critics are right to raise concerns about what this means in practice. There are certainly a lot of valuable lessons learned by Google over the past few years with its AI ethics board fiasco, wrongly flagging Black people in photos as gorillas, and others. Yet, there might be a better way to direct these lessons and resources rather than spinning it off as a consultancy.

Specifically, it might be useful to share these ideas openly with others through various channels and creating courses that might be undertaken by people who are in the field deploying these systems. Another avenue might be to fund efforts by other non-profit organisations that are doing work in this space rather than duplicating efforts. But, as the article points out, it might just be a mechanism that the company is using to distinguish itself in a crowded field of actors. Placing this in the context of all the developments that have taken place with the firing of Dr. Timnit Gebru and now the action taken against Margaret Mitchell, it calls into question the real motivations behind this effort. The *Abuse and Misogynoir Playbook* at the start of this report also provides more context around this.

## How Tech Companies Can Advance Data Science for Social Good
(Original *Stanford Social Innovation Review* article by Nick Martin )

This piece identifies some of the ways that the landscape has changed since and catalogues some of the initiatives in this space that are helping to equip non-profits with the technology and skills to apply AI to their work. Specifically, there is a great deal of enthusiasm on the part of technology companies to aid with this but the article urges them to ask critical questions such as what might be some of the incentives and disincentives for an organization to utilize AI, what skills gaps exist, and what gaps can funders help to fill.

**4 key takeaways:**
1. Focus on what it takes in the preprocessing stages of AI to make projects successful. Specifically, an emphasis on the work that is required to clean and prepare data is often invisible in the final result and equipping data scientists so that they can carry this out more effectively is crucial.
2. Helping the organizations build internal capacity so that they can carry out the work rather than having to continuously rely on external partners for skills and technology.
3. Supplementing that through skills transfer by having experienced data scientists work with the local teams is a great way to share knowledge and build internal capacity. Those who have a proven track record in being good communicators and teachers are great candidates for this.

4. Finally, providing insights such that the use of data is done in an ethical and responsible manner is also important, especially as a lot of these organizations work in areas that have significant impacts on human lives.

## What Does Building a Fair AI Really Entail?
(Original *Harvard Business Review* article by David De Cremer)

This article presents an oft-ignored perspective towards building AI-enabled systems that are fair, namely, the organizational scaffolding that is required to create AI fairness beyond just technical measures. In essence, it is not enough to have technical solutions that achieve fairness according to some definition as specified in academic literature. When rubber meets the road, there are a lot of other factors that go into the making of what is perceived as a fair AI system.

In terms of adoption of these systems, both by employees and users of the product, their perception of the degree of fairness of the organization plays an equal importance role. The more fair the organization is deemed to be, the more willing the users and employees are to accept the system. One recommendation made by the author in the article is to treat fairness in these systems as a cooperative act where you have a human devil's advocate to validate and check for the fairness of the system. Demonstrated through prior research, it is evident that humans are more likely to spot biases in other people than themselves and utilizing that trick here can lead to fairer systems.

The second recommendation made by the author is to look at the tradeoffs between utility and humanity of the system. This requires articulating the values that are important to the organization and squaring them with the technical goals as specified for the system. Finally, being transparent in communication, in treating the larger community with due respect also play an important role in the perception of the organization being fair and ultimately amplifying the impact of their work in building fair AI systems.

## Ethics in Tech: Are Regular Employees Responsible?
(Original *Welcome to the Jungle* article by Lenka Hudáková)

There is a strong preference expressed by employees to not only understand the societal impacts of the systems that they are building but also to have meaningful control over the decision-making process. But, the first step to being able to do that is to have open discussions within the organization and not facing unnecessary censure for expressing dissenting opinions.

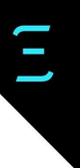

If critical feedback about products and services is actively discouraged, employees will over time "learn" a sense of helplessness; the employees in especially high-demand positions like machine learning do have considerable market power and should use that actively to voice their concerns. From an employee review perspective, including efforts made in bringing a more ethical lens to the development process should be rewarded rather than being perceived as an "extracurricular" activity. It must also be included in the formal compensation for the employees.

Having positions that are designated for disseminating ethics-related information across the organization and forming bridges between different efforts within the organization will also be crucial to the success of the adoption of ethical principles.

### Technology is Easier Than Ever to Use — and It's Making Us Miserable
(Original *Digital Trends* article by Shubham Agarwal)

Friction - a dreaded word in the world of technology design. So much effort is expended in making everything frictionless for our consumerist experiences. This stems from our desire to get everything at the touch of a button, or more accurately, at the touch of fewer buttons. Hence, the one-click ordering on Amazon, auto-play on YouTube, requesting a ride back home on Uber with a single tap after a night out on the town, and more.

But, the dark side of this removal of friction is that it leads to addictive behaviours that are subsumed in the convenience fading into the background of our digital existence. We rarely realize that we have spent many more minutes scrolling through photos on social media when all we wanted to do was to quickly look up how to center CSS elements on the webpage (yes, rejoice fellow web developers!).

But, friction is used ingeniously by platforms: in cases that they want to extract money from you, say getting a YouTube premium subscription, you will be shown "unskippable" ads twice in a row nudging you to make a decision to purchase the subscription to get rid of the friction. Or ads on Spotify every few minutes to disrupt your listening experience. On the other hand, when they want to collect as much data from you as possible, they can obscure privacy settings across a variety of pages so that you have trouble tuning the settings to your privacy desires.

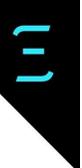

The smoothening of our experiences on these platforms obfuscates the complexity that the platforms engage in to keep you addicted. Additionally, we forget to ask critical questions about it because the interface is so simple, what could possibly go wrong?

### India's Internet Shutdowns Function Like 'Invisibility Cloaks'
([Original *DW* article](#) by Nehal Johri)

The success of disinformation lies in its ability to deceive the intended audiences into believing the content is trustworthy and truthful. While it might be hard to generate content and have it spread via social media to gain enough legitimacy to move the needle, there is tremendous potential for harm if such content gets wrapped in the veneer of respected news outlets. In particular, information operatives have leveraged the technique of hacking into content management systems (CMS) of media houses to plant fake news stories.

Researchers identified the use of this technique leading to divisiveness in eastern Europe, targeting NATO related news items along with relations of those countries with the US. While there are reporting mechanisms, if the content stays up long enough, it has the potential to be copied and impact people nonetheless. Emphasis on cybersecurity becomes critical in combating such attacks. In addition to fighting disinformation along the lines of content, provenance, and other dimensions, all of these efforts get subverted if appropriate guards are not in place.

### Selfies and Sharia Police
([Original *Rest of World* article](#) by Mehr Nadeem)

One of the last bastions of a free-to-access social media platform in Iran, Instagram has become home to people expressing themselves with freedom and without prying eyes. In a country that restricts how women express themselves on the streets, many young women have taken to Instagram to showcase their style and unique talents. But, a platform that was seen to be frivolous at first is increasingly being policed where a recent law in the country forbids women from appearing without a hijab. One of the teenagers interviewed in the article laments that since the lockdowns began, the authorities have taken to social media to continue to exercise the same influence they used to because women are staying home, out of sight.

Instagram has also become politicized and is being used to spread awareness and messages around some of the political injustices and perhaps this is one of the reasons for increased scrutiny of people's activity on the platform. Some of the more famous accounts have put up notices declaring that they conform with the latest religious diktats in an effort to evade any coercive actions against their accounts.

Some expats utilize the platform to stay in touch with the ongoings of their local communities and this self-censure has limited their ability to get a true glimpse of what is happening in their own countries. The effects of this have resonated across the world in another sense as well whereby people residing outside of Iran have also started recalibrating their online presence to conform with the latest laws so that they don't run afoul of them.

### Best Practices for Indigenous Keywording for Stock Images
(Original *Shutterstock* article by Julia Crawford)

This article provides clear guidance on how language creates power dynamics and what we can do to avoid intentionally and unintentionally weaponizing language. The article details the Shutterstock team's efforts in including more inclusive and respectful language for the images associated with Indigenous Peoples. The article provides great detail in terms of the phrasing to use, the capitalization, punctuation, and other language specifics so that we respect the wishes of the people we are trying to represent.

One thing that we really liked was how it paid heed to geographical differences and provided clear guidelines in terms of phrases that are appropriate to use and those that are not. In cases where there might be ambiguity, the golden rule of asking those you are trying to represent about the phrasing that they would like to see used about images or any other assets is the best way to go.

While not reproducing the list here (we strongly encourage you to read the original article if you are interested), what we found relevant for our discussions is the impact that such inclusive labeling will have downstream on systems that scrape metadata to train machine learning systems. Also, this has impacts from an archival perspective in terms of what future generations will have as artifacts when they seek to understand the state of a community in a bygone era, an idea that was touched upon in this work titled "Comprehensiveness of Archives: A Modern AI-enabled Approach to Build Comprehensive Shared Cultural Heritage".

## Australia's Artificial Intelligence (AI) Future: A Call to Action
([Original *Lexology* article](#) by Lesley Sutton, Luke Standen)

Australia has made access to high-quality data that is desensitized and ethically and legally sourced as a core tenet within their Action Plan for the adoption and use of AI. Especially encouraging is the fact that practitioners and researchers are going to be working together on addressing these issues. Without such collaboration between the theoretical and practical, a lot of the initiatives fail to meet their targets.

One of the shortcomings though identified in the article is the inadequacy of the support from a financial perspective which is also a key component to the success of any such work. Numerous self-funded initiatives peter out over time because of funds drying up and volunteers being forced to make a choice between sharing their expertise without compensation and being fiscally viable at a personal level.

Also noteworthy was the article's emphasis on having "AI translators", people who are able to navigate different domains so that the capabilities and limitations of these systems can be adequately communicated. Often, we are overzealous in our expectations with what AI-enabled systems can do which leads to unfortunate outcomes from an ethical standpoint. Both upskilling of existing staff to take advantage of AI advances and clear guidance for businesses on the ethical considerations will be critical for the success of a responsible AI ecosystem.

# 5. Labor Impacts

**Opening Remarks** by Ryan Khurana (Writer-in-Residence, Montreal AI Ethics Institute)

The COVID-19 pandemic has accelerated the adoption of AI technology globally, both in efforts to combat the virus and to help keep economies productive in the face of strict lockdown measures in many of the world's largest countries. This has brought to the forefront the challenges coming ahead as millions of global workers harmed by the pandemic seek to return to work in a landscape that looks vastly different than the one to which they were accustomed. In this context, the labour market impact of AI gains a new focus, both expanding scope across industries and geographically.

The first few papers look at the broader impacts of digitisation across the economy and the implications of this for workers. The first one proposes a model for the inequality that has become a politically salient issue in recent years by looking at the change in resource scarcity that digitisation has brought. While an abstract model, a theory explaining the economic impact of digitisation is essential in engaging with the challenges up ahead and proposing solutions. The second item follows a similar vein, making the case that digital workers bear disproportionate risks that were historically borne by the corporations that employed them. The precariousness of these arrangements provides a lens into growing inequality. Looking for more salient explanations of what is happening is a key task for those exploring the ethics of automation and these papers provide valuable contributions to mapping the territory.

The last two research papers look into the diverse implications of digitisation across geographies. Both look at the impact of digital work on the Global South, where international institutions have made a priority for future digitisation. The mobile penetration rates across Africa, which has been able in some ways to leapfrog Western legacy technologies, make gig economy work much easier to access. The psychological impacts of these precarious work arrangements are often ignored. In the African context, it is argued that digital labourers are suffering from an imposition of new technologies alongside the economic gains they bring. A similar insight is found in a paper on India's automation, which highlights the social benefits that digitisation has brought through the inclusion of workers typically marginalised by India's caste system, while simultaneously bringing new forms of surveillance and monitoring that curtail their dignity.

In the year ahead, these twin expansions of automation's scope — sectoral and geographical, should guide those working on remedying pressing issues such as unemployment, inequality, and the indignity of certain work arrangements. The post-pandemic landscape provides a significant opportunity for reimagining the future of work, but complacency and ignorance may produce barriers to realising a positive vision. We at MAIEI hope that our contribution in highlighting the critical work in this space may help guide researchers in a positive direction.

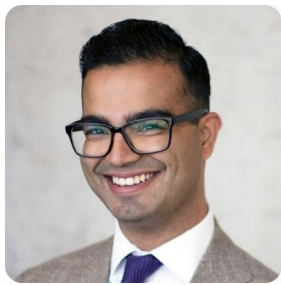

**Ryan Khurana (@RyanKhurana)**
Writer-in-Residence
Montreal AI Ethics Institute

In addition to being a writer-in-residence at the Montreal AI Ethics Institute, Ryan is an Analytics Program Manager at CBRE. Previously, he was the Executive Director at the Institute for Advancing Prosperity, which he founded to advance Canadian technology policy discourse and improve Canada's global reputation for technology and innovation. He graduated with a Master's in Management Analytics from University of Toronto's Rotman School of Management.

# Go Deep: Research Summaries

**Digital Abundance and Scarce Genius: Implications for Wages, Interest Rates, and Growth**
([Original paper](#) by Seth Benzell, Erik Brynjolfsson)
(Research summary by Ryan Khurana)

Traditionally the dynamics of capitalist economies are thought to be a clash between labor and capital. In competitive markets with low barriers to access to capital and a skilled labor force that is able to access opportunities, we would expect to find a good equilibrium where total income is growing and labor and capital each have reasonably large shares. When one is growing at the expense of the other, social unrest would be expected. It is surprising then that both labor and capital have had declining shares of income in recent years. Real interest rates are low, investment is low, and real incomes have barely increased in decades. The causes of this paradox are an active debate among economists, and the stakes for a good understanding couldn't be higher given the current political climate.

Economists Seth Benzell and Erik Brynjolfsson in Digital Abundance and Scarce Genius aim to resolve this paradox by pointing towards the changing nature of the economy due to digitization. The digital world has removed barriers to work and increased access to capital in a way that was not possible in previous eras. The speed at which software scales and the gargantuan revenues digital giants command does not take advantage of traditional labor or traditional capital in the way economists expect. The returns to these superstar firms in the digital economy, the authors argue, result from the scarcity of "Genius", a factor to which they show a sizeable chunk of total income has appreciated. "Genius" is not some innate superiority among a certain class of people, but rather the returns to a certain type of talent and intangible asset that the digital economy makes more valuable. One can be agnostic to the origins of genius, but the "10x programmers" that are mythical among the tech community do indicate an acknowledgment of the excess returns to a certain type of talent. The authors argue that genius is not just the raw talent, but the configuration of this in certain companies that can scale that talent. Organizational capital, such as being "ML-first" or knowing how to manage technical teams, is an aspect of genius that is rare outside of the tech giants. Virtual real estate, such as intellectual property or platform dominance due to network effects, is another aspect of genius that constricts its supply to a few firms.

The surprising findings of the paper are that in an economy where fewer firms are capturing scarce genius, which in turn commands a greater share of income, traditional means of boosting productivity would prove counterproductive. Upskilling low and medium-wage employees, should the models turn out to be correct, would decrease wages as more workers would be qualified to fill roles that contribute to a declining share of total productivity. Counter-intuitively, targeting productivity improvements at the top would increase wages at the bottom more rapidly. The logic is as follows: if the bottleneck on genius is lifted by increased top-end productivity, then the market for genius is more competitive and the cost of genius decreases, but given the necessity of genius to digital success, productivity would increase much more, returning a greater share to lower and medium-income workers. Efforts such as increasing high-skilled immigration, increasing access to top schools, loosening IP laws, and improved education that provides interdisciplinary management skills and creative thinking would all serve to boost genius.

This is where the authors leave their discussion and the point where AI ethicists need to pick up the mantle. The returns to genius would likely accelerate in an AI-driven world where only a few companies have yet to realize value from AI investments. While traditional companies seek out unicorn data scientists to do everything, and rarely realize value, "ML-first" companies are able to capitalize on specialized AI talent managed correctly. Democratizing AI needs to reckon with the scarcity of genius.

The solutions to this problem, however, require grappling with the institutional environment that allowed genius to become scarce. The current class of top earners whose salaries result from membership in the genius class have strong incentives to restrict access. Ivy league schools are selective not due to physical constraints, but in order to restrict the supply of people with their credentials, allowing them to command more income. Upskilling the middle class is far more politically popular than creating more elites because it presents less of a challenge to the current elites. Navigating this political minefield was briefly touched upon by the authors but presents an interesting area of exploration for AI ethicists. The question "who is benefitting from new technologies?" is at the heart of AI ethics research, and one that observing the scarcity of genius may help to answer.

# Risk Shifts in the Gig Economy: The Normative Case for an Insurance Scheme against the Effects of Precarious Work

([Original paper](#) by Friedemann Bieber, Jakob Moggia)
(Research summary by Anne Boily)

Authors Bieber and Moggia examine the issue of the gig economy from a political philosophy perspective. It is the notion of "risk shifting" that is central to their analysis, and which is relatively unexplored in the discipline with respect to labour research. It is also from the point of view of workers themselves (especially low-skill workers) that the article is written, rather than from the point of view of the interests of firms and employers.

The gig economy refers to the phenomenon of hiring workers for concrete and temporary tasks: the supply of labour, therefore, depends on the demand: think of companies like Lyft, Uber, or even Google. The central thesis of the article is that the authors are rather critical of the deleterious effects of the gig economy on workers. They propose a policy framework to policymakers that would allow them to reduce the risks and compensate for the disadvantages caused to workers and, by extension, to society, without entirely prohibiting the way in which the gig economy operates, which is not entirely detrimental, but whose harmful sides are never entirely erased.

Biber and Moggia present their argument in 3 sections:

1. Their diagnosis of the situation of the gig economy and its "risk shift". This "risk shift" is expressed in the fact that the firms are abandoning the risk and placing it on the workers, for whom such a risk (overly flexible and therefore precarious working conditions, no guaranteed income, difficulty in making long-term plans, the impossibility of finding time for further training, high levels of stress) becomes personal. Moreover, disparities are aggravated in the population, which can make social solidarity more difficult to achieve. Firms "externalize their risk" to third parties through 5 strategies, and encourage each other to remain competitive in the market:
    a. Short-term contracts
    b. Flexible number of hours
    c. Flexible remuneration
    d. A flexible schedule
    e. Less insurance coverage

2. The normative analysis of these "risk shifts", which weaken the workers themselves and open them up to situations of domination and exploitation. Workers seem to be free to make their own choices, but as the market changes, they are in reality less and less so. They depend too much on the supply side to be able to position themselves with freedom).

3. A proposal to policymakers, the "Principle of Inverse Coverage (PIC)", which contains two key aspects:
    a. "A contribution side": introduction of a Pigouvian tax (following the same principle as a carbon tax), which forces employers to financially compensate for the deleterious effects of their mode of operation.
    b. "An expenditure side": the profits from the Pigou tax could be used to finance social insurance for gig workers, to ensure a more stable income stream without discouraging them from working. If their total working hours decrease, the insurance income will also decrease.

This policy would stabilize the working conditions of the gig workers and allow them to project themselves into the future, by giving them back this agility. Compared to the "UBI", the "PIC" does not generalize the risks of the firms to the whole population but compensates all the same for the risks to which the workers are exposed.

## Between a Rock and a Hard Place: Freedom, Flexibility, Precarity and Vulnerability in the Gig Economy in Africa

([Original paper](#) by Mohammad Amir Anwar, Mark Graham)
(Research summary by Alexandrine Royer)

For many among us still caught in the throes of a global pandemic, the concept of a regular nine-to-five office job seems like a distant past. COVID 19, among its many societal lessons, has both made apparent and accelerated the digital transformation of labour, where work can continue to be conducted anywhere, at any time, so long as we have ready access to a computer and internet. While remote internet-based employment may be a new component of our lives, the advent of digitally mediated work has long-been making its mark in Africa.

Digital labour is a key part of a strategy of accelerated economic development in the Global South. Politicians, NGOs, and global economic institutions have confidently stated that participation in the digital economy can overturn decades of economic stagnation, high unemployment rates and offer profitable job opportunities for its citizens. Part of the political and economic discourse around the gig economy rests on the idea of the digital nomad, who has the power of freedom and flexibility over his work. Against this narrative, Mark Graham and Mohammad Amir Anwar point to the economic vulnerability and precarity facing digital gig workers in Africa.

The digital economy's growth in the Global South is attracting greater scholarly attention, although there remains a limited understanding of the quality of gig employment in Africa. Anwar and Graham are among the prominent pioneers in the field. By conducting a four-year study of 65 digital gig workers employed by freelancing platform UpWork in South Africa, Kenya, Nigeria, Ghana and Uganda, Anwar and Graham demonstrate how these workers' livelihoods are emblematic of the new digital wave reproducing the international division of labour. UpWork is a commonly used remote work platform in the Global South and the go-to for African workers. As part of their analysis, the authors consider four key dimensions to gig work: freedom, flexibility, precarity, and vulnerability.

Freelance work has garnered considerable media hype in Africa, and it is portrayed as a disruptive innovation by organizations such as the Rockefeller Foundation that will positively impact millions of Africans by providing them with a pathway out of poverty. It allows workers an escape from dysfunctional local labour markets and gives them a sense of connectivity to the wider world. Alongside the flexibility offered to firms, which can quickly meet their demand for quick digital work at lower prices, the digital gig economy appears as a win-win for employers and their employees. For Graham and Anwar, the supposed individual freedom of

The State of AI Ethics, January 2021    88

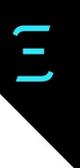

gig workers is part of a neoliberal discourse that masks structural inequities and the limitations posed by the algorithmic controls of digital work platforms. As part of an easily hirable yet disposable workforce, digital workers are left to bear the risks of fluctuating markets and labour demands.

Graham and Anwar refer to the work of Guy Standing, who describes the emergence of a new working class, 'the precariat,' which falls in the cracks of the social security net. However, they underline how African workers often seek jobs in the informal economy where standard employment relations are mostly absent; precarious work forms are more commonly the norm than the exception. In contrast to the North American and European context, gig economy jobs in Africa constitute an entry point into standardized employment. Workers find themselves subject to the organizational management strategies and reputational scoring system of freelance work platforms. While platforms like Upwork remove traditional work entry barriers, earning a successful bid for a first job can be a long and arduous process. Several workers interviewed noted that platform work paid better salaries but added that these earnings went into securing faster broadband internet. Freelance work is also subject to social stigma, particularly in Nigeria, due to the prevalence of online scamming networks. Only workers from educated and wealthier socioeconomic backgrounds benefitted from the ability to adapt their schedule and work at their own pace.

A far cry from the fulfilled and independent digital nomad, digital platform workers in Africa suffer from high stress and fatigue levels related to income insecurity and irregular working hours. Several of Anwar and Graham's interlocutors mentioned feeling lonely and socially isolated. While gig work remained an attractive option in countries where the local job market's constraints meant difficulties securing employment, it rarely brought the economic opportunities vaunted by politicians. Competing against workers from across the globe, Africans faced competition from low-bidding Indian, Bangladeshi, and Filipino workers, leading to a race to the bottom. Workers are at the whim of their employers, who can cease the contract without warning and refuse to pay wages if the work is deemed unsatisfactory. They must also remain glued to their screens, sometimes at all hours of the night, as Upwork will take screenshots of the worker's laptop to ensure they keep up with the task.

Through their insightful interviews, Anwar and Graham offer a much more nuanced picture of the quality of gig work among African labourers. The on-the-ground realities gleaned from gig work tarnishes the rhetoric of freedom and flexibility in gig work pronounced by various international institutions, private firms, internet-based platforms, and nation-states. For those who have now transitioned to working from home due to global circumstances, the social isolation and psychologically damaging effects of remote work expressed by African gig workers certainly ring true. It is in the interest of governments within and outside Africa to regulate these digital work platforms, whose headquarters are stationed in North America, which clearly will not guarantee fair working conditions out of their own volition. If we want to avoid creating a global digital underclass and end the proliferation of precarious work, immediate action towards decent digital labour standards must be undertaken by policymakers from local to international levels.

## Automating Informality: On AI and Labour in the Global South

([Original paper](#) by Noopur Raval)
(Research summary by Abhishek Gupta)

**Definitions**

- **Informal work:** This includes atypical, non-standard, self-generated, and home-based work that is unregistered (or too small to be registered) and is not typically accounted, taxed, or regulated because of a lack of clear employer-employee relationship.
    - This is particularly prevalent in emerging economies like India where there is a lot of "hidden" labor that faces a disproportionate burden of labor harms because of a lack of being folded into labor protections from a legal standpoint.
- **Artificial Intelligence:** This report uses AI as a term to describe algorithmic platforms and how they alter the work arrangements for those who demand and supply services and products on those platforms.
- **Heteromation:** Positioning human labor alongside a machine rather than a machine replacing humans. The thing to note here is that such a positioning makes a natural segue into human-in-the-loop (HITL) conversations.

**The Indian Context**

Unsurprisingly, the most common uses of algorithmic labor assignment are in the domains of ride-hailing (Ola, Uber), food delivery (Zomato, Swiggy), and in logistics and retail (Flipkart, Amazon).

The particular issue of *double marginalization*, as we will explore further in this summary, is that these platforms are typically, from a supply-side, staffed by those who are already marginalized and that they are often testing grounds for new automation technologies. Specifically, this refers to the use of surveillance techniques in the workplace and monitoring workplace productivity which might violate labor laws, but because this being experimented on in the informal sector with little to no regulation, it is a place where these platforms fine-tune their algorithmic approaches.

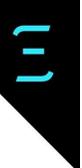

**Societal Context in India**

<u>The Positives</u>

Traditionally, because of strong hierarchical norms and caste-systems in India, certain occupations for the domain of people from certain backgrounds and even geographic mobility was restricted in terms of who you knew in the place that you are trying to migrate to and if they have the right connections to get you set up. This meant that there were ceilings on social and financial mobility that continued to reinforce social hierarchies. So, in a sense, such platforms have allowed these workers to bypass gatekeepers that prevent the inclusion of workers from different backgrounds into various subsegments of the labor market. Thus, it offers a pathway to reintermediation.

Another positive outcome from this is the framing of the work that is offered by the platforms that offers a higher degree of dignity to the labor compared to the very strong biases in the Indian context against some occupations as not being dignified labor. The rebranding of some of this labor has allowed workers to have higher levels of dignity, with of course the stated benefits of flexibility in work schedules and more. Though, this is not without the many, many harms that come from an always-on and available workforce that has other negative implications.

The participants that were surveyed as a part of this report indicated that they were doing such on-demand work as a temporary measure to meet short-term financial needs like paying off loans, gathering money for a wedding, etc.

<u>The Negatives</u>

This is not without consequences. Most of these platforms cater to the rising middle-class in India that skews to a certain sociodemographic and hence those who were hitherto viewed as "risky subjects" are now well-policed through these platforms when they are providing services to this mobile, urban, nouveau-riche class. That is, the venture capital dollars (or rupees) are funding the reinforcement of social hierarchies through platforms by placing some people above others and justifying workplace surveillance and other unethical uses of automated technologies to pander to this demographic in the interest of turning a profit.

**Conclusion**

As is the case with any new technology that has the potential to disrupt the incumbents, it doesn't come without unexamined risks. We need to think critically about the second-order effects of the deployment of such technologies and if we are preying on those who are already marginalized in the interest of large corporate interests.

**Implications**

- As a practitioner, it is more important than ever to have someone on your team who understands the local context of where the system is going to be deployed. Without that you risk entrenching societal inequities unwittingly.
- In the creation of some of the automated systems, having meaningful controls for those who are on the supply-side of the platform is important so that they are not exploited.
- In the design process, I also believe that it is important to present the workers in a humane way so that the demand-side understands that there are real humans on the other side fulfilling their requests and it doesn't abstract them away as just some cogs in a giant wheel.

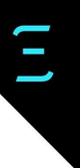

## Go Wide: Article Summaries (summarized by Abhishek Gupta)

### Are Robots Eating Our Jobs? Not According To AI
(Original *Forbes* article by Toby McClean)

Main takeaway: humans aren't going anywhere since the current dominant paradigm of machine learning requires large swathes of human efforts to enable their functioning and even if ML systems are good at making predictions, we still need humans to make judgements and act on those predictions. While there are widely varying estimates of the number of jobs that will be lost, a thing that is important to keep in mind is that these were estimates that were made prior to the start of the pandemic and didn't anticipate the pace of adoption of automation which has quickened.

This might mean that more jobs are lost but some labor economists and keen watchers of the space argue that there is room for new jobs that will be created to support the development of AI systems, through things such as data labelling services. One would be remiss to not mention how forecasting a decade away has severe flaws. There isn't a doubt that the estimates made on the labor impacts of AI are underscored by research but as Philip Tetlock mentions in Superforecasters, anything past 18 months severely degrades in quality because of the complexity of the world that we inhabit coupled with interactions and technological changes that have the potential to disrupt even the best-laid plans. As we emphasize at MAIEI when it comes to the future of work, the one definite thing that we can do is to remain adaptable and adopt life-long learning to the extent that is possible given our individual circumstances. Beyond that, it is hard to accurately forecast and plan for the pace of technological progress and therefore the accompanying societal changes.

### Insuring the Future of Work
(Original *Stanford Social Innovation Review* article by Deanna Mulligan)

Providing some insights into the world of actuaries, this article emphasizes how some of the transformation in how we work will take place over the next few years. Specifically, early on it makes the important point that it will not just be the so-called soft-skills that will be essential to the success of workers in the future but also a baseline of technical skills too so that they are able to effectively work with new-fangled tools that permeate their domain. Taking from the example of actuarial work, data science has certainly boosted the ability of the actuaries to make more accurate forecasts and help pricing different instruments in the space.

Insurance is a domain that is heavily dependent on data, large quantities of it, in their operations in an asymmetric market where they want to be able to determine the risks appropriately so that they can manage them well. Utilizing automated techniques also has the benefit of cutting down on the expenses of such work in the sense that the firms can serve a larger number of customers with the same amount of human staff. This means lower costs for customers making some of these instruments potentially more accessible while also reducing the lead time for getting back decisions on their applications.

They also touch on the importance of having more consistent job descriptions that can help to better inform what gets taught in the education system and what competencies future workers focus on, ultimately creating a workforce that is more aligned with the fast-changing environment.

# 6. Misinformation

**Opening Remarks** by Marianna Ganapini, PhD (Faculty Director, Montreal AI Ethics Institute)

The spread of misinformation online has been one of the most debated topics of 2020. In particular, the dissemination of conspiracy theories and false information about the US Presidential Election has raised questions about the role of social media companies in moderating the content on their platforms. In 2021 we face the new challenge of making sure that enough people get vaccinated for Covid-19, and the rampant misinformation surrounding vaccines is an element of constant worry for health officials around the World. So now more than ever, we need to engage in a fruitful discussion on how to best understand and address the problem of misinformation.

There are three main questions at the center of any debate about misinformation, fake news, and the like:

1. Why do people share fake news online at all?
2. What are the political and social consequences of the spread of misinformation?
3. How can we stop misinformation without infringing on people's freedom of speech?

The pieces presented in this chapter touch upon each one of these questions. The first two papers in this chapter tackle the question of why some people are inclined to share and trust fake news and debunked sources. Though some of the fake stories and conspiracy theories are the result of coordinated disinformation campaigns, it is surprising to see 'regular' citizens happily spreading ideas that are false and at times obviously so.  The answer to the why-question is more complicated than it may seem at first. It turns out that some people share fake news because they actually *believe* it to be true. The first two papers in the chapter point to recent work done in psychology that suggests that there are two main, alternative psychological forces behind people's trust in fake news. There are studies that indicate that we often fall prey to our motivated reasoning tendencies: we believe what we want to believe or what our social group believes. As a result, some of us accept and share stories that are unlikely to be true because and when they fit a narrative that we subscribe to -- they 'make sense' given what they already believe.

However, psychological studies indicate that there is another possible explanation for why we are susceptible to misinformation: when on social media, we are lazy reasoners and fail to engage our best analytic skills. As one of the papers suggests, this tendency might also partially explain why older adults are more prone to share fake news than younger folks: because of some memory or cognitive deficit, older people might be more susceptible to accepting fake stories if these are repeated and presented to them enough times.

Though quite compelling, these psychological explanations don't tell us the whole story because some people don't in fact believe what they share. In contrast, they spread fake stories because they do not understand social media as a means to share true information. The communicative function of social media is in fact a very complicated topic of discussion, but one thing is clear: for some people the point of online-sharing is to have fun and entertain one's audience or show their support for a certain group or ideology without trying to deliberately inform or misinform anyone.

Let's turn to the second question: what are the consequences of rampant disinformation? There is quite a bit of discussion on the political implications of fake news as many believe, for instance, that widespread misinformation represents a threat for our democracies because fake stories and conspiracy theories sow distrust in our system and in the values it is meant to represent. Similarly, this chapter focuses on how fake news can be used for political purposes by focusing on the situation in the Philippines and in India. The piece "In the Philippines, fake news can get you killed" points out that in the country Facebook guarantees access to the internet and that, though Facebook, the president spread fake news about his political opponents making them a target for intimidation and aggression.

Also in India, social media platforms are used to coordinate lynching and attacks against enemies. As pointed out in the piece "India's Lynching Epidemic and the Problem With Blaming Tech", social media bear some responsibility in all this as they have clearly made worse an already very tense situation. Although platforms need to promptly tackle the violence that spread on their watch, we also need to understand the social and political roots of this violence and avoid the temptation to simply blame the internet for what is a long-standing social problem in India. In other words, we need to better understand the role misinformation plays when, for instance, it is used as a tool to incite violent riots or as a means for political propaganda.

What can be done to stop online disinformation then? In this chapter, the reader will find a number of pieces that offer solutions to fight this widespread problem. For instance, Microsoft has designed technical tools to detect deep fake videos. On the educational side, more and more effort is spent to include media literacy in the K-12 school curricula, while also trying to inform the public on how to best guard themselves against misinformation. An important outstanding question concerns the role of social media platforms in doing content moderation: how much intervention and content moderation should we expect and want from social media companies? Is content moderation a threat to freedom of speech? We expect concerns to become more and more central in 2021 and we will address them. At the same time, we also firmly believe that a comprehensive, well-rounded (psychological, social, and political) analysis of the problem will be needed in order to successfully understand and fight misinformation.

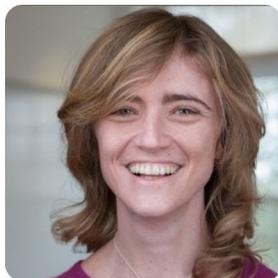

**Marianna Ganapini, PhD (@MariannaBergama)**
Faculty Director
Montreal AI Ethics Institute

Dr. Marianna Ganapini manages curriculum development and public consultations at the Montreal AI Ethics Institute. She is also an Assistant Professor Of Philosophy at Union College. Her main areas of research are Philosophy of Mind and Epistemology. She has numerous peer-reviewed publications and she is currently working on several projects on disinformation, content moderation, reasoning and human irrationality.

# Go Deep: Research Summaries

**The Cognitive Science of Fake News**
([Original paper](#) by Levy, Neil L., Robert M. Ross)
(Research summary by Andrew Buzzell)

**Do people believe fake news?**

We tend to study fake news by surveying people, and yet there are known challenges determining beliefs from behaviour – there are studies that show behaviour often fails to track assessed political beliefs, and the assertion of political beliefs is often a form of cheerleading rather than sincere agreement, a phenomenon that has been called "expressive responding" (Berinsky, 2018; Bullock et al., 2015). There is empirical evidence that reports of political belief frequently are instances of expressive response.

Other challenges to self-reporting are the extent to which motivated inference affects our responses – where we use heuristics and biased sampling to engage in belief construction in the context of the survey or interaction in which the belief is sampled.

There are substantial challenges to determining the real extent to which people truly believe fake news.

**What explains this belief?**

A tempting form of explanation for belief in fake news is the *deficit model*, that given limited cognitive and epistemic resources we become susceptible, but empirical evidence shows that similar deficits do not yield similar tendencies to believe fake news when there is a partisan framing. Kahan (2016, 2017) argues that we can explain this by appealing to *identity protective cognition* – the problem isn't a limitation of our cognitive resources, but the values that inform our deployment of them. The paper assesses empirical evidence supporting and challenging this view and suggests that this is still an open question.

**How might belief in (and spread of) fake news be prevented or reduced?**

There is a substantial empirical literature on the efficacy of correction and the perseverance of belief in the face of interventions such as fact-checking and warning labels. These efforts can have there kinds of negative consequences:

1. **Backfire effects:** the presentation of corrective information can result in increased belief in the false proposition, however, there is conflicting evidence for the strength and prevalence of this effect.
2. **Implied truth effects:** fake news that is not labelled or corrected becomes more convincing, a significant problem given the challenges of deploying corrective measures at internet-scale.
3. **Tainted truth effects:** erroneous corrective efforts can reduce belief in veridical news

Another kind of intervention tries to nudge the consumer of news into a cognitive state that is less likely to be influences by identity protection and motivation, either by inducing deliberation (Bago et al., 2020) or nudging the consumer to evaluate content in terms of its accuracy (Pennycook et al., 2020. This approach has some evidence that demonstrates efficacy. *Inoculation theory* is another approach to preventing belief in fake news, by exposing them to less persuasive forms of it, for example in the form of games.

**Summing up**

The article concludes its survey of the cognitive science of fake news by observing that even where we might find some evidence that analytic cognition reduces belief in fake news, there are further questions as to the relation between belief and the behaviour of sharing and distributing it. Empirical research on the relation between credence and sharing behaviour is inconclusive.

A particularly interesting takeaway is the need for researchers to critically appraise their laboratory results, and, in particular, to attend to more nuances propositional attitudes we can adopt towards news, such as cheerleading, trolling, and other forms of expressive response.

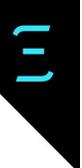

**Aging in an Era of Fake News**
([Original paper](#) by Nadia M. Brashier, Daniel L. Schacter)
(Research summary by Alexandrine Royer)

Older adults are popularly accused of being the first to fall for fake news. Statistics from the 2016 U.S. elections confirm this widespread belief, with older adult's Twitter feeds containing the highest counts of fake news. Those aged above 50 vastly overrepresented among fake news "supersharers." Blaming it on an older adult's cognitive decline is only part of the answer. As pointed out by the authors, people of every age will use mental shortcuts to evaluate incoming information's veracity. When seeing false statements repeatedly, it becomes easier to believe them. Older adults are more source-memory deficient, at risk of forgetting the details about the source of information and whether it was fact-checked. If seen numerous times, the original false statement will be fresher and feel truer in their minds (i.e. fluency) than the corrective information accompanying it. Fact-checking information does not necessarily shift people's belief in fake news.

We must rethink our strategies for coping with the unrelenting influx of fake news beyond adding corrective fact check measures. The authors point to research by Skurnik et al., which suggests that older adults, when shown statements identified as false repeatedly, tend to paradoxically list them as correct if later asked to evaluate the claim. With their accumulated knowledge throughout the years, older adults will reject statements that go against facts they know about the world. The authors refer to a study by Allcot and Gentzkow (2017), where older and younger adults were presented with fake headlines following the U.S. elections. Older adults performed better than younger counterparts in discerning true versus false headlines at first glance. Their successful performance suggests that information repetition, with viral news stories popping up regularly in their feeds, along with some memory failures, are more likely the cause of older adults' tendency to believe in fake news.

Other factors to consider are the makeup of older adults' social media networks and their goals when using these platforms. As you grow older, your social circle tends to narrow, yet your interpersonal trust in people will grow. Older adults will be more susceptible to bots and questionable pages made to appear as real accounts. They will assume that the information shared by social media by friends or acquaintances is factual unless they are given cues about a person's character. The paper points to a study by Skurnik et al. that social context, and the character of a given person, can have a longer and more lasting impression than "true or false" tags. Instead of debunking each of Donald Trump's "alternative facts," stating that the President averages 15 false claims per day in 2018 may be more beneficial to older adults. The authors also suggest that older adults, when interacting online, may cast aside the questionable factual elements about a statement or article or candidate, to pass on a moral message to their younger followers. Older adults can perform well in analytical thinking tasks, but this may not, along with their own social motivations, guard them against misleading content on social media.

A final factor to consider is the digital literacy divide. Older adults are still new to the internet, with Americans over 65 using social media going up from 8% to 40% in less than a decade. They are still learning their way around social media. Fake or sponsored new stories and manipulated images can be difficult to discern across all ages—only 9% of readers spot sponsored news stories. The authors mention research by Fenn et al. and Derksen et al. that claims which appear alongside photographs, even if they do not confirm the claim, are more likely accepted, and this truthiness effect will persist across the life span. Pictures also tend to incentivize users to share information. Older adults do not purposefully intend to share false information. The authors refer to a study by Pennycook et al. that shows that older adults self-report themselves as less willing to share fake news than their younger counterparts. This discrepancy between their online behaviour and their actual intentions may reflect, according to the authors, a misunderstanding of how algorithms work and what sharing on the platform implicates.

Brashier and Schacter's review of psychological studies relating to older adults' online behaviour should guide all those working in the field of disinformation. While it may be tempting to blame older adults' cognitive deficiencies as the main culprit for fake news sharing, the reality is much more nuanced and complex. Fake news sharing is likely to intensify in the future with even more sophisticated technology, and in combination with America's aging population, there will be a growing population susceptible to spreading disinformation. Psychological science will allow us to glean meaningful insights into how to stop the current misinformation crisis with more tailored strategies that account for both the social contexts, the goals and motivations behind older adult's online behaviour.

## The Political Power of Platforms: How Current Attempts to Regulate Misinformation Amplify Opinion Power
([Original paper](#) by Natali Helberger)
(Research summary by Alexandrine Royer)

Since the advent of social media, lawmakers have struggled to keep up with new communication technologies and the whirlpool of informational chaos they create. The risks posed by misinformation to democracies have been well-documented, and recent regulatory initiatives, according to Helberger, have failed to cause sufficient friction in the well-oiled fake news machine. Digital media platforms erode the watchdog function of denouncing political power excess ascribed to traditional media. For Helberger, current regulatory frameworks only seem to add fuel to the fire by failing to view social media platforms as "political in their own right." Rather than focusing on the users, governments should treat digital media platforms as political entities capable of exercising opinion power over a global online audience.

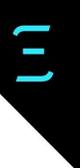

For Helberger, much of the current tactics aim at reining in the power of big tech are grounded in the realm of antitrust and competition law. Facebook, Alphabet (i.e. Google), and Twitter have annual revenues standing respectively at $70.7 billion, $161.9 billion and $3.46 billion, which go well beyond the yearly GDP of small nation-states. The global economic force of Big Tech is manifested into efforts to prevent any new forms of legislation that might disengage its users and/or hinder their growth. This year, in its *Open Letter to Australians*, Google actively encouraged its users to speak out against a proposed Australian law that would require search engines such as Google to pay Australian media companies for using their stories on their site. Google even went so far as to insert a yellow hazard warning below the main page search bar, with the accompanying message "the way Aussies search every day on Google is at risk from new government regulation."

Digital media platforms have not shied away from exercising their political muscle. Yet, we tend to treat these interventions in the political space as examples of lobbying corporate entities rather than governmental or political actors showcasing their right to population control. Helberger encourages us to view digital platforms as governments, who have their pool of citizens and own legal corpus (i.e. the Terms of Use, Privacy Policies and community guidelines), and as wielders of opinion power. Helberger draws on this concept of opinion power from the German Federal Constitutional Court (i.e. a translation of Meinungsmacht). It is defined as "the ability of the media to influence individual processes and public opinion formation." Germany uses this legal notion of opinion power to argue that any imbalances in public opinion formation, whereby one discourse predominates over another, poses a severe threat to the pluralistic media landscape and places democracy, and human lives, in peril.

Germany's 20th-century history is a sober reminder of the destructive power of widespread propaganda. Yet, the same patterns of disseminating hate speech, monopolizing public discourse against identifiable minorities and curtailing citizen's access to verifiable information were used via Facebook during the genocide against the Rohingya in Myanmar. Facebook did not only mediate users' opinions and serve as a digital infrastructure that influenced the political processes; its very existence enabled the movement to gain traction, and its algorithms sped up and shaped the unfolding of the violence. For Helberger, social media platforms ought to be treated as political actors with their power of opinion – they are political forces in and of themselves.

Germany, France, and the UK have all introduced legislation requiring greater accountability and oversight from big tech in monitoring the content on their platforms through considerable fines and penalties. By doing so, these countries legitimize digital media platforms' rights to act as governors of public opinion, allowing them to become "the new self-government of the online global population," hence increasing their power to dictate and determine how civic discourse takes shape. According to Helberger, the European Commission's *Shaping Europe's Digital Future* strategy is insufficient in dealing with opinion power. Such proposals fail to cover Facebook and YouTube's capacity to commission content and conclude deals with rights holders. One potential instrument to curtail these platforms' opinion power is through media concentration law, but it will require adjusting the traditional measurements of audience reach and ownership limitations.

Helberger concludes by asserting that "dispersing concentrations of opinion power and creating countervailing powers is essential to preventing certain social media platforms from becoming quasi-governments of online speech, while also ensuring that they each remain one of many platforms that allow us to engage in public debate". Indeed, the concept of opinion power and its associated legal workings are a welcome and refreshing addition to the public debate on how to regulate misinformation. It also asks lawmakers to critically assess whether their policy efforts are simply strengthening and cementing the political powers of Big Tech.

## Go Wide: Article Summaries (summarized by Abhishek Gupta)

**New Steps to Combat Disinformation**
([Original *Microsoft Blog* article](#) by Tom Burt, Eric Horvitz)

This announcement unveils both technical tools and educational efforts to enhance our defenses when it comes to combating problematic information. On the technical side, Microsoft has partnered with a number of organizations to create a video authenticator that detects deepfakes by analyzing the boundary between the deepfake elements and grayscale elements and subtle fading that would be imperceptible to a human. The system was trained on the Deep Fake Detection Challenge dataset and the Face Forensics++ dataset.

Another technical tool created by the team helps users identify whether the content that they are watching is authentic or not. It adds digital hashes and certificates to the content that travel along with the content as metadata allowing creators of the content to stamp their productions. On the other hand, there is an accompanying consumption tool that can help users verify those hashes and certificates to assert whether the content is authentic or not, and to provide details on the creators of the content.

Finally, given that detection and evasion are an inherently adversarial dynamic, technical solutions alone are not enough. To that end, the team has also partnered with organizations to release media literacy tools that will help to educate the public on signs they should look out for, current capabilities, and other critical thinking skills that will be essential in this fight against problematic information. An implementation of NewsGuard has also been expanded to increase the efforts of rating news and media sources along nine journalistic integrity criteria, essentially creating nutrition labels and red/green marks that can help consumers discern the veracity and authenticity of content coming from those sources.

*(Full disclosure: Our founder Abhishek Gupta works at Microsoft. However, the inclusion of this article in the newsletter is unrelated to his employment and not paid for or endorsed by Microsoft)*

## A Grassroots Effort to Fight Misinformation During the Pandemic
(Original *Scientific American* article by Lindsay Milliken, Michael A. Fisher, Ali Nouri)

The WHO mentioned that we are facing an infodemic at the same time as the pandemic that has hindered our ability to effectively combat the spread of the virus because of fragmented and conflicting advice on how to best protect ourselves and those around us. Specifically, the rampant misinformation has caused a lot of grief in terms of lives that could have been saved. This initiative mentioned in the article by the Federation of American Scientists (FAS) has been a public and successful demonstration in the power of grassroots initiatives.

With their service to answer the FAQ on this issue through automated means that have answers sourced from domain experts, they have made a true positive impact in the information discourse on this topic. By supplementing this approach with the ability to "learn" over time as more and more people parse the existing database and then asking questions directly for those that haven't been answered, the information only becomes richer over time.

Finally, the approach undertaken by the FAS and their partner organizations showcase how crowdsourced mechanisms can be mobilized and scaled to meet emergent community needs, something that the field of AI Ethics can also borrow from.

## India's Lynching Epidemic and the Problem With Blaming Tech
(Original *The Atlantic* article by Alexis C. Madrigal)

We've previously covered how bias and other problems in algorithmic systems aren't just technical problems and require a more in-depth understanding of the surrounding socio-technical ecosystem within which the technologies are deployed. In the case of the proliferation of technologies like Whatsapp in places like India, it is easy to jump to conclusions on how violence can be mobilized quite easily in an online context through forwarded messages that translate into real-world harm.

An important consideration here is to think about technology as being merely an enabler that amplifies the problems that the community faces, a lot of the violence in the Indian context can be attributed to a distrust in the governmental mechanisms and hence people taking justice into their own hands and mobilizing support through platforms like Whatsapp. A historical analysis reveals that the number of violent incidents has been affected significantly through higher levels of communication. Perhaps we are more aware of them now since smaller incidents and distant incidents are also able to pop up on our radar, but research points to the absolute numbers staying about the same.

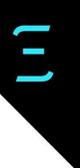

And it is not just new-fangled tools like Whatsapp that can be put to blame, in other places SMS has also been blamed to be a "weapon of war", though with Whatsapp, the potential to use richer media than just black-and-white text does have ramifications. While technological interventions, for example limiting the number of people to whom you can forward messages at a time does serve as a starting point, laying all the blame on the technological tools erodes the responsibility and actions that we need to take to affect larger societal changes that can help us build a more peaceful and just society.

## How to Guard Your Social Feeds Against Election Misinformation
([Original *Vox* article](#) by Rebecca Heilweil)

This article provides some very handy tips for those who want to navigate this polluted information landscape relatively unscathed. Setting the stage up front in the article, we can't rely on platform-led interventions since they are mostly focused on specific instances and handling them rather than sweeping changes that are required to clean up the state of misinformation on the platforms. As a starting point, users can move away from questionable sources by unfollowing them on these platforms. They can also be more careful of so-called trusted sources if they have been spewing misinformation.

For example, reliance on labels of whether something is true or not can be a way to navigate the system, it only means that you have some indicators pointing to the veracity of the content, it doesn't always mean that the content itself will be taken down. Echo chambers confirming our beliefs are too powerful to resist and can overshadow the effectiveness of measures such as truth-labeling. Even though platforms like Facebook now prioritize content from your friends and families over those from third-party sources, it doesn't stop that information from showing up in your feed when one of those connections share it on their pages.

To the credit of Facebook, they do have initiatives like "Why am I seeing this?" and interstitials that help to provide a bit more context around a piece of content, they might not be as widely used as we might imagine - akin to our aversion to reading terms and conditions prior to using software. Relying on markers like the verified checkmark doesn't mean that the entity is an authority on a particular subject — it is just an indicator that they are who they claim to be. Finally, helping out people within your network can also help to make our information ecosystem healthier.

### In the Philippines, Fake News Can Get You Killed
([Original *Rest of World* article](#) by Peter Guest)

The Philippines has been an interesting case study with press freedom and freedom of speech: extreme outcomes like lynching and death are a possibility if you're suspected of spreading misinformation on social media. Since the rise of the new Presiden, government instruments have been utilized to inflict death upon people who are dealing drugs, often in violation of human rights. In this article, we get a glimpse into what happens when someone is suspected of spreading fake news or is critical of the government.

One thing that is worth noting about the state of the information ecosystem is that Facebook is the primary gateway for people to access the internet. Facebook offers free access to their site without incurring data charges which makes the internet experience for a lot of people limited to their walled garden, often to the point of unawareness that there is an internet that exists outside of the Facebook platform. This also means that the information on Facebook has an outsized impact on the people in terms of what they learn about their country.

While the platform disengages from fact-checking political information on the platform, there is very subjective enforcement of community standards that are applied unevenly. Journalists are sometimes unfairly penalized: some veterans remember carrying toiletries with them in case they are captured and put in prison. After the Cambridge Analytica incident, Facebook has limited third-party access to the data on their platform, which has also made it hard for researchers to audit the platform for concerns. Often, they have to jump through many hoops just to get basic access. If we're to combat this problem, we will need more transparency and accountability from the platform. Initiatives like the Oversight Board might seem like a good step in a positive direction but without real power, they might be just signalling of virtues without making any meaningful impacts.

# 7. Privacy

**Opening Remarks** by Connor Wright (Partnerships Manager, Montreal AI Ethics Institute)

In this chapter, we reflect on one of the mainstay topics in the AI space, and for good reason. Privacy in today's world carries an especially precarious nature, stemming from the cybersecurity threats being like nothing we've ever dealt with before in cybersecurity, its effects then being multiplied by the COVID-19 pandemic. With accessibility and the ways forward being offered proving the two other major themes of the chapter, the other centres itself around the pandemic and the loss of control that comes along with it.

Given the nature of the times we live in, a severe lack of control has almost been normalised in our daily lives, especially in the privacy arena surrounding COVID-19. The chapter observes how the requirement of technological innovation has meant that privacy concerns have been given a back seat in their development, both in terms of the need for the technology and the possibility of profit. The fear of data collection being tailored to business' interests, as well as questionable tech innovations being justified under the banner of "the public good", are reflected upon, elaborating on the severe absence of an opportunity for the population to control what happens to their data. One shining light exposed by the chapter can be seen in Portland banning all facial recognition technology (FRT) in public spaces, but simply banning the technology doesn't eliminate the need to tackle its current and potential privacy concerns. To do this, the chapter is focused on increasing accessibility.

As mentioned in the chapter's piece on voice phishing, user defences against cybersecurity attacks is one of the most effective ways of combating said attacks. However, this becomes impossible without making the data both accessible and intelligible to the public. The chapter makes reference to FRT being used at U.S. border crossings where limited information was available to passengers about when and where the technology was being used, as well as a lack of understanding about what happens to their data once it is captured. Even when such information is offered, it needs to be done in a way that's more perspicuous to the wider public, instead of requiring a deep understanding of the legal vernacular used. Without accomplishing both accessibility and intelligibility, the dangers can be seen within the chapter where not only can we not understand the privacy issues surrounding our data, but we also cannot help to protect it.

Without focussing solely on the negatives, the chapter does offer what we here at MAIEI are very passionate about, practical ways forward. It highlights how there is no requirement to blindly adhere to the obligations surrounding consenting to the use of FRT, demonstrated in the social liberty group Liberty's case against the use of FRT by the South Wales Police. Sandboxing, defence in depth, and providing policymakers first-hand voices can all be seen as concrete ways of progressing against the pitfalls of the privacy arena. India's personal data privacy law, as well as the aforementioned Portland, can both be deemed as steps in the right direction, striving for meaningful privacy that's tailored to the user, rather than blanket statements about the lack of privacy considerations being "for the public good".

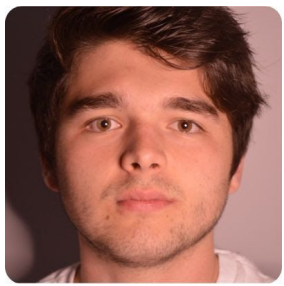

**Connor Wright (@Csi_wright)**
Partnerships Manager
Montreal AI Ethics Institute

Connor manages partnerships at the Montreal AI Ethics Institute, working with organizations from all over the world to collaborate on public consultations and other events around AI ethics. Previously, he was a facilitator for academic conferences by the European Association for International Education.

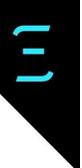

# Go Deep: Research Summaries

## "I Don't Want Someone to Watch Me While I'm Working": Gendered Views of Facial Recognition Technology in Workplace Surveillance

([Original paper](#) by Luke Stark, Amanda Stanhaus, Denise L. Anthony)
(Research summary by Alexandrine Royer)

The worldwide outbreak of COVID-19 has seemingly changed the future of work by forcing millions to leave the office and transform the home into their designated working space. The sudden transition to work from home has brought its onset of challenges, from being confined to a one-bedroom apartment, juggling childcare, battling feelings of loneliness, and losing track of a healthy work-life balance. As one Twitter user, Neil Webb, adeptly [pointed out](#), "you are not working from home; you are at your home during a crisis trying to work." Amid this crisis, to keep tabs on their employees, companies are deploying digital monitoring systems to ensure that worker productivity remains high. However, the use of digital monitoring is not as new a phenomenon as it may seem, as revealed by this December 2019 study by Luke Stark, Amanda Stanhaus, and Denise L. Anthony. The authors assess whether gender differences exist when it comes to facial recognition technology (FRT) in the workplace and offer a more general discussion of current digital workplace surveillance.

No one likes their boss breathing down their back when trying to accomplish a task. With digital surveillance, you are constantly under the watchful gaze of your employers, yet this monitoring happens in insidious ways, often unbeknownst to you. To understand women and men's perspectives on digital surveillance, the authors drew on survey results from the Pew Research Center to produce a multivariate registration analysis of self-identified female or male respondents' opinions of FRT-enabled workplace cameras. The authors acknowledge that the male versus female binary is a partial and exclusionary definition of gender but note that they were constrained by the categories used by Pew. The survey, dating back to 2015, was distributed among a sample of 461 U.S. adults over the age of 18.

Despite the long history behind factory tools aimed at controlling workers' time and effort, Stark et al. note the drastic changes in the pervasiveness and extent of workplace control brought on by new technologies. Today, employers use facial analytics, workplace screenshots, email and keystroke analysis, and online searches to keep workers in check. Around 75% of U.S. companies monitor worker communications and activities. Surveillance is justified on the grounds of productivity and workplace safety or security. Still, such discourses obscure the power asymmetries present in digital workplace surveillance and its tendency to amplify racial, gender, and class inequalities. The authors offer examples of women who face increasing scrutiny in the workplace through surveillance categories and yet do not benefit from additional production from harassment and assault. They also point to the expanding use of FRT in automated hiring questionnaires by firms such as HireVue designed to rank a candidate's body language and emotional expression irrespective of the racial and gender biases encoded in the software. Excessive workplace monitoring, in addition to creating an environment of distrust, can frequently lead to counter-productive results by generating "anticipatory conformity" among employees or encouraging employees to utilize resistance tactics to avoid managerial scrutiny.

The authors point to a correlation between industries that subject employees to high workplace surveillance and the overrepresentation of low-wage minority and female workers, such as the retail sector, hospital administration, and childcare. They cite earlier research by Bell et al. in 2012 that found women in call centers were more likely to complain about excessive or intrusive forms of personal information collection via emails and CCTV cameras than their male counterparts. From their own data analysis, Stark et al. found that "women are 49% less likely than employed men to say workplace surveillance via cameras with facial recognition software is acceptable". Feminist and gender scholars have pointed to the heightened awareness of women in the workplace, often being the subject of unwanted male gaze and advances. Surprisingly, there were no statistically significant differences between men and women in perceptions of privacy; in the Pew survey, this was measured by a declaration of not wanting to be monitored at work. Both women and men expressed similar concerns over the intrusiveness- the erosion of individual autonomy-, fairness, and totalitarian aspects of workplace surveillance, arguing that FRT could quickly lead to power abuses.

Stark et al. recognize the apparent limitations of their study, noting that the Pew research survey took place before the emergence of the #MeToo movement, which empowered women to denounce in higher numbers workplace sexual harassment and assault. The questions asked in the Pew survey involved scenarios relating to workplace theft, rather than concerns over civil liberties or sexual misconduct, and did not inquire about respondents' occupations. This is a considerable failing as the authors note that "given the distribution of power in the workplace in which managers and supervisors are more likely to be male, the application of workplace camera surveillance technologies would be controlled by precisely those likely to be the harassers." The Pew survey also makes the crucial omission of allowing respondents to self-identify as nonbinary or genderqueer, with digital systems often misgendering and adversely impacting queer and trans people.

While academic research on digital workplace surveillance is likely to emerge in the coming months, there remains a need for more in-depth studies and inclusive studies of gender-based attitudes and concerns towards digital monitoring. Media outlets are already [reporting](#) how COVID-19 work from home monitoring poses a dangerous precedent that risks nullifying employee privacy, sour employee and employer relations, and threatens mental well-being by incentivizing overwork. In the Canadian context, Calgary-based Provision Analytics saw a boost in clientele, and in the US, San-Francisco tech startup Pragli reported similar growth. With the increasing implementation of FRT and emotion recognition technologies in corporate, commercial, and retail settings, it is high time lawmakers intervene to regulate workplace surveillance. Whether the use of such systems is ethically permissible also ought to be a broader civic debate.

## Snapshot Series: Facial Recognition Technology
([Original paper](#) by Centre for Data Ethics and Innovation)
(Research summary by Connor Wright)

The snapshot series provides a very readable, intuitive, and engaging overview of the facial recognition scene in the UK. The paper splits into 6 sections, stretching from what facial recognition technology (FRT) is, how it works, what the risks and benefits are, and what the future looks like. I will now summarise some of my highlights from the paper, including the difference between facial verification and facial identification systems, and how bias can be sewn into these systems.

Being from the UK, what first struck me was how the report made note of the South Wales Police's (SWP) use of FRT systems being ruled as unlawful by the Court of Appeal. Here, the SWP's use of the technology was taken to court by the civil liberties group Liberty, and was initially ruled as lawful by the Supreme Court, having been deemed as following all the necessary regulation. However, Liberty then won the appeal with the SWP being ruled to have breached the Human Rights Act, the Data Privacy Act, and the Equality Act. Not only does this show the precarious nature of FRT, but also how this technology is not immune to civil protest. Such technology is often surrounded by a false obligation to commit to its usage by authorities, where civil society has no question on the matter. Yet, Liberty has shown that no such obligation exists, and how civilians can feel empowered to call out governments and big corporations on their use of seemingly untouchable technology.

There is nonetheless still an important distinction to draw in the use of FRT. Facial verification systems utilise a template of an already scanned face (such as on iPhones), and scan the face being presented to see if it matches the template. Facial identification systems on the other hand, are not looking for a face in particular. Instead, it operates on a one-to-many matching basis, whereby a face template is utilised to sift through millions of images in order to reveal which faces match. Furthermore, facial verification is more likely to be automated, with a match proving enough to warrant an action (such as unlocking your phone), whereas facial identification is more likely to be augmentative, being overseen by a human before a decision is made. Facial identification is then further split into live and retroactive recognition. Here, when talking about FRT and its problems, this is more likely to be centred around live facial identification, as opposed to retroactive systems, or the system on your phone.

One problem faced by live FRT in particular is that of bias, and one of the key ways this is woven into the system is through a non-representative data set. What gave me a lot of food for thought was how the paper highlighted that the FRT can be as accurate as it wants, but accuracy does not guarantee the elimination of bias. Even if the data set was accurate, and the data set contained over 10 million images, this will still not eliminate the presence of bias if the data set is homogeneous. Such bias can only then be exacerbated by private data collection efforts, which will be tailored to the company's interests.

Despite all this doom and gloom, the paper did shed some positive light on the use of FRT. The technology is able to scale the use of security infrastructure by improving efficiency, as well as aid the already swamped police forces around the world. Not only this, but there are multiple laws governing the use of the technology within the UK thanks to the GDPR agreement spanning both private and public uses of FRT, as well as the laws broken by the SWP mentioned above.

These reassurances certainly provide a welcomed respite in the FRT debate, and the paper offers a positive image of the ongoing conversations. Nevertheless, the paper makes sure to emphasise how this is unfortunately not the only aspect to take note of. FRT can be of great benefit to society, but the elimination of bias and the new legislation proposals still have a long way to go to guide FRT to this destination.

## Project Let's Talk Privacy

([Original paper](#) by Anna Chung, Dennis Jen, Jasmine McNealy, Pardis Emami Naeni, Stephanie Nguyen)
(Research summary by Abhishek Gupta)

One of the things that I really liked about how this study was conducted and how the results were communicated is that they choose to avoid using the term *user* which, as they explain, carries with it a connotation of someone being a *research subject* to be studied, whereas we are talking about real people with rich lives that face impacts from these systems.

**Recommendations for policy makers**

If you are a policy maker who is reading this, my key takeaway from this is the importance of obtaining first-hand accounts from stakeholders. Often, in policy making we rely on voices to gain an understanding of what the community thinks about something, but we fall short of gathering first-hand responses and instead relegate our insights gathering to second- or third-hand voices who might not be truly representative of the concerns of the community.

Collaborating directly with the people in the field will help to *sandbox* and test the policies iteratively to arrive at something that works well for the needs of the community, not what we imagine their needs to be.

As is the case with all laws, a balance needs to be struck between the specificity of the language describing certain pieces of technology but also being open enough so that future evolutions meaningfully fit within the framework that is defined in the legislation.

*Communication by policymakers should happen in a fashion that makes their message perspicuous to the audience. Too often we have this communication in the form of speeches and press releases that require deep familiarity with their vernacular to make any sense of it.*

This doesn't mean *vulgarization* to the point that we lose nuance, it just means having a framework in place that emphasizes the need for such communication practices.

A premortem, or analyzing all the ways that the policy can go wrong is an engaging and fruitful way to surface pitfalls when the policy goes into effect. Especially surfacing concerns that are otherwise not covered in the theoretical policy making space can come alive from this speculative process and lead to the creation of more robust policies.

**Recommendations for design practitioners and technology organizations**

- Be upfront about the rights that the user has and then map those to appropriate controls in a transparent way so that they can exercise those rights.
- Having a shared vocabulary to explicate data governance will be important across activists, technical, and policy stakeholders.
- We can't place the onus of upholding rights on those who are impacted by their violations, this mantle needs to be picked up by those who have power, which is often companies and governments.
- Reducing *click fatigue* and creating empowerment that people can really act on is a tangible goal that industry and regulators should work towards.

**Relation to existing work**

A lot of work has already been done in the space on the subject of dark patterns that seek to subvert the autonomy of people to nudge them towards behaviours that benefit the platform, more than they do the individuals. An example mentioned in this paper talks about controls for privacy might be hidden in obscure menus discouraging their utilization but still meeting compliance requirements for providing the option to the user.

**Privacy bills covered**

The researchers looked at the Consumer Online Privacy Rights Act (COPRA), Online Privacy Act (OPA), and Social Media Addiction Reduction Technology (SMART) to meet the goals of their research.

They picked ones that focused on strengthening privacy controls, and advocating for platform design changes as the beachhead for this study.

**What was common to all the bills?**

- Having a multi-layered approach, akin to the *defense in depth* principle from cybersecurity where you have multiple checks and balances so that you don't have one single point of failure.
- My personal preference for consent mechanisms is the notion of progressive disclosure, the idea that you reveal requisite concepts and implications just-in-time (JIT) so that you don't overwhelm the user into making an uninformed choice.
- Related to the above point, we need to make sure that we don't *desensitize* the user to privacy concerns by bombarding them with unnecessary notifications and controls.

- The need for more clarity on the terms used within the bill.

They prototyped the bills to take a look at how the clauses within the bill could actually map to features in platforms and how people perceive those, and judge their efficacy. It also in the qualitative aspect of the study paid attention to *proof quotes* to bolster arguments and *provocative quotes* to illustrate points using examples.

**What was unique to each of the bills?**

SMART
- While heavy on the notions of control and notice, an interesting tension to highlight here was the tradeoff between maintaining user agency and coming off as being too paternalistic.
- There were some highly specific recommendations, for example the 30 minutes timer on browsing a feed that felt too specific and could lead to clashes as the platform and preferences evolve in the future.

OPA
- Enforcement of individual's rights and considerations of marginalized populations in case they are left out of the discussions was an important facet of this bill.
- In advocating for the rights in the first place, the proposal of having a new agency might be too much of an ask, perhaps expanding the powers of a body like the FTC is a better approach.

COPRA
- If we take the view that privacy is a fundamental right, then we must embark on removing as many of the cost barriers as possible in making privacy attainable for everyone.
- The notion of *Duty of Loyalty* was ambiguous at best: there are many interpretations for it and it requires more clarity for it to be actionable.

**What is privacy anyways?**

So, one of the things that particularly caught my eye was the expression of privacy in 4 forms as: solitude, reserve, anonymity, and intimacy. Most conversations that seek to trivialize the notion of privacy being *something that is an artifact of the pre-digital era* or those who support the argument of *I have nothing to hide* can benefit from this framing.

While the researchers had surveyed people from many walks of life and experiences, they did find some commonalities: control of your data, privacy as a mechanism to achieve fairness, and risks vs. benefits of having privacy in terms of the products and services that we have and we could have in the future.

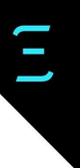

Such a multi-dimensional approach to privacy requires that the definition also have multiple dimensions and some of the related work and background talked about in the paper does a good job of familiarizing the reader about it. Such a formulation also has the benefit of *meeting different people where they are at* in terms of their needs and what matters to them when it comes to the notion of privacy.

Some of the people surveyed as a part of this study did frame privacy in the context of human rights as well which I think is a powerful approach since it can use all the levers that come with the enforcement of human rights to push the agenda for having meaningful privacy.

**Generational differences in the perception of privacy**

We often get to hear that young people don't care about privacy because they share their life's ongoings so freely on the internet. That is only a single dimensional view and this is talked about in the paper to the extent that different generations have different perceptions of what privacy means. Keeping that in mind will help us make better decisions in terms of how we should communicate the policies and their impacts on the design of the platform.

**Where does well-meaning legislation fail?**

Reiterated in the report, and something that I have experienced in my own work is a lack of interdisciplinary collaboration that leads to conflicting definitions and recommendations that ultimately fail to hit the mark from a technical perspective. In [work](#) that I did with a former colleague [Mirka Snyder Caron](#) for the Office of the Privacy Commissioner of Canada, we combined legal and technical expertise to make recommendations that weren't limited in their applicability, precisely because we were cognizant of the shortcomings of a uni-dimensional approach.

**Where do well-meaning technical implementations fail?**

This is an interesting point of discussion because sometimes apps that are secure don't have the fantastic and slick user experience that we have come to expect from apps that might be made by companies that exploit user data. To a certain extent, if we can *hack* this design problem, we can reach a place where people will choose these alternatives because they offer everything and give little to no reason to stick to less secure and non-private solutions.

**Conclusion**

What I liked about this study was that it really paid attention to the inherent tension (at least what I believe to be the case) between technical practitioners and policymakers and how consumers get stuck with subpar interactions from a user experience and exercising of their democratic rights, including that of privacy.

The recommendations made here and the approach taken, I think, has applications beyond just the field of privacy as well and this study can be a great instrument for people seeking to learn more about how to effectively communicate policy decisions and deliberations to their constituents.

**Implications:**

- Concretely, from an AI ethics standpoint, one should pay attention to the ideas of progressive disclosures of privacy notices and consent, and put in place simple-to-use controls to allow users to exercise their data rights.
- Stakeholder consultation, especially those whom you can identify from the premortem process for policy analysis should become an integral part of the early stages of your AI lifecycle.

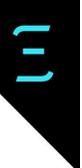

# Go Wide: Article Summaries (summarized by Abhishek Gupta)

## China: Big Data Program Targets Xinjiang's Muslims
(Original article by *Human Rights Watch*)

The Chinese government has a big data surveillance program called the Integrated Joint Operations Platform (IJOP) which flagged thousands of citizens, who were then detained and sent to "political education" camps in Xinjiang where they are being detained. A recent list of these citizens of over 2000 detainees from Aksu prefecture was provided to Human Rights Watch (HRW).

Those on the list, including Xinjiang's Turkic Muslims, are being detained without justification. The Chinese government, however, frames it as predicting and preventing criminal behavior. This mass surveillance and unjustified detention goes against both China's constitution as well as international human rights law.

Based on anonymously sent data that HRW received in 2018, 80% of the residents in Aksu prefecture are Uyghurs. The list also suggests that Chinese authorities have been cracking down on Islamic religious activity, including reciting the Quran or wearing a veil.

"The Chinese government should immediately shut down the IJOP, delete all the data it has collected, and release everyone arbitrarily detained in Xinjiang", said Maya Wang, senior China researcher at HRW.

## Voice Phishers Targeting Corporate VPNs
(Original *Krebs on Security* article by Brian Krebs)

Voice is the next frontier where phishing attacks are going to be mounted. *Vishing* is the application of a likeness of voice to craft phishing attacks. In a time where work is within the home front, and people are using VPNs to connect to work resources, attackers are taking advantage of new attack surfaces that open up. The article mentions how new hires are particularly susceptible to such attacks because they are not yet familiar with legitimate corporate resources and those that are not.

The article provides a lot of details on how such actors mount their attacks, including the use of multiple actors working in cahoots, one doing social engineering over the phone and another directing the unsuspecting employee to a phishing website to steal their credentials. In cases where the attackers fail, they have a chance to try again with another employee, each time gaining a bit more information on the colloquialisms and tooling used by the organization. The researchers also found that the attackers have grown more and more sophisticated in penetrating through the networks but are still learning the most effective ways to cash out.

One of the things that we think will become more prevalent in the future is the impersonation of someone's voice, someone that you trust, and then stealing credentials and other valuable information that way. With the advent of more AI-enabled tools that democratize this ability, user awareness is one of the best defenses that we have.

### COVID-19 Surveillance Strengthens Authoritarian Governments
(Original *CSET Foretell* article by Jack Clark)

An intriguing article that applies a different lens to comprehend when we will head towards surveillance infrastructure that gets established during the COVID-19 pandemic. In particular, the use of indicators is exactly the kind of tangible insight that we relish at MAIEI. Adopting a "canary in the coalmine" approach, the article mentions a few things of note. The incentive structures for businesses as revenue streams dried up in various places has helped to fuel their advances into working with players who are looking to establish this infrastructure in place. The research domain is similarly pivoting to doing work in this space due to other funding resources becoming limited.

While the mileage in how these systems get deployed and used will vary in different countries, the more authoritarian regimes will see overt deployment whereas the more democratic regimes will see pushback and calls for fairness, privacy protections, and accounting for other ethical considerations. Finally, the flywheel of AI kicks in where regimes that don't have hindrances to the deployment of these technologies can gain massive competitive data advantages and train more proficient systems. Ultimately as the technology becomes more effective and cheaper, these regimes will become exporters of the technology to the rest of the world, both in subtle and not-so-subtle ways.

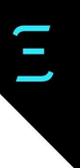

## Whatever Happened to Those Coronavirus Contact Tracing Apps?
(Original *The Markup* article by Sara Harrison)

As the pandemic raged on, many states rushed to implement their own digital contact-tracing solutions. Then Apple and Google announced a joint proposal that offered a unified API for others to build their solutions on top of it. The key benefit being consistency in the application of security and privacy standards which were being created in an ad-hoc manner for other apps. The Exposure Notification System (ENS) from Apple and Google has the benefits of not storing data in a centralized repository and not tracking location information, two of the primary privacy concerns that people speculate has led to low rates of adoption of this technology. Though the ENS is not without problems - for example, researchers have pointed out that it is vulnerable to Bluetooth spoofing and until recently, there was also no requirement for verification of positive test results which triggered the notification flow. The latter has been addressed now with the requirement to have a verification server but the former is still an unsolved problem though researchers are quick to point out that this hasn't been done in practice yet so it remains unclear how much of a risk this is.

While one of the studies cited in the article mentions that even with adoption rates of 20-40%, we could reduce the rates of daily infections, it is still not clear what the actual degree of efficacy of these solutions is. Specifically, some argue that investment in other basic infrastructure instead and raising awareness on the proper use of masks and following social distancing would actually do more to reduce the spread of infection.

## Border Patrol Has Used Facial Recognition to Scan More Than 16 Million Fliers — and Caught Just 7 Imposters
(Original *OneZero* article by Dave Gershgorn)

Facial Recognition Technology (FRT) continues to be featured prominently in discussions on AI ethics, specifically around the issues of bias and privacy. Not much of a surprise in this article that talks about a report published by the Government Accountability Office (GAO) in the US on the use of FRT in border crossings. Specifically, the report found that in scanning 16 million passengers only 7 were found to be imposters. While there might be gains to be had in terms of enabling the border patrol to operate in a leaner and more efficient way, such small gains when compared to the potential bias and privacy pitfalls raise concerns.

There was limited information available to passengers with respect to when they were being subjected to automated FRT. Additionally, the opt-out process placed a higher burden on the passengers and the choice was often framed as a negative for the passengers thus discouraging them from opting out. They were told that they would experience delays and additional security checks. Also, when passengers did choose to opt-out, often there was a shortage of staff to process them manually and limited information on the expectations from the alternate process. While the use of AI brings a lot of potential, it is important to consider who bears the brunt of the negative consequences and if the actors who face those burdens have adequate recourse in case things go wrong.

### Explainer: What Do Political Databases Know About You?
(Original *MIT Tech Review* article by Tate Ryan-Mosley)

When looking at things like the mosaic effect and the ability to collate large datasets about a person, we notoriously have very little information to inform us as to how we are targeted. With the 2020 election in the US, knowing the targeting mechanisms on the internet is essential to ensuring the integrity of our democracy. Political campaigns engage in persistent messaging to modify behaviour, relying on parallel and continuous texts, calls, and social media messages to sway voting patterns.

Research has shown that most US adults are present in these datasets, with data collated from social media, credit card records, and other public information. Some attributes are inferred from statistical norms but their accuracy is questionable, sometimes even to the point that it hinders their usefulness to the political campaign targeting. But, such data still through mechanisms like A/B testing can be refined over time to the point that it is hyper-specific to the user and ostensibly successful in tweaking their behaviour.

In terms of who is allowed to use the targeting tools on platforms like Facebook, there isn't complete clarity on that, but it would appear that the rules apply in an uneven fashion and some political organizations are able to skirt scrutiny and evade the consequences of breaking some of the regulations around campaign finance spending. The polluted landscape of information also plays into the hands of those who want to utilize these tools in a savvy manner. This is only worsened by the disparity between the candidates with varying levels of technology savvy who are able to use this pollution to their advantage.

**Face-mask Recognition Has Arrived—for Better or Worse**
(Original *National Geographic* article by Wudan Yan)

The article makes for a fascinating read because it highlights some of the tensions in the field when it comes to using facial recognition technology, especially when it is framed in the context of the benefits that it might provide as it helps to combat the pandemic.

Some developers are arguing that mask recognition, since it differs from facial recognition, doesn't have many privacy concerns since they are only considering the presence or absence of masks and are doing it purely for the purposes of combating non-compliance. Yet, pervasive monitoring, be that for detecting crime or wearing masks still creates an unnerving context in a physical space, altering the relationships that people have with the spaces around them. This is also problematic since it has the potential to open doors to other forms of surveillance which can be guised under the pretence of doing public good.

One of the companies mentioned in this article has talked about deploying their technology in stealth mode. This is concerning since it inherently creates an asymmetric power dynamic that encourages one person/organization's viewpoint as superior to others. Asking critical questions should be something that is done more publicly and companies who are in the design, development, and deployment of such technologies should be comfortable in responding to these questions in the interest of transparency, especially when they frame its deployment as a public good.

**How It Feels When Software Watches You Take Tests**
(Original *NY Times* article by Anushka Patil, Jonah Engel Bromwich)

While the benefits of digital technology have certainly helped to alleviate some of the concerns around education in a remote environment because of the pandemic, it isn't a panacea and the effects are unevenly distributed. Specifically, those who faced struggles in a regular environment before are facing more aggravated versions of the same because automated solutions aren't inclusive of differing needs. Solutions to proctor exams in a remote environment rely on many intrusive pieces of technology like continuous monitoring of faces, sound analysis in a room, looking at objects present in a room, whether there is someone else entering the room, etc.

For students who have accommodation needs that would typically be met by their educational institution, in a remote environment they are at the mercy of the company providing the software solution and this has led to some horrendous experiences as documented in this article. Students who have tics, who have to take care of siblings, who have darker skin, or any other traits that are not the "norm" on which the proctoring solutions were trained on lead to red flags that exacerbate the exam-taking experience for these students.

While these companies have been making amends as problems are being reported, it begs the question as to why educators and administrators weren't consulted beforehand (or if they were, why some of these problems arise in the first place) that would have made for a more proactive rather than reactive approach. The educational system already places undue pressure on students who request such accommodations, to put them through the wringer where responsibility is outsourced to a third-party is especially unfortunate.

## The Real Promise of Synthetic Data
([Original *MIT News* article](#) by Laboratory for Information and Decision Systems)

Synthetic data certainly has been talked about on and off for the last few years when it comes to privacy protection and enabling research and model development in machine learning without having to sacrifice the rights of people, or doing research in the face of deeply siloed datasets, due to legislative or technical reasons. An apt comparison made in this article talks about the requirements for what makes a good synthetic dataset: it should be like diet soda in that it has all the qualities (appropriate correlations, structure, richness, diversity, etc.) with none of the calories (resemblance to real people for example that can compromise their privacy). For a long time, there were many disparate approaches to synthetic data generation and the work featured in this article called Synthetic Data Vault provides a handy toolkit to bring some of those techniques together under one roof.

An additional contribution made by the authors of this research is that they have provided constraints to manage the creation of the synthetic data: things that might not be explicitly captured in the statistical relations but are important nonetheless. For example, the landing time of a flight occurring after the takeoff time. This leads to datasets that are ultimately more realistic in their representation of the real-world data and hence more usable which will hopefully enhance the uptake of synthetic data over time.

### India's Personal Data Privacy Law Triggers Surveillance Fears
(**Original *DW* article** by Manasi Gopalakrishnan)

More and more nations are coming up with their own privacy legislations, and India being home to more than a billion people has also thrown its hat into the ring with the Personal Data Protection (PDP) Act that is due to be enacted in 2021. As per researchers, it is broadly based on the EU's GDPR, which is the case with many other privacy legislations around the world. Some of the key highlights of the PDP include: laws around storing consumer data, asking for consent before using private information, and periodic audits for companies storing personal data and protocols for reporting any breaches that they experience.

Storing data locally seems to be one of the most significant implications of the PDP; this will impact the business models of multinational companies that serve Indian users. Personal data is categorized into 3 tiers with each requiring different degrees of protection. As one of the researchers quoted in this article points out, there are concerns with requiring companies to store data locally since it opens up that data to potential state surveillance.

The intelligence community in India wasn't created through an Act of Parliament and hence their scope is nebulous. But, for a company where progress across the nation has been uneven, rising digitalization has the potential to bridge the gaps that exist between those that have and those don't.

### Data Audit of UK Political Parties Finds Laundry List of Failings
(**Original *TechCrunch* article** by Natasha Lomas)

Misuse of personal data in the context of political targeting is one of the premier concerns when it comes to the application of privacy legislation within any nation. The Information Commissioner's Office (ICO) in the UK published a list of failings on part of the British political parties on how they handled citizens' personal data. The parties need this data to campaign effectively and reach those who they feel are most likely to vote for them. But, such practices can quickly conflict with privacy legislation.

While ICO provides an exhaustive list of failures, the recommended actions are soft, and the language around follow-up actions are also weak in their severity as pointed out by researchers quoted in the article. As an example, some of the concerns raised by ICO include lack of transparency and clarity in the privacy notices, lack of appropriate legal bases for collection of personal data, lack of lawful consent for the collection of data, lack of clarity in how this data is being combined with other data for voter profiling, and lack of auditing whether data obtained from third-parties has been obtained legally and continual checks on whether vendors for this data are meeting their data protection requirements.

After the Cambridge Analytica incident, ICO called for an "ethical pause" on the collection of personal data for political microtargeting. One of the implications with the GDPR has been on joint controllership whereby the responsibilities of the parties that are using social media platforms for campaign targeting need to clearly define their roles and responsibilities so that there is clarity amongst the various actors for what protections need to be undertaken and who is responsible for what.

### How China Surveils the World
([Original *MIT Tech Review* article](#) by Mara Hvistendahl)

A little bit akin to gaining an inside hold on the infrastructure being built in countries through massive aid provided to emerging economies and the rapid expansion of the Belt and Road project from China, the use of partnerships with universities, access to data from social media platforms, and apps that store data in the PRC provide CCP with a convenient mechanism to gather data on the entire world.

As pointed out in the article, the collection of data sometimes doesn't carry any immediate purpose but is being done proactively so that it might be used in the future, perhaps gleaning insights that aren't possible with technical solutions today.

The storage location of the data is also a hot-button issue, because of the ability of the CCP to requisition data as it sees fit, something that was a sticking point in the TikTok bans in several countries and the fiasco with trying to find US buyers to avoid this very concern amongst others. The article also mentions a company GTCOM, similar to Palantir, that ingests a lot of data and provides intelligence from that to various actors. The kind of data that is utilized runs the gamut: voice, text, images, video, metadata, and everything else. Given that the agenda for this data gathering is ushered by the CCP, this is an area of severe concern as it wrests a great deal of power in the hands of an authoritarian government concerning people from around the world.

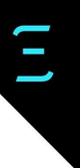

### Activists Turn Facial Recognition Tools Against the Police
(Original *NY Times* article by Kashmir Hill)

An interesting take on demonstrating the power of facial recognition technology, using it against the very people who strive to oppress everyday people. This article documents the efforts of developers who are foraying into utilizing facial recognition technology to identify and act against authoritative figures who conceal their identities when they are engaging in suppression during protests for example.

In an interesting outcome on the ban of facial recognition technology in Portland in public spaces, the developer of this program wanted to check if his development of this system would be prohibited and it was deemed to not be the case because this was a private use of the technology.

As aptly summarized in the article, the lack of anonymity is not what strikes fear in the hearts of erring officers, it is the imposition of infamy through recognition by facial recognition technology. Power dynamics are fickle and for once, it seems that there is an opportunity for people to take back control from their oppressors.

### Why Amazon Tried to Thwart Portland's Historic Facial Recognition Ban
(Original *Salon* article by Matthew Rozsa)

Portland has jumped front and center in the battle for regulating facial recognition technology and how we think about this going into the future. While they restrict the use of this in areas that are open to the public, other places have taken less restrictive approaches to regulating facial recognition technology.

Given that some organizations can stand to lose a lot of money if the demand for their systems goes down amidst new regulations, it shouldn't come as a surprise to the readers of previous reports that there were attempts to subvert the efforts made by Portland in regulating this technology. However, bringing about transparency into where interventions are made by firms that lobby to thwart such regulatory efforts and other initiatives that seek to dismantle such efforts are much needed.

Despite Portland creating a historic push on this front, there remain several holes, especially as it relates to how such technologies are used in settings like public schools which fall under

The State of AI Ethics, January 2021    127

different bodies and until we get a consistent and coherent regulatory mechanism in place, the patchy efforts are going to continue to place people in harm's way leaving them with few options for recourse.

# 8. Risk & Security

**Opening Remarks** by Abhishek Gupta (Founder, Director, & Principal Researcher, Montreal AI Ethics Institute)

One of the things that the Montreal AI Ethics Institute has been strong advocates for is the focus on the importance of machine learning security as a field that will have meaningful impacts in actually achieving our goals of building responsible AI systems. We have covered these areas extensively in our previous reports and in [The AI Ethics Brief](). In the opening item of this chapter, we get a glimpse into some of these risks as we look at the problem of memorization in large-scale models as they pick up sensitive information like social security numbers, credit card details, and other personally identifiable information (PII). The paper proposes some metrics to help measure the exposure and also hints at some methods that can help counteract this memorization. But, for the most part, these are stop-gap measures and we need to rethink our strategies when it comes to how we scoop up data to create models to be used in practice.

Building on this need for having more robust systems, the chapter also looks into the adoption pathway for automation in the aviation industry and draws parallels between other safety-critical domains like nuclear power generation while highlighting some key differences in the degree of exposure that the domain has to end-consumers. Given some of the specific market dynamics and the large incentives to maintain safety records exacerbated by the razor-thin operating margins which have been put to the test with the COVID-19 pandemic, the paper makes a strong case for how aviation organizations can preemptively build for critical safety on the automation front so that when regulations are enacted, they are already ahead of the curve; definitely, something that other domains can also borrow from.

Risk is a core consideration when thinking about regulations surrounding AI systems, something that is codified in the guidelines from the EU. In this chapter we also take a quick peek at the call that was put out late last year by a few countries including Denmark that asked for reducing the onus in terms of assessing the AI systems lest too many of them are classified as high-risk, hence warranting higher scrutiny, and stifling innovation. A core concern is that this further marginalizes those who are negatively affected by the systems and won't have the opportunity to raise questions and seek answers.

When we talk about risks, we have to give due consideration to autonomous vehicles (AVs) which struggle from a lack of robustness when faced with adversarial examples. In addition, the current regulatory atmosphere is ill-equipped to handle cases when accidents occur involving AVs as was the case with Uber where blame wasn't appropriately allocated. This is important not only because it sets a precedent for future cases but also because we need to be realistic about the expectations from human drivers who are in the AV who might not have enough time to respond when they are presented with information by the system if there is a hand-off.

For some hidden gems into the history of CAPTCHA and how they can serve as secret portals to access forbidden communities on the internet, I strongly encourage you to read this chapter to the end! Risk and security are pervasive and fundamental issues that will be critical if we're to evoke higher degrees of trust from people when we use AI systems in real-life.

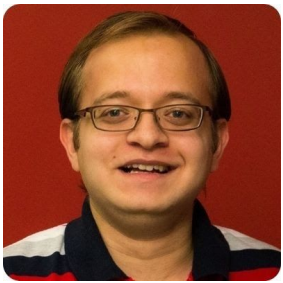

**Abhishek Gupta (@atg_abhishek)**
Founder, Director, & Principal Researcher,
Montreal AI Ethics Institute

Abhishek Gupta is the founder, director, and principal researcher at the Montreal AI Ethics Institute. He is also a machine learning engineer at Microsoft, where he serves on the CSE Responsible AI Board. His book 'Actionable AI Ethics' will be published by Manning in 2021.

# Go Deep: Research Summaries

## The Secret Sharer: Evaluating and Testing Unintended Memorization in Neural Networks
([Original paper](#) by Nicholas Carlini, Chang Liu, Ulfar Erlingsson, Jernej Kos, Dawn Song)
(Research summary by Erick Galinkin)

Neural networks have proven extremely effective at a variety of tasks including computer vision and natural language processing. Generative networks such as Google's predictive text are built on large corpora of text harvested from various locations. This poses an important question – to quote the paper: "Is my model likely to memorize and potentially expose rarely-occurring, sensitive sequences in training data?"

Carlini et al. used Google's Smart Compose in partnership with Google to evaluate the risk of unintentional memorization of these training data sequences. In particular, concerned with rare or unique sequences of numbers and words. The implications of this are clear – valid Social Security numbers, Credit Card numbers, trade secrets, or other sensitive information encountered during training could be reproduced and exposed to individuals who did not provide that data. The paper assumes a threat model of users who can query a generative model an arbitrarily large number of times, but have only model output probabilities. This threat model corresponds to, for example, a user in Gmail trying to generate 16-digit sequences of numbers by starting to type the first 8 digits and then auto-completing.

Carlini et al. use a metric called perplexity to measure how "confused" the model is by seeing a particular sequence. This perplexity measure is used with a randomness space and a format sequence to compare the perplexity of a random sequence selected from the randomness space with a predetermined "canary" sequence placed in the training data. The canary sequence's perplexity and several random sequences a small edit distance away from the phrase are compared are used to compute the rank of the canary – that is, its index in the list of sequences ordered by perplexity from lowest to highest (e.g. the lowest perplexity has rank 1; the second-lowest has rank 2; and so on). Given this rank, an exposure metric is approximated using sampling and distribution modeling. Based on the Kolmogorof-Smirnov test, the use of a skew-normal distribution to approximate the discrete distribution seen in the data fails to reject the hypothesis that the distributions are the same.

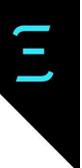

Testing their methods on Google Smart Compose, Carlini et al. find that the memorization happens quite early in training and has no correlation with overfitting the dataset. Exposure becomes maximized around the time that training loss begins to level-off. Taking all the results together, there is an indication that unintended memorization is not only an artifact of training, but seems to be a necessary component of training a neural network. This ties in with a result of Tishby and Schwartz-Ziv suggesting that neural networks first learn by memorizing and then generalizing.

Carlini et al. also find that extraction is quite difficult when the randomness space is small, or when exposure of the canary is high. For the space of credit card numbers, extracting a single targeted value would require 4,100 GPU-years. Using a variety of search mechanisms, a shortest-path search algorithm based on Djikstra's algorithm allowed for the extraction of a variety of secrets in a relatively short amount of time, when the secret in question was highly exposed.

A variety of methods were considered to mitigate the unintended memorization. These include differential privacy, dropout, quantization, sanitization, weight decay, and regularization. Although differential privacy did prevent the extraction of secrets in all cases, there was meaningful error introduced when using differential privacy. Sanitization is always a best practice, but did manage to miss some secrets since it then becomes the weakest link in the chain. Dropout, quantization, and regularization did not have any meaningful impact on the extraction of secrets.

Carlini et al. conclude by saying: "To date, no good method exists for helping practitioners measure the degree to which a model may have memorized aspects of the training data". Since we cannot prevent memorization – and if Tishby and Shwartz-Ziv are to be believed, we would not want to – we must instead consider exposure and mitigate exposing secrets or allowing secrets to be extracted from our model.

### The Flight to Safety-Critical AI
([Original paper](#) by Will Hunt)
(Research summary by Abhishek Gupta)

The paper takes a critical look at how automation is being deployed in practice with a focus on aviation which is known for its excellent safety records and highly stringent regulations that uphold that safety record. While a lot of arguments might be made about how the current framing of AI development across different regions constitutes a race where ethics and safety might fall by the wayside, the paper arrives at a different conclusion based on interviews and of practitioners in the field and in-depth analysis of the current state of deployment of automation in aviation. Given the nascent stage of development of technical safety measures in AI, automation is being used in aviation in less safety-critical areas like predictive maintenance, decision support, and route planning rather than handing over full control of the flight.

Something that is unique about aviation is that the firms will unprompted invest in mechanisms like Testing, Evaluation, Validation, and Verification (TEVV) compared to research labs and other industries, especially given how open the aviation industry has been to firms opening up about their mistakes and helping the entire ecosystem move towards higher levels of safety.

Preliminary findings in the paper show that military avionics is even more conservative in adopting automation measures compared to commercial avionics, allaying some of the concerns around warfare automation. The paper applies the "race" paradigm, given that it is a popular trope, to analyze the trends in the space today and where things might be heading. For example, self-regulation is often touted as a way to prevent a race to the bottom when it comes to things like facial recognition technology with firms like Microsoft advocating for both industry and government action in helping the ecosystem reach greater levels of ethics and responsibility. Lower levels of regulation are of potential interest to firms since they offer an easy way to cut on costs of compliance and offer them competitive advantages, more so in domains where profit margins are thin.

The "Delaware effect" is an epitome of this whereby firms choose to incorporate in Delaware because of favorable circumstances offered by the state. In contrast, there is the "California effect" or the related "Brussels effect" where firms encourage the state to levy higher levels of regulation. So, when we are talking about safety-critical systems, we are talking about systems where errors can be quite costly, especially in terms of human lives. This distinction is important because it allows us to analyze in a more nuanced way where AI might be deployed first to "iron out the kinks." Compared to other industries that have high safety requirements like nuclear power, what is different about aviation is that it is subject a lot more to market forces and doesn't have natural monopolies that are as strong as in the case of nuclear power. Secondly, being directly exposed to end-users also makes it different from nuclear power and hence a good test-bed to see the impacts of automation and its implications for the future. One of the driving forces for automation adoption in aviation is the volatility of the industry based on demand and supply forces, including fuel prices and razor-thin margins because of a high degree of competition. This along with the requirement from regulators to have very high standards of safety gives a push for the potential consistency that automation can offer over human errors.

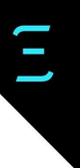

An additional factor that supports the potential adoption of widespread automation is the large amounts of structured data collected by air traffic controllers (ATCs). Yet, because of a lack of consistent standards in terms of what constitutes the requisite safety demonstrations, there is a bit of fragmentation in the understanding of what needs to be done. A point highlighted in the paper is the need for regulators working across different countries and jurisdictions to develop a shared understanding and standards so that safety can be evaluated in a comparable manner not disrupting the important flows of capital and labor across borders. Beginning this exercise in non-critical areas is a good testbed in ironing out the process so that when automation is applied to the more safety-critical aspects, there is a well-established process for enabling the transition to happen smoothly. The reciprocity in agreements around safety standards is essential.

The Max 737 crashes demonstrated how large the backlash can be from errors in this domain, especially in the case where the firms pay massive fines, lose future contracts, and suffer long-term reputational damage. Given the high degree of public scrutiny, one can only imagine how chilling the effects of premature and untested automation in safety-critical areas of aviation will be. Firms are also hesitant in deploying this technology commercially till standards and regulations are put in place. Some systems have been tested for making landings in low visibility scenarios, but these are still quite limited, especially since there isn't an established process for evaluating the performance of the system in varied conditions. Aviation is also noteworthy for its emphasis on discouraging competition on safety-standards since any missteps lead to a decrease in traveler confidence and have a negative impact on the entire industry. An Airbus ad that tried to imply that their planes were safer than those made by Boeing faced massive backlash from many airlines that led to a retraction. Since failures in airlines are catastrophically expensive, in human lives and financially, there is very little room to experiment to arrive at the "failsafe" configurations which makes the implementation of safety-critical testing of automation all the more crucial but also difficult to achieve. A reason also put forward for the slow deployment of automation is the arduous and long process of certification where returns on investment take a while to show up, potentially imposing financial hardship on the firms that are already pressed for cash flow.

Some of the recommendations made in the paper are as follows: calling on policymakers to make more investments in TEVV-styled mechanisms for AI systems and calling on regulators to collaborate on creating standards in AI safety and encourage information sharing on system failures. Firms that are willing to pay this upfront cost in terms of investments will reap the benefits in being safety compliant that will allow them to transition into being able to use AI in safety-critical scenarios offering them a competitive advantage. From the analysis done in the report, there is little evidence that competitive pressures will overwhelm the safety imperatives for automation adoption in the aviation industry.

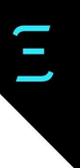

## Go Wide: Article Summaries (summarized by Abhishek Gupta)

### Why Wasn't Uber Charged in a Fatal Self-Driving Car Crash?
(Original *Wired* article by Aarian Marshall)

It would appear that we can't go about the conversation on autonomous vehicles (AVs) without talking about the fatal car crash that happened when a safety driver in the Uber vehicle in Arizona took their eyes off and hit a pedestrian. And rightly so! Going into the future, if such AVs are to become a common sight, we need to have strong precedents for handling cases and inadvertent accidents that will arise. But, does the way that this Uber crash was dealt with, set such a precedent?

The National Transportation Safety Board (NTSB) did a detailed analysis and found many parties at fault in the accident. In harrowing detail, the report talks about how the major culprit was that the Uber self-driving system didn't account for the possibility of pedestrians outside of the designated crosswalks, as well as a misclassification error and confusion until a mere 1.2 seconds before the collision. When the alarm was finally sounded for the safety driver, perhaps there was too little time left for the safety driver to act.

But, fully laying the blame on the safety driver as is being advocated leads to a false separation of concern where there are multiple folks who need to be held accountable for the final outcome and liability must be shared. Yet, once the hearings proceed and the dust settles, we might have created the precedent that will set the stage for any future collisions and how they are dealt with.

### Attention EU Regulators: We Need More Than AI "Ethics" to Keep Us Safe
(Original *Access Now* article by Daniel Leufer, Ella Jakubowska)

This article calls to attention the recent position put forward by Denmark (along with 14 nations) that request the EU to not place too onerous requirements on AI systems from an ethical standpoint lest it stifles innovation. The reasoning behind that is that too many systems might be categorized as "high-risk" and hence limit the potential deployment of innovative technologies prematurely. In effect, they ask for a more objective evaluation methodology that would prevent such transgressions.

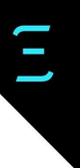

But, such an approach strips even more power from those who are already marginalized by disenfranchising them from the ability to question the deployment of a system that is obscured behind these assessments. Additionally, ignoring the human rights implications of these systems means that, say your city uses live facial recognition, you don't even have the ability anymore to protest in a safe manner without potentially facing repercussions. If the risk assessments themselves are not publicly examinable, then you further entrench the power asymmetries between those that build the system and on whom it is used.

The report mistakenly assumes that AI ethics principles will be enough, which as we've covered in the past are certainly not, including the importance of virtues in this discussion. By not drawing any red lines on cases where there is consensus on more misuse than beneficial uses, the report misses an important opportunity for steering the conversation in the right direction.

While it doesn't say that there aren't any cases where red lines will not be considered, it lauds the potential benefits over the actual harms taking place right now. It also advocates for purely technical fixes as a way to address the problems. Soft law, self-regulation, and principles are inadequate, we need to have stronger regulations and consultations with those who are affected if we're to do better.

## Split-Second 'Phantom' Images Can Fool Tesla's Autopilot
([Original *Wired* article](#) by Andy Greenberg)

ML Security is a burgeoning field that is slowly gaining traction with practitioners in the security and the ML domains. This article points out how semi-autonomous vehicles (AV) can be compromised by flashing phantom images, something that a human driver would potentially ignore, to trigger crashes, halts, and unwanted behaviour from the self-driving system. One of the ways the researchers do this is by injecting a few frames of a road sign into the video of a billboard that can confuse the camera system on an AV. Testing it on the Tesla and Mobileye systems, they were able to elicit unwanted behaviour like triggering a communication of the incorrect speed and halting when flashed with these phantom images. These images are not persistent and appearing for as little as 0.42 seconds befuddled the Tesla system and 0.125 seconds for the Mobileye system.

To push their research even further, they attempted to find the least noticeable areas within a video to inject these phantom images to evade detection from the human driver. While there have been previous demonstrations of attacks in the form of placing cheap stickers on the road and confusing the lane-following systems in the AV, such attacks leave behind forensic evidence. The novelty in this attack is that it can be executed remotely and leave behind little evidence. Additionally, such attacks don't require as much special expertise compared to some of the other previously executed attacks.

Other researchers and operators at companies like Cruise have offered that the vehicles with higher degrees of autonomy rely on multiple sensors to make decisions, including LIDAR which is unaffected by such an attack. There is also an argument made by AV manufacturers that the present version of the technology isn't meant to be used without human supervision but that is not what happens in practice. Hence, such attacks are still valuable and demonstrate the brittle nature of the existing systems, driver and pedestrian safety is paramount before we can have AVs roaming the world freely.

## Why Security Experts are Braced for The Next Election Hack-and-Leak
([Original *MIT Tech Review* article](#) by Patrick Howell O'Neill)

Building on the importance of having a cleaner information ecosystem, it is not a surprise when well-timed dumps of leaked documents and information can lead to problematic situations whereby the more important issues get sidelined by an exclusive focus on the sometimes banal (at least compared to the more important issues of the day) contents of these leaked documents. Given this vulnerability, newsrooms around the internet have been advising their reporters and journalists to be wary of this and apply caution and circumspection before adding more oxygen to the leaked documents.

A particular problem with this is that by nature they are anonymous and don't provide avenues to hold actors accountable. While information from official sources is tightly regulated close to the election timeframe, leaked information from anonymous sources can't be held to the same standards and can hence be used as an effective tactic to distort elections and their results.

France with the Macron leaks did provide a model perhaps that is worth contrasting with the US model whereby they showed restraint in what was released so as to not unduly influence the election. Ultimately, having good media literacy in terms of knowing when to trust which sources is going to be an increasingly essential skill that we need both consumers and producers to have.

## Bot or Not
([Original *Real Life Mag* article](#) by Brian Justie)

Providing a fascinating background into the origin story of the CAPTCHA, the distorted letters that we have to identify to prove our humanness, and selecting boxes that contain specific objects in an image, this article showcases how these tollgates for discerning humans from machines might now be flipped and how the dynamics of what constitute bots online has changed since the conception of this "human-processing power" harnessing tool that was created many years ago.

The inventor of the CAPTCHA wanted to help curb the problem of spam and bots online and decided that a gamification approach would be one that is appropriate to incentivize participation and at the same time create microtasks which could be utilized for other purposes. Fast forward to today, they do serve as perfect training grounds for machine learning systems getting free labeling from humans to create datasets. But, of course as the machine learning systems have been getting better, these checks are no longer sufficient and researchers point to how such tests are merely decoys and the services that are verifying our humanness actually run more background risk score calculations by analyzing everything from your IP address, browser history, cookies, mouse movements, and other markers to ascertain whether you are a machine or not. If there is doubt after that evaluation, then you are presented with the 9 boxes over an image that acts as a further check.

The most interesting tidbit in this article centered around how CAPTCHAs could be used as gateways to access portals where there is prohibited activity or speech taking place. By identifying racist elements in a seemingly innocuous image, it might grant you access to a hidden community that discusses those ideas while a regular person would just pass through transparently none the wiser. In that sense, they can start acting like memes which have an in-group sensibility to them in that they are targeted to specific demographics but are veiled in broad daylight from those who don't have all the nuance and context to decipher the hidden meaning. Finally, the way we interact online and how we police for what constitutes bot-like behaviour might itself evolve over time since this can just become an easy scapegoat to evade responsibility, relegating any bad actions to the background by blaming it on bots when in reality it could be humans just masquerading as bots.

# 9. Social Media

**Opening Remarks** by Abhishek Gupta (Founder, Director, & Principal Researcher, Montreal AI Ethics Institute)

As we start up 2021, we see the potency of information operations and how that can be used to manifest real-world harm as was the case with the insurrection and attack on the Capitol in the US that led to tragic losses and a shattering of confidence in the security of our public institutions. The opening piece in this chapter talks about cyber troops delineating the different structures and forms that information operations can take, especially shedding light on the diversity of motivations, funding mechanisms, and tools at their disposal. Though the case of the US attack takes a bit of a different form in the sense of deliberate spreading of misinformation with intentions to incite others to take action, we do see that such events are not isolated and spread across both time and space.

The second item in this chapter raises an important counterweight when trying to combat the potency of information operations using social media to organize and spread misinformation and violence. A lot of such efforts are organized through the use of closed messaging groups and often they are inscrutable due to privacy and the lack of cooperation from the social media companies in opening those up to research and analysis. There are definitely questions around the privacy implications of such an opening and questions that surround data ownership. The authors of that piece shed some light on different ways that research has been carried out on closed messaging groups including perhaps the most insidious method of such collection relying on some users exporting their message logs and sharing them with researchers and researchers directly participating in those groups without explicit identification. As information operations become more prevalent on social media platforms, especially those that are closed groups, the questions raised in this piece will become important for us to address.

One of the things that particularly rubbed us the wrong way in 2020 was all the attention that was accorded to *The Social Dilemma* and how it continued to provide technologists who were firmly implicated in the crises in the first place with an easy path to redemption while marginalizing the work of those who have been on the ground and working on these issues for many years. An interesting throwback to the days before we had quantified every aspect of our online interactions, an article in this chapter talks about the effects of removing metrics associated (at least publicly) with how we interact with each other in the hopes that it would help to clean platforms of clickbait styled articles and bot accounts. Perhaps some things before we entered our currently scrutinized and gamified environment did serve us better from a mental health perspective.

One of our favorite organizations for 2020 was The Markup that has invested in creating and making public tools like Blacklight that empower users with the ability to scan websites to see what kind of tracking is being applied to them. We firmly believe that having more such tools will help to develop expertise over time and empower people to be more conscious with their choices when it comes to picking different platforms and apps to use. Another example of great work from The Markup is investigative journalism that helped to highlight how the Biden campaign was charged more money for the ads that they placed on Facebook which hampers democratic processes and create a profit-driven intermediary that is able to control the efficacy of political spending to a certain extent. While the article does caveat that it probably doesn't explain all the variances, it certainly raises interesting questions that warrant further research in the field.

As we try to explore novel models for how to regulate information flow on social media platforms, an interesting take on moderating for "freedom of reach" might be more effective and targeting the right thing compared to "freedom of speech" — the idea being that you can regulate how content is distributed on a platform to better achieve your goals of limiting the negative impacts of problematic information on the platforms.

Ultimately, over the past quarter, we have seen some interesting developments both in terms of technical tooling and thought processes that will lead to a more healthy information ecosystem. This also requires vigilance and informed engagement on our part to bolster this work and take action to implement these ideas in practice where we have the ability to do so.

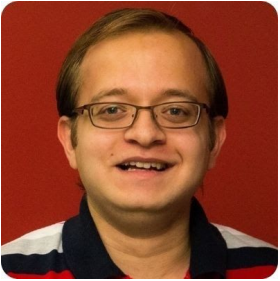

**Abhishek Gupta (@atg_abhishek)**
Founder, Director, & Principal Researcher,
Montreal AI Ethics Institute

Abhishek Gupta is the founder, director, and principal researcher at the Montreal AI Ethics Institute. He is also a machine learning engineer at Microsoft, where he serves on the CSE Responsible AI Board. His book 'Actionable AI Ethics' will be published by Manning in 2021.

# Go Deep: Research Summaries

**Troops, Trolls and Troublemakers: A Global Inventory of Organized Social Media Manipulation**
([Original paper](#) by Samantha Bradshaw, Philip N. Howard)
(Research summary by Nga Than)

Social media has played an increasingly important role in shaping public life, and public discussions. It helps form public opinion and serve as a place of information acquisition across the globe. Governments, and political actors increasingly have taken advantage of these communication platforms to their own advantage. They spend an increasing amount of financial resources to employ people to generate content, influence public opinion, and engage with domestic and foreign audiences. Bradshaw and Howard gather and create a unique dataset of organized media manipulation organizations to understand this global trend.

The authors start out by defining the term: "cyber troops," which refer to "government, military or political-party teams committed to manipulating public opinion over social media." The authors maintain that these groups play an increasing role in shaping public opinion. Then, they describe the process of gathering information to construct a unique dataset that they created to study those organizations to analyze the size, scale and extent to which different kinds of political regimes deploy cyber troops to influence and manipulate the public online. The authors rely on mainly news media sources written in English to find information such as budgets, personnel, organizational behavior and communication strategies. They further corroborate and supplement this information by consulting countries experts, and with reports from research institutes, civil society organizations.

The authors found that cyber troops adopt a wide range of strategies, tools, and techniques for social media manipulation. These strategies include commenting on social media posts to engage with citizens, targeting individuals, using both real and fake social media accounts, and bots to spread propaganda and pro-government messages, as well as creating original content. Messages to users range from positive, to harassing and verbal abuse, to neutral language to distract public attention from important issues. Individual users are targeted to silence political dissent. This method is considered most harmful to targeted individuals because they often receive real life threats and suffer reputational damage. Cyber troops create original content such as videos, blog posts under online aliases.

Cyber troops have a wide range of organization forms, structures, and capacity of cyber troops. The authors observe that some governments have their own in-house teams, while others outsource these activities to private contractors, and sometimes galvanize volunteers, and hire private citizens to spread political messages on the Internet.

In some countries, organized media manipulation is done through a small team, while in others a large network of government employees is involved. A notable example is China, which has more than 2 million individuals working to promote the party ideology. The research team also found that these different groups have different operating budgets, yet they encountered the problem of incomplete information because such information is not readily available. In authoritarian regimes, governments tend to provide funding for these activities, while in democratic regimes, political parties tend to be the main drivers of organized social media manipulation.

Bradshaw and Howard show that cyber troops have heterogeneity in terms of organizational structure. Such organizations could have 5 different types of structure:

1. A clear hierarchy and reporting structure
2. Content review by superiors
3. Strong coordination across agencies or team
4. Weak coordination across agencies or teams
5. Liminal teams

Cyber troops also engage in capacity building activities from training staff to improve skills and abilities associated with producing and disseminating propaganda to providing rewards or incentives for high-performing individuals to investing in research and development projects.

This cross-country comparative research highlights the heterogeneous nature of cyber troops' activities across the world. They are increasing in size, scope, and organizational resources and capacity. This paper has important implications for researchers, civil society organizations, and private citizens to question how their online activities are shaped and influenced by government-funded groups. Furthermore, the paper raises important questions about the social media environment where government cyber operations could operate, shape public opinion, and sometimes divert public attention from important issues.

## Considerations for Closed Messaging Research in Democratic Contexts
([Original paper](#) by Connie Moon Sehat, Aleksei Kaminski)
(Research summary by Khoa Lam)

The use of closed messaging apps has grown in recent years and their impact on public elections-related discussion poses challenges and ethical conundrums for researchers. Sehat and Kaminski present a review of four research practices within these apps and explore various key questions to clarify the ethical considerations faced by researchers.

The popularity of messenger apps has risen globally, where WhatsApp, Facebook Messenger, and WeChat have garnered a number of users ranging in the hundreds of millions to billions per month. These apps offer a plethora of features, including instant speed messaging, asynchronous reach, opt-in encrypted privacy, in addition to transnational texting, phone, and video service without additional costs.

It comes as no surprise that closed messaging apps can act as a powerful political tool for spreading political information. However, due to the nature of encrypted privacy embedded in their design, investigations, and studies of misinformation within these apps are often met with difficulties in professional ethics.

The authors lay out 4 models typically used in practice:

1. **Voluntary contribution**

In the first model, researchers do not enter the chats and instead receive message texts from users with consent. This approach was implemented either as tip lines (e.g., during the 2018 Brazilian elections or the 2019 Indian elections) or via voluntary submissions in one-to-one and one-to-many broadcasts (e.g., in the 2015 Nigerian presidential elections).

2. **Collection through focused partnerships**

In the second model, researchers enter directly into chat spaces themselves, as part of a collaborative election tracking project. Analysts collect messages for a period of time, and, as a result, examine both the message texts, sender details, along with their conversational contexts. This model was implemented during the 2016 Ghanaian general election, where a collective of organizations established the Social Media Tracking Center (SMTC) to monitor messages on social media for violence and election threats.

3. **Entrance with announcement or identification**

Researchers who employ the third model leverage the ambiguous nature of private chat invitations and publicly available links. They entered the chats with researcher identities with or without announcement and allowed for removal or withdrawal when requested. This approach was used in some studies of the 2019 Indian elections.

4. **Entrance without identification**

The fourth model, in which researchers entered the chats without disclosing their researcher identities and purposes, raises the question around the notion of a public chat group and its implications for research. This approach was used during the Brazilian 2018 and 2019 elections by various academic organizations.

To conclude, Sehat and Kaminski explore the ethical considerations the researchers implicitly or explicitly decide prior to the collection and analysis of closed messaging texts:

1. Exactly when is a closed message chat "public"?
2. Who does the data belong to?
3. What are the obligations for researcher disclosure and or informed consent?
4. When should researchers inform or report back to the groups involved the findings of their studies?
5. Are these the questions that researchers can ask the public?
6. Are these the questions that researchers should discuss with companies?

The authors further analyze the intricacies within these questions. Conditions including indexed invites, discussion topics, scope, group size, and expectations are considered. In addition, answers to these questions are also guided by laws, regulations, along with historical and recent court rulings. The last two questions in particular also address concerns for conflict of interest and data access while, at the same time, open up opportunities for cross-organization and interdisciplinary collaboration.

# Go Wide: Article Summaries (summarized by Abhishek Gupta)

### The Social Dilemma Fails to Tackle the Real Issues in Tech
([Original Slate article](#) by Pranav Malhotra)

If you haven't watched *The Social Dilemma* yet, it is perhaps worth at least some of your time to give it a look. It does get some things right, and again a lot of those things might not come as surprises or novel information to the readers of previous reports but at least it serves as a wake-up call for a lot of other people who are still new to the idea of the invasiveness and violations of privacy and other issues that are rampant in social media platforms.

But, as pointed out in this article, there is something that is deeply problematic about a documentary like this which valorizes the "woke" technology workers who get an "easy ride to redemption" while others who have been laboring for years, doing the hard work of mobilizing communities have been largely ignored. The documentary perpetuates the problem of biases and lack of diversity, things that had the companies from which the interviewees hail had taken more seriously, at least some of the problems would have been mitigated.

While yes the design patterns used in the creation of this technology can be cast as something that is meant to solicit addictive behaviour, this doesn't mean that a singular, pathological framing is adequate in how these issues need to be thought about. Finally, as we mentioned, continuing to offer a spotlight to those who already had a seat at the table, who blew their chance at making a difference, at the expense of those who have always been marginalized is something that is deeply concerning, especially as more people become aware of these issues and think that these now "enlightened" technology workers are the avenues for change when there are intrepid and indefatigable workers who have been working on these issues for years.

### Inside TikTok's Killer Algorithm
([Original *Axios* article](#) by Sara Fischer)

While this tool might still fall under the realm of being "cool", a lot of people outside the target demographic of this tool have been paying attention to it ever since the announcement on requiring the parent company to sell the tool to a US-based entity if it were to continue operating in the US, supposedly one of its most profitable markets; TikTok has since closed a deal with Oracle and Walmart. The reason for the success of the platform has been its underlying platform algorithm which is optimized for engagement (as is the case with all other platforms too) taking into account factors such as the expressed interests that evolve over time, the user location, type of device, etc. They use machine learning to cluster users into groups that might have similar preferences but also supplement that with some deterministic rules to prevent showing repeated content that might bore the user.

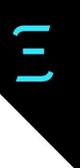

But, as is the case with all other platforms too, there is a massive implication in terms of creating filter bubbles that can negatively affect the quality of the information ecosystem. At the time this article was published, this posed significant challenges for problematic information to spread on the platform because the 2020 US Presidential Elections were coming up. The policy team at TikTok takes that seriously and mentioned how they have been in contact with lawmakers to ensure that they don't run into issues and how they might prevent the spread of problematic information on the platform. Ultimately, from an international policy standpoint, whether or not it's possible will depend on the interpretation that the Chinese government has made on the fine print of the relationship between Oracle, Walmart, and TikTok. This would include whether their famed algorithm would be allowed to be transferred outside of China in the face of new rules imposed by the Chinese government.

## Would the Internet Be Healthier Without 'Like' Counts?
(Original *Wired Magazine* article by Paris Martineau)

Mental health problems due to addiction to internet usage is something that is discussed in waves under varying circumstances. This article dives into the intricacies of the efforts made by different platforms in terms of reorienting their services (potentially) towards a metric-free experience. The argument goes that this radicalizes the consumption patterns for the users who pay more attention to the metrics rather than the content. On the creators end, it creates obsessive behaviour patterns which skew them towards putting out content that is apparently optimized for garnering higher counts of these metrics on the platforms.

Even before these talks began, artists and activists have put tools that have sparked a movement called demetrication, including extensions in your browsers to help you do that. One of the other positive impacts of this is to discourage the purchase of bot accounts that help to artificially boost counts and depress the sales and prevalence of shady companies on platforms who try to trick users into buying low quality products or services. Experiments on Instagram, Facebook, YouTube, and Twitter have shown mixed results thus far and a lot of keen watchers of the space have argued that this might not eventually come to be, especially in the case where one might experience lower rates of engagement on the platform since that would hurt the bottomline in a highly lucrative and competitive domain.

## I Scanned My Favorite Social Media Site on Blacklight and It Came Up Pretty Clean. What's Going On?
([Original *The Markup* article](#) by Aaron Sankin, Surya Mattu)

We have certainly enjoyed using the Blacklight tool from The Markup to scan some of the places where there is intrusive tracking being done by companies, at least ones that purport to provide ethical solutions and have found (un)surprisingly that they don't fare so well. Some users have reported that they didn't find anything wrong with some of the most flagrant violators and were surprised (rightfully so!) that there wasn't any intrusive tracking on those websites.

In this article, the creators of the tool walk through what the tool is actually testing for and showcase to people, even veterans of the field, that it is not just third-party tracking that we need to be wary of, but also the first-party tracking that happens on these websites and how there is a morphing of some of the third-party tools into a form where they appear to be first-party such that they evade detection by tools such as Blacklight.

What the Blacklight tool does is to scan for data being sent to third parties — it can't see what tracking is being done because it doesn't log into your account, e.g. when you are scanning Facebook. The article uses the following analogy to explain further: Blacklight drives you around the parking lot of a store and see if there are any people in cars outside who are noting your activities but that doesn't tell you anything about what a cashier at the store might be doing — say, noting down your credit card information. Giving some more technical details on the different types of cookies that Facebook uses, the article concludes with some information on how cookies are moving from third-party to first-party like behaviour limiting the efficacy of tools like Blacklight, which is certainly a blow to our ability to demand transparency from these companies.

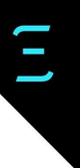

### Facebook Charged Biden a Higher Price Than Trump for Campaign Ads
(Original *The Markup* article by Jeremy B. Merrill)

No matter which side of the political spectrum you fall on, when it comes to preserving the integrity of our democratic institutions and upholding legislations around campaigning, we can't let third-party corporations become arbiters of what flies and what doesn't.

In this excellent piece of investigative journalism by The Markup, they uncovered that there was a discrepancy between how much Biden and Trump were charged for each of their political ads that they placed on the platform. Not only that, but it also found that this difference was much more pronounced in the swing states where votes matter much more than the rest of the country. Such discrepancies have cost the Biden campaign millions of dollars more to try to achieve the same impact in their social media strategy in terms of the number of impressions compared to Trump.

Facebook ads work through an auction method where costs are determined dynamically and it is hard to ascertain all the factors that go into determining that price (as many researchers have lamented from a transparency perspective). But, according to campaign finance regulations, advertisers are not allowed to charge different amounts to different candidates, for example for TV and radio ads. But, Facebook apparently is able to evade this through the loophole that the pricing is not set by them in particular and they are determined algorithmically due to market forces taking into account factors like the reach and engagement that the content will generate and the relevance of the content.

While this may be a technicality, it creates severe concerns when it comes to how candidates are able to reach their constituents in the last mile of the election cycle. Alas, surfacing the problem is a start and we hope that more transparency around this and clarity in how campaign finance regulations are applied will help to improve the situation over time.

### Social Media's Struggle with Self-Censorship
(Original *The Economist* article)

In some categories, automated content moderation on social media platforms is surprisingly effective, it takes down offending content even before a human has a chance to flag it. But, for the most part, this works on items that are egregious in their violation of the community standards and content policies of the platform. In earlier days, this problem was splintered across a series of platforms making the problem quite hard to address. With the agglomeration of most of our online activity, at least there are fewer platforms where such policing needs to be done well to have the largest impact.

Where things get dicey is when we have content that falls in the gray zone and companies have large incentives to not err in pulling down content lest they fall on the wrong side of the law. Of course, this comes with freedom of speech ramifications. So, while a judicial review can incorporate nuance, say in the case of a German citizen exercising his right to forget to have some prior information removed from Google search results, the companies on their own have no incentives to risk actions that can draw the ire of regulators.

Privatising freedom of speech comes with obvious problems and we need to be careful to what extent we are relegating control over to entities that are not beholden to the public good. An approach that companies have been adopting is to look at "freedom of reach" instead of "freedom of speech" by which they can reduce the prevalence of the content on the platform as a way of treading this fine line.

There are also concerns when companies are asked to take down content by authorities and off late we have seen large demands from organizations to hold the corporation accountable. Transparency on the number of requests received, their nature, and the actions taken will help to at least regain a portion of the trust from people.

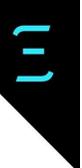

# 10. Outside The Boxes

**Opening Remarks** by Jade Abbott (Machine Learning Lead, Retro Rabbit)

It's the places, people and things at the edge of our hegemonic "default" awareness that tend to be left out of the technological conversation. In 2021, it's vital that we see who and what is not catered for - and to include those individuals and ideas which seem to not fit into the boxes.

Much of the AI community originates from a computer science background, where we've been indoctrinated that what we learn is "superior" and "rational" - a topic explored in the following chapter. When I studied computer science 10 years ago, our course on technology ethics was a security course in disguise, and the idea of an ethics subject was laughed and groaned at. We are taught in binary, and the gray areas of ethics were thought of as "fluffy", "imprecise" and "impractical". Ultimately, this rejection of the social and philosophical sciences has contributed to the state of AI today -- ethics as an afterthought at most. Parts of this chapter explore how we could introduce ethics into the AI curriculum in a way that is both exciting and practically relevant - a necessary step in solving the "tech ethics pipeline" problem.

This chapter challenges us to decentralize our own narratives as AI practitioners, and allow ourselves to be hand-held through expertise-spaces that we're solving for - whether they be health, education, development, or politics. As a software engineer, I've seen our techno-superiority constantly flaunted in every sphere, ignoring domain experts and solving the wrong problems - sometimes to dangerous effect. It's time to get off our elitist high-horse and admit that we do not know the society using our tools better than the society themselves

Technology should be driven by the stakeholders and not the technologists -- the communities where technologies are to be deployed should be in the driver's seat. In fact, I would argue that our AI ethics positioning for 2021 should be even more drastic -- AI practitioners themselves should originate from the societies and spaces where such tools should be deployed. In my experience in Masakhane - a grassroots open research NLP community for African languages - we face the fact that hegemonic approach simply does not solve for African societies. To solve this, we've employed broad societal participation in the scientific process as a means for spurring AI research and implementation that is less blind, more grounded, more inclusive, more impactful. Through close participation, we achieve a deep knowledge sharing and sense of community, which appear to enable us to create tools that build up our societies.

I will leave you to read the following chapter, with a quote by Kenyan author Ngugi Wa Thiong'o, where he talks about an open theatre group in rural Gorki, Kenya.

> "Auditions and rehearsals for instance were in the open. I must say this was initially forced on us by the empty space but it was also part of the growing conviction that a democratic participation even in the solution of artistic problems, however slow and chaotic it at times seemed, was producing results of a high artistic order and was forging a communal spirit in a community of artistic workers. PhDs from the university of Nairobi: PhDs from the university of the factory and the plantation: PhD's from Gorki's "university of the streets" - each person's worth was judged by the scale of each person's contribution to the group effort. The open auditions and the rehearsals with everybody seeing all the elements that went into the whole had the effect of demystifying the theatrical process" ~ Decolonising the Mind, Ngugi Wa Thiong'o

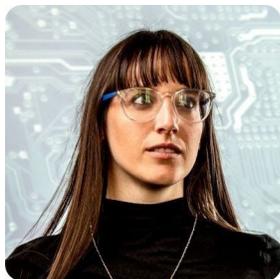

**Jade Abbott(@alienelf)**
Machine Learning Lead
Retro Rabbit

Jade Abbott is the Machine Learning Lead at Retro Rabbit in South Africa. She has an MSc Computer Science from the University of Pretoria and works as a software engineer across Africa in every field from fintech, to NGOs, to startups. Currently, she trains and deploys deep learning systems to perform a variety of tasks for real world systems. In 2019, she co-founded Masakhane, an open research grassroots natural language processing initiative for Africans, by Africans, which aims to spur research into NLP for African languages, currently boasting over 400 members, from 38 African countries, and 13 affiliated publications.

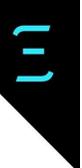

# Go Deep: Research Summaries

**Policy Brief: AI's Promise and Peril for the U.S. Government**
([Original paper](#) by David Freeman Engstrom, Daniel E. Ho, Catherine M. Sharkey, and Mariano-Florentino Cuéllar)
(Research summary by Connor Wright)

The influence of Artificial Intelligence (AI) in government procedures has the potential to not only reduce costs and increase efficiency, but also to help the government make fairer decisions. The authors of this paper have thus scoured federal agencies across the U.S. in order to see how the AI (if being used at all) is being implemented. Their study produced 5 main findings, which I shall introduce and expand on now.

1. **Nearly half (45%) of federal agencies had implemented some form of AI toolkit:**

Federal agencies have been taking advantage of the benefits AI brings. The agencies use the technology for aspects such as customer communication, as well as extracting huge amounts of data from the government's data stream. In this sense, AI is slowly becoming the norm in the federal sphere. However, how do the public and private sectors compare?

2. **The public sector lacks the technological sophistication possessed by the private sector:**

The authors found that only 12% of the technologies used in the public sector could be deemed equivalent to that of the private sector. Without significant public investment, the sector will lag behind, finding it harder to see gains in accuracy enjoyed by the private sector. In order to guide such investment, how are federal agencies to go about designing AI systems?

3. **In house AI systems are the way to go:**

In house AI systems were found to be more adequately adjusted towards the complex legal requirements, as well as being more likely to be implemented in a compliant fashion. In fact, some 53% of the agencies studied had utilised in house AI systems. Proving a safer bet than calling on external contractors who do not know the company's requirements as well as those from within, this brings me on to point number 4.

4. **AI must take into account the unique legal norms and practices of the US legal system:**

Assurance of adherence to aspects of the U.S. legal system such as transparency, explainability, and non-discrimination must be had surrounding AI systems. This will prove essential to the safe proliferation of AI in society as it creeps into more and more areas of society, which allows me to introduce point number 5.

5. **AI has the potential to augment social anxieties and create an equity gap within society:**

Here, the fear is that bigger companies with the human resources and purchasing power that they possess will find it easier to 'game' any government AI model to be able to be compliant, unlike smaller businesses. Without the same resources and potential expertise, they will not be able to keep out of the cross-hairs as easily as bigger businesses. Such inequity could then translate to society, building a culture of discontent and distrust with a techno-government. For an AI-adoptive government to survive, such problems need to be properly considered.

Questions surrounding these findings are then pondered by the authors. For example, how much transparency will be required in order for AI systems to be compliant with U.S. legal norms? What does an 'explainable AI' actually look like? Answers to questions like these will need to be answered in due course, which will go on to shape the ultimate overview of whether the federal agencies will manage AI policy poorly, or well.

Overall, AI has the potential to aid the government in making a fairer society, but also to make it more unjust. While the uptake of AI systems in government is encouraging, this creates a greater need for caution and scrutiny over how AI systems are to be implemented in order to not exacerbate society's level of inequality. AI is the pen to our notebook world, and how we use the pen to write will either convert the notebook into a glorious adventure novel, or a terrifying horror.

## Repairing Innovation – A Study of Integrating AI in Clinical Care
([Original paper](#) by Madeleine Clare Elish, Elizabeth Anne Watkins)
(Research summary by Alexandrine Royer)

Health care is one of the areas in which AI promises to bring celebrated advances to patients' and medical practitioners' lives. Outspoken medical professionals such as Eric Topol have stated that, paradoxically, machine-led and digital medicine can bring back the central elements of human interactions in care by giving hospital workers the "gift of time." AI technologies can alleviate the time-pressed schedules of medical professionals through assisting in the early detection and treatment of disease, effectively acting as a safety net against fatigue-induced human oversight, error in judgement and mistaken gut instinct. Sepsis, a highly prevalent and deadly infection in US hospitals, requires early detection for treatment to be efficient. According to the Centers for Disease Control and Prevention, one in three patients who dies in hospitals has sepsis. However, as underscored by Elish and Watkins, sepsis is notoriously challenging to diagnose consistently in its early stages when patients are most receptive to treatment, making it a prime target for AI-based medical interventions.

Elish and Watkins have produced an in-depth study of Sepsis Watch, an AI-powered monitoring tool launched by Duke University and Duke Health system that uses deep learning and hospital protocols to improve sepsis treatment quality. Rather than a simple technological intervention or a more reductionist view, a machine learning model, Watkins and Elish, underline how Sepsis Watch is a complex sociotechnical system. To elaborate, social context, relationships, and power dynamics are central to creating, designing, and implementing this technology. In their introductory statement, the authors demonstrate how AI technologies are often uncritically conceived as the solution to medical challenges yet are rarely tested "in the wild," and their disruptive effects within the environment they are launched frequently escape analysis. In their words, "technology-driven innovation is disruptive; it will always require corresponding repair work to complete the process of effective innovation." Part of this repair work entails creating new sets of practices within existing workflows that allow for the successful integration and application of medical technologies.

Sepsis Watch was piloted at Duke Emergency Department as part of routine clinical care for two and a half years. The Sepsis Watch Model was built on a dataset curated by Duke Health clinicians, who identified the necessary variables, and contained over 32 million data points. It can only identify correlations in patient characteristics and the likelihood of sepsis -it works to signal if a patient is at risk, and not why. To describe how Sepsis Watch works, when a patient is admitted to the emergency department, their personal health record data is processed by the Sepsis Watch system, used on an IPad application, which signals whether or not they are high-risk. This information is received by a rapid response nurse who transmits the information to the patient's attending ED physician, with the latter being responsible for making the final call whether to begin sepsis treatment. If so, the treatment by ER clinicians is tracked on the application until it is completed.

Sepsis Watch's success in patient treatment relies on the fluid interplay between human actors, technical infrastructure, and expert medical knowledge. However, none of these interactions are as straightforward as one would expect. As the application's monitoring was left in the primary hands of a specialized team of rapid response nurses (RRTs), they needed to develop new communication procedures with ED physicians. Part of this involved reversing normative working hierarchies, as nurses were now responsible for making diagnosis calls to attending physicians, causing some initial tensions in the workplaces. The exchange of information was made especially difficult as RRT nurses would not see the patients and only responded to Sepsis Watch alerts. Working patterns among ED physicians had to be modified to ensure that sepsis risk was well-communicated. The app also revealed the hospital managerial staff's different perceptions, who saw ED as the focal point for sepsis diagnosis, and ED physicians who did not consider the disease a priority in the department and complained it hindered their professional discretion. RRTs had to find new strategies and techniques to ensure that doctors would follow up with their sepsis alerts and minimize their annoyance.

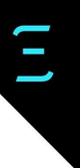

Although the pilot project results will only be released next year, the authors report that hospital managers found that Sepsis Watch had "dramatically improved the care of patients who are diagnosed with sepsis." However, the application's success did not rest solely on its technological capabilities, but rather the skills of RRT nurses who were tasked with setting the system in motion and attenuating its disruptive effects through repair work. The authors wonder whether improvement in patient care was due to the RRTs reminding calls over sepsis, keeping it fresh in physicians' minds, rather than alerts from the app itself.

Ellis and Watkin's report reveals the importance of studying medical technologies' implementation within hospital settings and closely attending to those bumps along the road. Such studies reveal how AI-power medical technologies must not be evaluated solely on their capacity to produce medical results but also their impact on existing workplace flows and professional norms. Seeing medical technologies as part of broader social networks means considering the power, often gendered, dynamics of hospital workplaces and recognizing the emotional labour that goes into ensuring the adoption of a new system. This also entails a rethinking of the division of labour within hospital settings and proper compensation for the tactical knowledge and new kinds of expertise required of healthcare practitioners. While Sepsis Watch may have given physicians more time to attend to their patient's risk of sepsis, it entailed putting additional responsibilities on RRTs. Indeed, referring back to Eric Topol, before celebrating the gift of time of AI medical technologies, there needs to be closer studies of the time-consuming and labour-intensive process of implementing these very technologies.

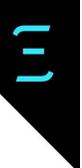

# Lanfrica: A Participatory Approach to Documenting Machine Translation Research on African Languages

([Original paper](#) by Chris C. Emezue, Bonaventure F.P. Dossou)
(Research summary by Alexandrine Royer)

English has become the lingua franca of machine learners and data scientists, yet a minority of fewer than 26% of internet users speak it. Against this trend, there have been a growing number of initiatives to include African languages in machine translation research, and in particular, natural learning processes for online platforms. Africa is the continent with the highest language diversity, being home to over 1500 documented languages, and over 40% of its population uses social media platforms. To keep track of these ongoing developments, Emezue et Dossou offers Lanfrica a participatory-led framework in documenting researches, projects, benchmarks, and datasets on African languages.

As Emezue et Dossou points out, there are already several existing online communities dedicated to promoting AI research in Africa, such as Masakhane, Deep Learning Indaba, BlackinAI and Zindi. These organizations reflect not only a desire to put Africa forward in machine learning but also to preserve the continent's distinct cultures within the digital space. Some limitations currently hinder the advancement of African natural language processes, including:

- A lack of confidence from African societies that their languages can be a prevalent mode of communication in the future
- A lack of resources for African languages
- A lack of publicly available benchmarks
- Minimal sharing of existing research and code

To redress these issues of lack discoverability, publicly available benchmarks, and sharing of resources, Emezue et Dossou created an open-source and user-friendly database system that documents machine learning researches, research-results, benchmarks, and projects on African languages. By surveying the Masakhane community, an open-source group of NLP researchers, the authors found that to build a neural machine translation (NMT) model, researchers had difficulty accessing model comparisons to guide them in data preparation, model configuration, training, and evaluation.

The soon-to-be-launched Lanfrica website will catalog ongoing ML research efforts based on the African language of interest and allow users to submit information on their projects, with contributions coming from both researchers and non-researchers alike. To improve ML reproducibility, links that provide access to open-source test data will be featured on the website.

Despite being a growing pole of ML research, Africa is underrepresented in discussions surrounding AI, often overshadowed by academic and corporate research labs in wealthy bubbles such as Silicon Valley and Zhongguancun. Digital assistants like Siri, Google Talk, and Alexa have yet to be programmed to accommodate widely-spoken languages such as Lingala, Oromo, and Swahili, and Google Translate only offers translations for 13 African languages. Unlike large databases such as Google scholar, Lanfrica is an initiative that is specifically tailored to African language researchers, allowing them to build networks in a digital space that reflects their interests and priorities. As the most linguistically diverse place on Earth, natural language machine learners in North America and Asia can also benefit from learning about the advances in Africa.

### Teaching AI Ethics Using Science Fiction
([Original paper](#) by Emmanuelle Burton, Judy Goldsmith, Nicholas Mattei)
(Research summary by Connor Wright)

It wouldn't prove shocking to say how science fiction has had its fair share in generating some of the AI fears of today. From iRobot to Terminator, science fiction has formed the basis of many people's fears and concerns with AI technology. However, the authors of this paper have turned this on its head. Now, science fiction is being proposed as a way to unlock and make more accessible the AI ethics space to computer science students, and they make a good point. I'll now explain why.

Given how technical practitioners are no longer being met solely with technical questions, the practice has now expanded into the realm of ethics. This has generally not been well-received by technologists such as computer science students, with ethics already being seen as something to tick off the lost after one class. The educational elitism stemming from ethics (and philosophy in general) in the Greek ages, and often throwing up more questions than answers, it's been viewed as out of place in the dualistic environment of computer science. The authors dully make this observation and draw an interesting comparison to the way the subject is taught. Here, in order to establish a firm grounding of the basics, computer science is generally taught with an authoritative overtone, leaving very little room for dissent. As a result, a mindset of 'absolute truths' is generated within the field, with the basics being taught seen as unquestionable. With this mindset, it's easy to see why considering multiple viewpoints on even basic concepts is considered as a best practice rather than a requirement. However, the authors see a way to solve this, and it goes by the name of science fiction.

Science fiction possesses the ability to represent current ethical problems without the veil of educational necessity, reducing the barrier to entry for the topic. Furthermore, the weird and unfamiliar representations of these problems make it more difficult to arrive at a viewpoint on the topic. Resultantly, students are encouraged to utilise their critical thinking skills and consider multiple viewpoints at the same time. For example, the problem of self-driving cars. While the decision to implement the technology was to save countless lives from road traffic accidents, even the best intentions have grave consequences. While preventing the deaths of many, the technology will not prevent mass unemployment resulting from the technology slowly replacing human-centered driving tasks (such as long-distance truck journeys and taxi drivers). Hence, tackling such an issue through a different, engaging, science fiction example will help to bring these concerns to light.

One of the most important points to consider in light of this proposal is the shift in mindset associated with the practice of ethics. Instead of being orientated around academic tradition, ethics is now being framed as 'practical ethics' which encompasses and can be applied to current situations. Instead of debating about the theory and then applying it to the problem, the problem is considered first in order to garner intuitions before launching into a moral framework. Above all, this problem-centered approach keeps the considerations relevant to the students, as well as creating a tangible motivation as to why an ethics course is important in the computer science space.

Creating engaging content without the educational barrier to entry, ethics is able to be re-marketed as an engaging and worthy topic in the computer science world. Permitting the consideration of other viewpoints in unfamiliar scenarios and its up-to-date nature, science fiction stands itself in good stead to help further ethics' cause. The proliferation of technology thanks to AI has now made ethics a priority in the space, and the more students that engage in the practice, the more refined our AI future gets.

### Algorithmic Accountability
([Original paper](#) by Hetan Shah)
(Research summary by Abhishek Gupta)

The public trust in algorithmic decision-making systems is at an all-time low. In the last 6 months we've seen three major companies abandon their pursuit of general-purpose Facial Recognition software and as of this week, a government scrapping the results of an Exam Grading algorithm. Given the immense economic interest in and the rapid adoption of this (arguably) budding technology, AI fiascos have become commonplace in recent times, but what is truly groundbreaking this time around is the power of the public in forcing the hands of powerful companies and governments to disavow this technology in settings where it's performance has been sub-par.

With this social backdrop, it serves us to remember some of the guiding principles for creating Accountable Algorithms laid out by Hetan Shah in his op-ed from 2018. In this piece, Shah argues that it is crucial to build public trust in a new piece of technology before pushing for its widespread adoption. He contrasts the speedy adoption of Stem Cell technology, that had a great deal of careful public dialogue around it during its development, with the paralyzing impact of public pushback on the advancement of genetic modification technologies. He then underscores that the best way to build trust is to implicitly improve the trustworthiness of algorithms, rather than explicitly pushing for more public trust.

Given the gravitas of the situation, there needs to be a coordinated effort by all the different stakeholders; namely: the practitioners (research labs, industry), the public sector and the policy-makers/regulators.

Practitioners need to be more conscientious about the creation of open source benchmarks by improving the diversity and representation in datasets. This is critical since benchmarks navigate the research direction of entire communities and so bias needs to be eradicated at its root. Shah also recommends piloting any model before it is deployed, using multiple datasets. In settings where in-house expertise is lacking, companies could engage bodies such as the Algorithmic Justice League to audit biases in models. Another approach could be to monitor for differential impacts, specially on vulnerable demographics, using causal models and counterfactuals. While Shah agrees that transparency would help only in a limited capacity, he recommends publishing models and the associated data and meta-data.

In addition to technical solutions, Shah outlines some important process changes that companies can make, such as improving diversity in the workforce, conducting ethics training and enforcing a professional code of conduct.

Regulation also has a key role in building trust. Following suit with the EU's GDPR, provisions such as the right to challenge an unfair decision from an algorithmic system and the right to redress, would go a long way in confirming a commitment to mitigating the negative effects of misbehaving models. Other recommendations from Shah include building capacity for regulators to be able to understand and close gaps across the several sectors in which algorithms are driving decisions.

Lastly, the public sector holds a key role in building trust. Shah envisions the emergence of a Data Commons, in which the ultimate ownership of data would go back to the public. This would allow for a much-needed balance of power, where the public could do away with exclusive contracts and have the bargaining power to enforce high standards of accountability and transparency from any contractor that wishes to use their data to create predictive models.

Keeping pace with evolving technology is an uphill battle, but it is one we must take on. As Shah eloquently argues in this piece, we need to approach the creation of accountable algorithms by pushing for systems that can positively impact society, and do away with our extant approach of negative screening/mitigating damage.

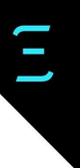

# Go Wide: Article Summaries (summarized by Abhishek Gupta)

### Why Kids Need Special Protection From AI's Influence
(Original *MIT Tech Review* article by Karen Hao)

Principles abound in the domain of AI ethics, ever since more people started paying attention to them. Yet, some disadvantaged or neglected groups who have special needs haven't been given due consideration. This article highlights a couple of such efforts: one from UNICEF and the other from the Beijing Academy of Artificial Intelligence (BAAI) to provide additional considerations from an ethical standpoint when going through the design, development, and deployment of AI systems that will be used for children.

AI-enabled systems are used for children in the context of toys with which they interact which might record their personal data, decisions about whether their parents are deemed fit, and most recently in the grades that they may have secured and thus determining their educational and career trajectories. The efforts from UNICEF are meant to augment existing guidelines with specific considerations for interactions with children. An interesting consideration was to make the guidelines explainable to children so that it can empower them with more agency in determining their futures. More education about how AI operates and what capabilities and limitations that it has will actually help to steer the development in a direction that does actually align with their interests and values.

The next steps in both of these initiatives is to run pilot development programs that actually battle-test these guidelines to see how they actually protect the rights of children in practice. Working groups with major industry players is another avenue to source feedback so that the guidelines are practical and will make a tangible difference.

### When AI in Healthcare Goes Wrong, Who is Responsible?
(Original *Quartz* article by Olivia Goldhill)

When it comes to taking a look at the use of automated technology in healthcare, there are many concerns that abound which need to be addressed proactively. In the use of automated systems, where it is clear that product liability applies, there the cases that need to be resolved are fairly obvious (with the usual caveats) as was the case for Da Vinci Robotics and mechanical errors with their surgical assistant robots. But, in the case where there might be software issues in terms of diagnostics, it is less clear on what can be done.

Some advocate that everyone in the value chain be held liable, others advocate for assigning strict liability on the manufacturer, and yet others who advocate for the use of liability mechanisms that place the onus on those who are utilizing the system since they are the final arbiters of use. In cases where automated systems have very high accuracies and a doctor makes another decision, there is an increased legal risk that they take which might disincentivize them from ever disagreeing with the machine. That said, if such legal holes continue to exist, not only will the adoption of these systems slow down but it will also erode trust and patients will be the ultimate losers in this scenario since they are stripped down to even fewer mechanisms for recourse.

As we dive deeper into integrating these systems, it is clear that we need to have stronger accountability measures in place that are enforceable and unambiguous.

## An Experiment in End-of-Life Care: Tapping AI's Cold Calculus to Nudge the Most Human of Conversations
([Original *STAT* article](#) by Rebecca Robbins)

Bringing the worst sort of dystopias alive from the series *Black Mirror*, medical institutions are beginning to use AI systems to nudge doctors to have end-of-life conversations with patients that the system deems are at risk of dying. Doctors are put in awkward positions bringing up the most intimate and sensitive thing up to a patient based on the recommendation of a machine.

In some cases, doctors agree with the recommendations coming from the system in terms of which patients they should broach these subjects, the article points to the case of one doctor who mentions that this system has helped to make her judgement sharper. Other doctors make sure to exclude mentioning that it was a machine that prompted them to have that conversation because it is inherently cold and often unexplainable in why it arrived at a certain decision which makes it even more challenging for patients to grasp why this is being discussed.

A particular challenge arises when doctors disagree with the recommendation from the system. In this case, even though the doctor does have the final say, they are weighed down by the consideration of whether or not they are making the right decision in not bringing up these advanced care options with the patient in case they are wrong and the system is indeed right.

Another problem to be highlighted is a potential over-reliance on the system for making these decisions and reducing the autonomy that doctors would have, related to the token human problem.

The designers of the systems have taken into consideration many different design choices, especially around the number of notifications and alerts to provide the doctors to avoid "notification fatigue". Another design consideration is to explicitly not include the probability rating with the patient list, given the understanding that humans are terrible at discerning differences between percentage figures unless they are on the extremes or dead-center. The labeling around the recommendations coming from the system are also framed as those requiring "palliative care" rather than talking about "will die" which can subtly create different expectations. One of the benefits of a system like this, as documented in an associated study, is that it has helped to boost the number of conversations around this subject that are being had with the patients which are essential and are sometimes ignored due to competing priorities and time burdens on the doctors.

## Artificial Intelligence Will Change How We Think About Leadership
([Original *Knowledge@Wharton* article](#))

An interesting article that talks about how business leaders should think about the impacts that AI is going to have on their employees and how they operate their businesses. One of the places where the article falls short though unfortunately is an unnecessary focus on the soul and why that must play an important role in how an organization is governed. The soul is a highly value-laden term that differs widely across cultures and geographies which limits the applicability of the arguments made by the author of the book that is featured in the article. In addition, some of the technical concepts are wrongly defined which warrants a question on the importance of the technical savvy of the business leaders. While an explicit argument is made by the interviewed author that not all business leaders need to have deep technical knowledge but at least a passing understanding of how AI works, the way the concepts are described in the article can lead to problematic conclusions that can misguide business folks reading the book and the interview.

Aside from some of the errors, there are some useful points made in the article around how both soft and hard skills complements will become even more important going into the future. Something that business leaders need to pay attention to is to be more cognizant of where they are getting their information from and how correct it is. If business strategy is to hinge on their understanding of some of these concepts, a flawed mental model will lead to erroneous results

The State of AI Ethics, January 2021                                                                                              163

that will ultimately harm the organization rather than help them make the best of the potential of AI.

### A Buyer's Guide to AI in Health and Care
([Original *NHSX* article](#))

The public sector gets quite a bit of flak when it comes to their procurement practices of AI solutions. This handy guide from the NHSX provides a few action items that officers in government and other public sector entities can utilize to make the integration of AI into their organizations better. A key consideration is to think about what problem one is trying to solve with the use of AI, and if it is really necessary or if non-AI automation methods could work instead. Related to this point is the availability of data, without which training an AI system would not be possible. This can be a problem in cases where the data is fragmented across different departments or units and poses a challenge for implementation.

Compliance with standards like NICE and being able to certify the software, especially when embedded inside medical devices in critical scenarios is also essential. This also needs to be an iterative process that recognizes that in the case of online learning systems, the behaviour of the system will evolve over time and needs calibration.

One of the things that really stood out in these recommendations was the adoption of a "no-surprises" mindset: being fully transparent about which data is being used, how the AI system is being utilized, and the limitations of the capabilities of the system. Utilizing techniques like data flow diagrams can assist with this. Completing a stakeholder impact assessment can also help to engender trust from those who are going to be responsible for using the system on an everyday basis.

In speaking with the vendors during the procurement process, due consideration ought to be given to the level of on-going support that one is expected to receive. Also, if data is going to be collected from the use of the system and sent back to the vendor for training or calibration of the system, this might have IP implications and data governance issues as well. And finally, paying attention to how the system might be decommissioned and disentangled from the remaining software infrastructure also forms a key consideration in making procurement decisions.

### Standing Up for Developers: youtube-dl Is Back
(Original *Github Blog* article by Abby Vollmer)

If you're not familiar with the youtube-dl tool, it is a nifty utility that helps you download videos from YouTube. The immediate concern that comes to mind is that it might be misused to gather copyrighted content right? Not entirely. There are many legitimate uses of such tools, say for fair use, demonstrating harm, gathering evidence, documentation on the part of journalists, downloading public-domain videos, changing playback speeds for accessibility, amongst other uses. So, the internet was in an uproar when GitHub decided to take down the popular repository in response to some DMCA notices (a notice from copyright owners when they believe there has been an infringement). This is not atypical for any website that hosts content generated by users.

This article provided much-needed clarity on how this particular request was handled, why it was different from others that they have received, and the new, more robust practices that are being instituted by GitHub to prevent unnecessary takedowns. The repository was finally restored after GitHub deemed that there weren't adequate grounds for the removal of the repository.

The crux of the argument from the people that requested the takedown of the repository was that the tools allowed the circumvention of technical protection measures (TPM) that are supposed to protect the content and rights holders as enshrined in Section 1201 of the DMCA. The original formulation of the DMCA from the 90s doesn't account for cases where such circumvention might not lead to copyright infringement and hence limits all such software.

In a nutshell, after the maintainer of the repository made the requisite changes around how some of the unit tests in the repository were utilizing copyrighted videos, and keeping in line with the ethos of supporting developers, GitHub reinstated the repository. Going forward, they have also set up more robust guidelines so that such incidents are minimized, and they will lean towards erring on the side of developers.

### Software is Trying to Change Your Habits. Make Sure You Agree With It.
(Original *Zapier Blog* article by Justin Pot)

While ultimately the article does talk about how Zapier might be used in a way that helps us break out of some habits, the thrust of the article is quite relevant to discussions on the societal impacts of technology and the role that design plays in that. Drawing on The Gruen Effect and talking about how malls are intentionally designed to be all-encompassing and labyrinth-like making it harder for us to stick to our goals, the article makes a strong point on the same thing that happens with the apps that we use.

When notifications on Facebook initially started off as a way to see if anyone had tagged you in something, commented on your stuff, or liked something that you had shared, we now get notifications for unrelated activity in the hopes that when we open the website/app, we find something else that catches our attention and we end up spending our precious time. While these notifications are tunable to prevent this kind of behaviour, rarely do we spend time doing that. While we might not consciously spend too much time thinking about these design decisions, the companies spend an inordinate amount of time doing so to harvest our attention.

Some habits are good and we might want to have the apps guide us in sticking to our goals of eating healthy and exercising more, but there are many more negative habits that the apps help to enforce and knowing the agenda of the apps and what tactics might be used to hold your attention in unwanted ways can help us make better decisions.

# 11. The Future of AI Ethics (Opinion)

**Opening Remarks** by Max Santinelli (Associate Director Data Science, Boston Consulting Group)

As AI systems are being deployed in more areas to address increasingly complex use cases, the nature of the ethical issues we face continues to expand. The body of work in this report reinforces the rapidly expanding depth and breadth of issues that are being tackled. In this context, it is difficult to know what comes next, but there are a few high-level themes that are emerging.

Customers are becoming increasingly concerned about how personal data are being used, the use cases being pursued, and the role AI can sometimes play in exacerbating societal division and unrest. We believe that customer demand for Responsible AI (RAI) will accelerate significantly over the coming years, to the point that ethical AI principles, which now serve as brand differentiators, will soon become nothing less than industry table stakes. As this evolution takes place, there is going to be a need for standards or certifications to ensure adherence to these principles. This community must be prepared to take a proactive role in shaping their structure and function.

To date, much of the work around RAI has focused on such fundamental concerns as privacy, bias, human interaction, and disinformation. But as AI penetrates deeper into society new themes emerge, such as the need for model transparency, explainability, and reproducibility as well as the call for more controls over AI solution's goals and training process. Companies will require practical guidelines, approaches, and tools they can use to implement programs that address the full breadth of potential concerns. They need broad assessment of AI development processes and organizational governance, guidance on how to motivate and reward staff for stepping forward to identify potential ethical issues, clear definitions of what the RAI organization within a company actually does, and clear metrics to measure RAI maturity or compliance. There are already some notable efforts underway in this area, but additional emphasis is needed from researchers and organizations to help provide practical guidance.

As the uses for AI expand exponentially throughout our global society, RAI principles are similarly expanding from a niche research area into the mainstream. We as a community of practitioners must be mindful that even as we work to implement these principles, we are not ourselves creating the conditions for continued bias. For years, we have operated as a small, close-knit community working together to tackle tough challenges. Looking ahead, our community is going to expand rapidly as more organizations begin to tackle these issues and more experts will be involved to ensure ethical implementation of AI in diverse domains such as healthcare, the judicial system, financial services, autonomous driving, and others. We must open our arms to people from ever-wider backgrounds and areas of expertise, including those new to the field. And as the diversity of our community grows, we must work together to maintain the spirit of collaboration that has been so powerful to date.

AI is, at this moment, at a point of inflection. We must act quickly to make sure that the algorithms that determine so much of how society will operate in the immediate future do as much good in the world as possible while minimizing unintended harms. We as a community can have a profound impact on how corporations and governments achieve this goal. The research contained in this chapter can help us understand where we are headed, increase our understanding of the challenges we face as we exert our power—and improve our ability to quickly take action.

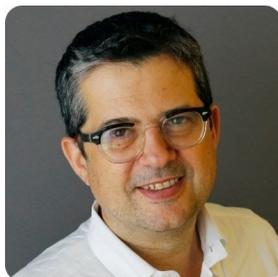

**Max Santinelli**
Associate Director Data Science
Boston Consulting Group

Max is an Associate Director Data Science at Boston Consulting Group (BCG). He has more than 15 years of experience in leading advanced analytics projects across multiple industries. His experience spans from the conceptual design of the analytics solutions to their integration in a company's technology and business environment. Max focuses on customer analytics and HR analytics as well as on the ethical implications of AI/machine learning solutions.

# Get In Early: New Directions for AI Ethics Research by Juan Mateos-Garcia (Director of Analytics, Nesta)

2020 was a mixed year for AI: GPT-3 wowed many with its generative capabilities, and Timnit Gebru lost her job for pointing at the risks of the kind of models it represents. AlphaFold 2's predicted the structure of proteins with unprecedented accuracy towards the end of a year when AI failed to make a significant contribution in the fight against Covid-19.

These disparate yet connected events point at AI's expanding influence and impact but also at its limitations, and at the complexity of its development and deployment. They also suggest new directions for AI ethics researchers looking to steer AI's trajectory towards the common good and respect for human values. I outline them in turn.

First, there is the question of the **directionality of AI research**: AI ethics tends to focus on the mitigation of ethical risks created by existing AI systems, in particular around fairness and privacy (the [Chouldechova and Roth](#) paper summary earlier in this report summarises the state of play and frontiers in fairness AI research). More recently, AI ethics researchers have started to [pay more attention to questions of power and inclusion](#) - for example under the rubric of participatory ML - and developed methods to involve relevant stakeholders and communities in the deployment of AI systems to make it more beneficial and empowering. But even this progressive work takes AI systems and technologies as exogenous.

[*"On the Dangers of Stochastic Parrots: Can Language Models Be Too Big?"*](#), the paper that ostensibly led to Timnit Gebru's dismissal from Google Brain, breaks away from this approach and questions the ethical desirability of AI's dominant technological trajectory along several fronts: today's state-of-the-art AI systems have to be trained on massive and hard to curate datasets that include racist, sexist and inflammatory content. They rely on big computational infrastructures that create environmental costs that are primarily imposed on vulnerable groups and communities. They provide an illusion of progress in AI capabilities ("to understand language") while diverting resources and talent away from alternative techniques.

The issue at stake is not how to mitigate the risks that these AI systems create, but whether we should develop them at all (or at least to the extent we are), given those risks. Perhaps unsurprisingly, this is a controversial thesis for Google, a company that has invested billions of dollars in the development of large language models and embedded them deeply in many of its products and services.

If some technologies are intrinsically more likely to create ethical risks, this also means that ethical considerations should be taken into account when AI technologies are starting to be developed as well as further downstream when their risks are identified but harder to mitigate.

As [Gebru pointed out in an interview](#) soon after being ousted from her position at Google Brain, *"You can have foresight about the future, and you can get in early while products are being thought about and developed, and not just do things after the fact"*.

This upstreaming of AI ethics would make it easier to raise a central ethical question about new technologies: when not to create them. Developing frameworks and processes to inform this decision in the context of pervasive uncertainty that characterises the early stages of an innovation, and given the intrinsic dual nature of many AI technologies should be a driving concern for AI ethicists in future work.

This brings me to a second potential direction for AI ethics research, concerning **the type of organisations that undertake it**. The Gebru case has cast serious doubts on the idea that private companies can be trusted to be impartial when they assess the ethical risks of the AI systems that they develop (it also questions the desirability of relying on them to offer "ethics-as-a-service" as [Google is planning to do](#), as mentioned earlier in this report). AI ethics researchers who have called for increasing diversity in the AI workforce with the goal of mitigating biases and blind spots that may lead to the development of discriminatory AI systems should reflect on how the organisational environments where they operate may, in a similar way, skew their research agendas implicitly and explicitly. This could create AI ethics meta-risks - the risk that AI ethics neglects some ethical risks. Perhaps the lack of attention to AI directionality which I pointed to above is an illustration of this.

One implication is that the AI ethics community needs to create an effective division of labour between researchers in industry, academia, government, and the third sector. This requires determining what AI ethics questions can be better explored outside of industry because they require a degree of impartiality that is unfeasible or unsustainable in the private sector and which ones are more suitable for an industry context, for example, because they require deep knowledge of specific technologies or development processes.

It will also be important to put in place institutional mechanisms and incentives to create and sustain a healthy flow of knowledge and ideas between public and private research domains while preserving independence and contestation. These efforts should be underpinned by a stronger understanding of the way in which different organisational and social contexts shape the development and deployment of AI ethics. This is a fertile ground for new AI ethics research in collaboration with other disciplines such as sociology, management science, and economics (in this vein, the [Hudáková article](#) summary included in this report explores practical steps to embed ethical thinking and responsibility in the day-to-day operation of technology companies).

I conclude with a third direction for AI ethics that will also require researchers to pay more attention to the organisational and social context where AI systems are deployed - **the ethical** **implications of the adoption of AI in scientific R&D** (We can also think of this as a research and application site for the two other directions I set out above and will draw their connections throughout.)

AlphaFold 2's success has demonstrated the potential of AI as a driver of scientific productivity and discovery. The ethical risks that it creates have received much less - if any - attention. This is probably explained by the fact that AlphaFold 2 makes predictions about the structure of proteins using open biological datasets so there are at this point, fewer reasons to worry about its implications for fairness or privacy, two focus areas for AI ethics researchers. This is also consistent with the idea that ethical considerations around new AI systems tend to receive attention downstream, reactively, instead of upstream, strategically.

I believe that this is misguided. The indirect impacts of powerful AI systems such as AlphaFold 2 on the scientific enterprise raise important questions that AI ethicists should study in tandem with philosophers of science, epistemologists and science and technology studies and "meta-science" scholars.

How will AI transform scientific norms around reproducibility, explainability and publicity of research findings? Will it increase the concentration of research on a small number of institutions - many of which are in the private sector - that have access to data and computation, perhaps neglecting the interests and needs of developing countries and vulnerable groups? Will it devalue some scientific problems less amenable to prediction and forms of knowledge such as theoretical and subject-specific knowledge and displace scientific labour in processes of automation that [economists like Anton Korinek](#) claim have an important ethical dimension?

[This assessment of the implications of AlphaFold 2 for structural biology and protein prediction by Mohammed AlQuraishi](#) highlights many of the issues at play. [My own research about the deployment of AI methods in the fight against Covid-19](#) suggests that AI researchers have focused on problems such as analysing medical scans that are amenable to existing deep learning techniques but of limited relevance to medical practitioners.

This brings us back to the idea that the current direction of AI research may be privileging some use cases for AI and downplaying others. It also highlights the need to ensure that public interests play a central role in shaping agendas of AI in science that are starting to be dominated by the private sector. Science (underpinned by the scientific method) deeply influences the pace and direction of human progress and our understanding of nature, so it is vital to carefully assess how it stands to be transformed by powerful AI systems developed by private sector companies. As I pointed out earlier in this essay, ethics has a vital role to play informing this process from the very beginning, which is now.

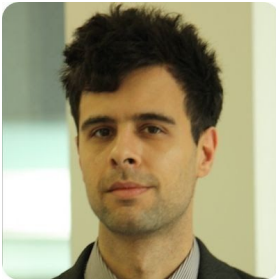

**Juan Mateos-Garcia (@JMateosGarcia)**
Director of Analytics
Nesta

Juan Mateos-Garcia is Director of Analytics at Nesta, the UK innovation foundation. There, he leads a team of data developers, data scientists and visualisation experts using novel data sources and methods to map emerging research, technology and economic trends. This includes a programme of research about AI's technological trajectory which has resulted in highly cited papers about the geography of deep learning research, gender diversity in the AI workforce, thematic diversity in AI research and AI researcher mobility between academic and industry. Juan is an economist with a MSc in Science and Technology Policy from the University of Sussex.

### Is the Global South ready for an AI invasion? by Dr. George R. Obaido, Dr. Kehinde Aruleba, & Dr. Lauri Goldkind

The recent rise in the adoption of artificial intelligence (AI) technologies worldwide can provide solutions to issues that impact the Global South driving development and growth in many sectors, such as education, financial services, medicine, and agriculture. Tech-forward actors and stakeholders have tried to promote the growth of AI systems, providing an ecosystem supportive of AI-enhanced everything in these nations. Despite these efforts, there remain significant inherent challenges in the Global South, which might affect a healthy AI ecosystem. Many challenges, such as lack of basic infrastructures, electricity, Internet, and access to quality education, continue to plague nations' economies in the Global South. As AI grows more ubiquitous and sophisticated, a commensurate clarion call of alarms regarding the potential harms of these systems is growing globally. The adoption of AI in the Global South has led to massive unemployment, data (in)accessibility issues as AI relies on a large volume of data, and ethical implications -- AI systems disproportionately impact these groups. Conflating these challenges is the growing acknowledgment of AI's environmental impacts, which disproportionately impact Global South countries.

Are Global South nations ready to mitigate the threats posed by these technologies? A 2018 study reported by Accenture estimated that due to AI invasion, about 5.7 million jobs (35% of all jobs) are currently at risk in South Africa by 2025; a country that has already had a staggering unemployment rate of almost 30%. AI-generated videos, audio, and images are increasingly influencing our decision on whether the news media is fact or fiction. In the Middle East, Ziad Nasrallah, a principal at consultancy Booz Allen Hamilton reported that 63% of Arab youth source their information on Facebook and Twitter. Deepfakes and other algorithmic media manipulation make significant numbers of the youth population vulnerable to fake news. A misinformation case would result in a severe violation of privacy since there are no federal data protection laws in place.

For example, the use of deepfakes led to a military coup in Gabon where a video was reproduced to show the Vice President announcing that the President — who had been ill and out of the country — was incapable of ruling. This, in fact, was false information. The implications of fake news are also being felt across public health circles, in the KwaZulu-Natal province in South Africa where four telecommunication radio masts were burnt down after a conspiracy theory linking the emergence of COVID-19 to 5G technology was circulated.

India and Latin America also face challenges, similar to those of the Middle East and Africa, including a lack of infrastructure, economic and social issues, as well as skills shortages, leading to a steady decline of AI-readiness. While the systems to support inclusive AI adoption are fragile in these regions, state actors are maximizing the use of AI technologies for surveillance, with and without transparency.

To develop unbiased, more inclusive, and affordable AI systems for countries in the Global South, we recommend a developmental approach, analogous to the *Crawl, Walk, Run, and Fly* framework found in the business world. Crawling requires that AI-enabled devices are readily affordable to promote learning. Internet Services Providers (ISP) must provide discounted data or special affordable packages for low-income groups. These ISP companies can take these initiatives to expand connectivity and improve lives. Once internet and data access are established, walking can begin. This might entail training a workforce of individuals in these regions to utilize these technologies fully. Additionally, 'walking' should include formal education on critical data literacy, an ability to understand algorithms that these technologies are built with will be of great benefit. Having a deep knowledge of these technologies would create avenues for innovation and catalyze entrepreneurship; leading to a 'running' stage of development where citizens from these regions can utilize technologies and AI systems to innovate new ideas commensurate with AI development in the Global North. With an educated workforce and digital infrastructure, 'flying' will not be far behind. Once flying, the Global South nations will equitably bring technological innovations to local communities, unlocking opportunities for positive impacts.

It is crucial to align AI initiatives, and the data used for training AI models to local communities in the Global South. Engaging these communities, offering training solutions, understanding the local issues, and their unique needs will help create an avenue for developing more inclusive AI technology.

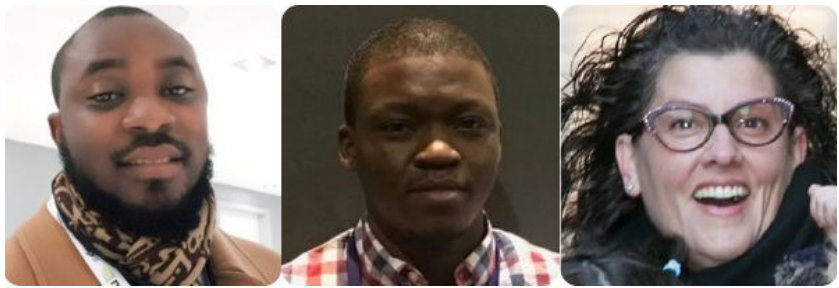

*(from left to right)*

**George Obaido** (@Geobaido) has a PhD in computer science from University of the Witwatersrand in Johannesburg, South Africa, and focuses on formal language and Automata applications.

**Kehinde Aruleba** (@arulebak) has a PhD in computer science from the University of the Witwatersrand and focuses on image processing and information retrieval.

**Lauri Goldkind** ([@brooklyn11210](https://twitter.com/brooklyn11210)) is an Associate Professor at the Graduate School of Social Service at Fordham University. Dr. Goldkind's current research has two strands: technology implementation and information and communication technologies (ICT) tools in human services and nonprofits and social justice and civic engagement in organizational life.

# 12. Community-Nominated Spotlights

**Founder's Note:** The community-nominated highlights in this section are generous contributions, recommendations, and pointers from the global AI ethics community, helping us shed light on work being done by people from around the world representing a diverse mix of backgrounds and research interests.

## Spotlight #1: Marie-Therese Png (PhD candidate, Oxford Internet Institute)

**How My Work Fills a Gap In The Current Discourse in AI Ethics**

I work in three different areas that intersect with AI ethics - academia, policy, and community organising. I am currently a PhD candidate at the Oxford Internet Institute, was Technology Advisor to the UN Secretary General's High-level Panel on Digital Cooperation in 2019, and have worked as part of various organising groups committed to resisting interlocking systems of oppression such as Radical AI, Black in AI, and Tierra Común.

The common thread across these areas has been centring excluded discourses which are by, from, and for the "margins" - people of colour, dis-abled, Black, Indigenous, queer, poor, womxn, and others pushed beyond the margins of safety. As a PhD candidate I am learning to use scholarly traditions of decoloniality, black feminism, and critical geography to unearth AI harms and propose restructuring practices.

One effort to articulate this was our paper *Decolonial AI: Decolonial Theory as Sociotechnical Foresight in Artificial Intelligence,* co-authored with Dr. Shakir Mohamed, and Dr. William Isaac. This built on my work as a DeepMind PhD Intern during my first year, when I worked on AI Value Alignment, moving beyond the epistemic and moral authority of Western political philosophy, and attempting to frame value alignment within Buddhist and Confucian ethics, Indigenous philosophies, Ubuntu ethics, etc.

In this way, the narrative of the 'unintended consequences' of AI technologies can be reframed not as unpredictable externalities, but as areas of risk that institutions are blind to because of knowledge deficits, and a lack of connection with marginalised groups they aim to benefit.

At the UN, I led efforts to centre the interests of "Global South" stakeholders in policy development areas such as digital divides, cybersecurity, lethal autonomous weapons, and AI governance. I now know first hand that though the UN is a necessary institution for international cooperation, material assistance, and conflict resolution, significant reform of internal structures and exclusionary dynamics are needed to confront the monopolisation of AI advantages.

The continued exclusion of "Global South" stakeholders, especially civil society, means they cannot act unilaterally in the protection of their interests, or build capacity without continued dependency dynamics. Governance policies are being replicated across jurisdictions in ways that are incompatible with the goals and constraints of developing countries. AI governance initiatives "for good" continue to enact what Prof. Ruha Benjamin terms "techno-benevolence" - interventions that intend to address inequalities, but instead reproduce or deepen dependency and extractivism.

Moreover, popular discourses such as the "Fourth Industrial Revolution" - which celebrates the transformative power of AI and the potential for developing countries to "leapfrog" - do not adequately recognise how first-mover advantages and exclusionary path dependencies still persist and are, in part, living relics from our colonial histories. This precludes an articulation and execution of developing countries' self-determined industrialisation and development.

**Why I Was Moved to Do This Work**

Not surprisingly, I entered this work because of an existing proximity with systemic oppression (Zuroski, 2018). I am of Afro-St Lucian and Singaporean heritage, and grew up in the UK, meaning colonial and imperial histories are especially alive in my family's experience, civic work, and diaspora politics. I therefore try to situate myself in discourses and organising with decolonial principles, whilst acknowledging aspects of my positionality which are privileged and complicit with colonial structures (language, education, nationality, etc.).

My initial training was in a dual biological and social sciences degree, and have always focused on epistemic, structural, and historic erasure/violence in the applied sciences. Initially in evolutionary biology, then biotechnology, neuroscience, and now AI.

Overall, I hope my work can, in some small part, contribute to the multitudes of communities resisting and reconstructing beyond interlocking systems of oppression.

**The State of AI Ethics**

I've been energised by the State of AI Ethics Report. I think it exemplifies how AI ethics research can be cross-geographic, interdisciplinary, and embody the spirit of "ethics as politics". I am grateful to have learnt from the contributors, and the team who created the report. Germane to my own work, I've personally been inspired by many of the contributors who discuss AI ethics in the context of macro-political inequalities, including Noopur Raval (Automating Informality: On AI and Labour in the Global South), Abeba Birhane (Algorithmic Colonization of Africa), Mohammad Amir Anwar and Mark Graham (Between a Rock and a Hard Place: Freedom, Flexibility, Precarity and Vulnerability in the Gig Economy in Africa), Stephen Cave and Kanta Dihal (The Whiteness of AI), and many more.

**My Outlook for AI Ethics in 2021**

*Climate justice*

Regarding my outlook for AI ethics in 2021, in addition to AI as an accelerator of Green Tech R&D, I've seen more academic and policy conversations on AI's deleterious climate impacts (Schwartz et al.,)

2019). The use of AI in rare earth mining, the energy (and financial) costs of machine learning, the carbon footprint of computation, etc. are increasingly identified as under-explored dimensions of harm. This identification underlines connections between racial justice, environmental racism, and colonialism; Caribbean intellectuals such as Glissant or Brand, for example, articulate the extraction of labour (or information) from black and brown bodies as synonymous with ecological extraction.

Given the scale of AI driven impacts, I'm also curious about what we can learn from climate justice and climate governance "inclusion" protocols.

*Ethics as politics*

Ethics as politics has been particularly acute within industry AI ethics research in 2020, with Dr. Gebru Gebru's being "resigned" by Google as one of many rallying points.

As many, including Dr. Gebru, have repeated - what it means to be ethical cannot be defined by those who are already in positions of power, it must be defined by those who know and experience the costs. For this to happen, there needs to be structural changes in research - the composition of organisational leadership, funding structures (Abdalla & Abdalla, 2020), or

The State of AI Ethics, January 2021    178

barriers to peer-reviewed publication for academic researchers in developing countries (Jeater, 2018) or outside of "elite" institutions.

I'm hopeful that in 2021 a prerequisite to building ethical AI systems will be the strengthening and safety of the people who ask hard questions about systemic injustice. For industry, this means more internal and public accountability, as well as worker coalitions and collective organising centred in care. An example of this that motivates me was both the rallying in support for Dr. Timnit, and the NeurIPS Resistance AI Workshop, for which I was a co-organiser. I learnt a lot from this workshop effort - it centred the margins, did not accept corporate funding, was open to the public, accepted submissions of academic papers, art, and poetry, and had open dialogues on realistic tactics for resistance, and how many of us occupy both positionalities of marginalisation and privilege.

**Call to Action**

*Navigating the in-betweens*

Many of us in AI ethics work at difficult intersections - whether it be industry and academia, academia and activism, activism and policy, etc. We are navigating (sometimes irreconcilable) differences in incentive structures, ideological assumptions, definitions of harm, theories of social change, strategies of harm reduction and standards of adequate protective mechanisms. My own trajectory in AI ethics is centred in communities such as Black in AI, Radical AI, and the Non Aligned Tech Movement, but also started in long-term existential risk / AI safety communities such as Effective Altruism, the Future of Humanity Institute, and the Future of Life Institute. Work such as *Concrete Problems in AI Safety, Revisited*(Raji & Dobbe, 2020) make useful connections between both.

At times the niches we occupy render us complicit with a harmful status quo, and participants in the co-opting of marginalised voices, which is something I have grappled with. In 2021 I would like to see even more challenging conversations around "who am I, and who is my work serving". I would also like to see more bridge-building across different research areas and geographic regions with a focus on collective wellbeing, and liberatory theory/practice.

**How People Can Support My Work**

I will be carrying out interviews for my PhD this year, and am interested in connecting with people who are working on:

- Political participation of the "margins" in AI governance
- AI ethics intersecting with environmental justice
- Navigating positionality and occupying the in-betweens of non-aligned communities in AI ethics

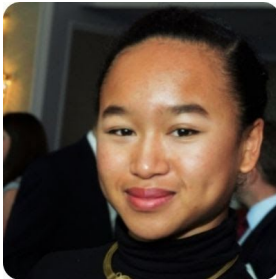

**Marie-Therese Png ([@png_marie](https://twitter.com/png_marie))**
PhD candidate
Oxford Internet Institute

Marie-Therese is a PhD candidate at the Oxford Internet Institute, researching AI governance and coloniality. She was previously Technology Advisor to the UN Secretary General's Digital Cooperation Initiative, is member of the IEEE Ethically Aligned AI Classical Ethics Committee, and co-authored Decolonial Theory as Socio-technical Foresight in Artificial Intelligence Research with DeepMind. Marie-Therese works in community organising with Radical AI, Black in AI, and was a co-organiser of the 2020 iteration of the Rhodes Must Fall Oxford movement. She completed an undergraduate in Human Sciences at Oxford and Masters in the cognition of racial prejudice at Harvard. She can be reached at marie-therese.png@oii.ox.ac.uk.

**References**

- Ruha Benjamin. 2019. Race after Technology: Abolitionist Tools for the New Jim Code.
- Inioluwa Deborah Raji & Roel Dobbe. 2020. Concrete Problems in AI Safety, Revisited
- Eugenia Zuroski. 2018. Holding Patterns: On Academic Knowledge and Labor
- Mohamed Abdalla & Moustafa Abdalla. 2020. The Grey Hoodie Project: Big Tobacco, Big Tech, and the threat on academic integrity
- Diana Jeater. 2018. Academic Standards or Academic Imperialism? Zimbabwean perceptions of hegemonic power in the global construction of knowledge
- Schwartz et al. 2019. Green AI
- Shakir Mohamed, Marie-Therese Png, William Isaac. 2020. Decolonial AI: Decolonial Theory as Sociotechnical Foresight in Artificial Intelligence

## Spotlight #2: Emanuel Moss (Researcher, Data & Society Research Institute)

The year 2020 saw tremendous challenges and upheavals across society, revealing how the risks we collectively bear can be shifted onto the most vulnerable among us while protecting those who already have the most resources at their disposal. For those who work in artificial intelligence, the COVID-19 pandemic appeared as a test case for whether the tools they build might be able to help society through one of the darkest years in recent memory. Indeed, with the sudden, dramatically increased reliance on services provided by tech companies – curbside grocery delivery, online entertainment, and video conferencing – it might have seemed like the backlash against the tech industry that had been mounting over previous years was abating, at least a little bit. But many of these services shunted risk of exposure to the coronavirus onto the lowest-paid, most-essential workers, while protecting those who could afford the luxury of having their food and amusements brought to them.

And often the highest risks accumulated for minoritized groups, as we saw with unequal death rates across racial groups in the pandemic (Moss and Metcalf 2020a). Grappling with the ethical implications of how AI-driven technologies distribute risk, therefore, remains a crucial concern. Inside tech companies, an important role for experts tasked with considering the ethical implications of their companies' products has emerged. In 2020, they have had their hands full "owning" the portfolio of ethics concerns across a company and translating organizational values into engineering practices. While "ethics owners" continue to face significant challenges navigating the tensions that characterize their work (see Moss and Metcalf 2020b), the past year has seen an increase in the opportunities people in these roles have to meet with each other (over video conference), share the strategies and tactics they've employed, and learn from each other about how to navigate those tensions.

Outside tech companies, there has been continued interest in developing ways of building algorithmic accountability through algorithmic impact assessment, along the lines that were proposed in the U.S. Congress in 2019 (Wyden 2019). True algorithmic accountability, as my collaborators and I have observed, consists of requiring an adequate description of technical systems that can be reviewed by an entity that is empowered to mandate changes to those systems that minimize and mitigate harmful impacts. It also requires developing ways of assessing algorithmic impacts in ways that map as closely as possible to the harms experienced by people "on the ground", and the best way of doing that is by assembling a broad range of experts across disciplines, including those most vulnerable to algorithmic harms (who are, themselves, experts on their own lives) to construct the scope of an adequate algorithmic impact assessment process (Metcalf et al. 2021).

In the coming year, I hope to see far greater collaboration across disciplines to better understand the stakes of AI ethics for all. This means interrogating what harms such systems produce, how those harms can be minimized or mitigated, and contexts of use that are entirely inappropriate for AI. It also means holding space for and acknowledging the value of those most vulnerable to algorithmic harms making such harms visible to us all (Sloane et al. 2020). I also hope to see continued movement acknowledging the power that all workers have in crafting AI, as we have recently seen in unionization efforts at [Alphabet](Alphabet).

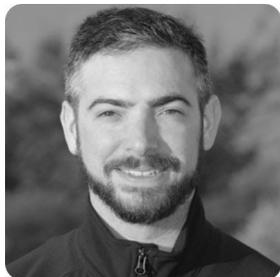

**Emanuel Moss ([@mannymoss](@mannymoss))**
Researcher
Data & Society Research Institute

Emanuel Moss conducts ethnographic research on issues of fairness and accountability in AI systems. He is a researcher for the AI on the Ground Initiative (AIGI) at Data & Society and a research assistant on the Pervasive Data Ethics for Computational Research (PERVADE) project. He is also a doctoral candidate in cultural anthropology at the CUNY Graduate Center, where he is studying machine learning from an ethnographic perspective and the role of data scientists as producers of knowledge. He is particularly interested in how data science, machine learning, and artificial intelligence are shaped by organizational, economic, and ethical prerogatives. He can be reached at [emanuel@datasociety.net](emanuel@datasociety.net).

# Spotlight #3: Emily L. Spratt, PhD (Postdoctoral Research Fellow, Data Science Institute, Columbia University)

**The Optimism of Ethical Frameworks for AI: Reconsiderations of the Binary Perspectives on Emerging Technologies and the Role of the Humanities in the Discourse**

The rise of AI applications empowered by deep learning techniques has fueled an approach to the creation, management, and interpretation of information that has been so transformative of our society that its effects could be described as a modern-day Renaissance. Accompanying this transformation, the burgeoning ethics of AI discourse has placed much recent focus on algorithmic design, yet less attention has been directed to the widespread reexamination of existing sociopolitical and economic structures that has been catalyzed by the technological boom. This phenomenon—driven by the often-exacerbating effects that machine learning and natural language processing have on the translation of information into knowledge—has resulted in a fundamental re-questioning of those things that are subjected to AI analysis. Consequently, AI is fostering a rebirth of interest in the fundamental structures and values upon which society is built and a revived humanism driven by the reflections that the mirror of AI provides to us, replete with its illuminating distortions.

Indeed, the processes of AI-driven analytics often create amplifying effects in the interpretation of data. Subsequently, subjects that the data has bearing upon are made more transparent, and the tenets upon which those subjects are built, or are assumed to be founded on, are revealed with less opacity. For example, some uses of facial recognition-based data in the service of the judicial system have been demonstrated to replicate racial and gender biases to such a degree that their suspension has been promoted by tech leaders. Notably, just this year, IBM halted research in facial recognition technology on account of the currently unresolved ethical issues it poses for society. In this regard, the deployment of AI-empowered tools for the analysis and generation of information has the ability to contribute to the increased attention being placed on the role that underlying social biases play in constructions of knowledge and power. This phenomenon is already at work in regard to the critical reexamination that the values of privacy, race, and gender hold in our society. The impetus to reflect on and challenge the underlying mechanisms that support societal structures with the approach of reason and an enlightened humanism has a long tradition whose revival is welcome. In the same vein, this is a phenomenological and epistemological issue that cannot be bottlenecked or stymied by methodological approaches to emerging technologies that are reductionistic and ill-equipped to address their own applications.

In the data science community, the phrase "the data is biased" is often heard as a catch-all statement that works as a convenient disclaimer when research results are found to be problematic, but perhaps too much is lost in this phrase. The general acceptance of this statement as a sufficient answer to the concerns raised by the tools we are building is untenable. At root is the question of data's interpretability, including what its selection at the point of a given dataset's formation is purported to mean—issues that cast light on the ethics of the translatability of any given aspect of society into quantitatively derivable measurements—and how those results are understood. Although ethically informed algorithmic designs typically acknowledge components of this phenomenon in compliance-focused pipelines built specifically for AI applications, too often they are unfathomably insufficient, as if recognition of just a few grains of sand could adequately account for the entirety of a beach, let alone its relation to the totality of the world's shorelines.

This is not to say that computer scientists have necessarily erred in their degree of acceptance of responsibility; rather, this points to the need for more collaboration is needed at the point that a decision is taken to examine an aspect of our world with AI-empowered tools. Instead of approaching discovery in data science through the all-too-common implicit method—"what dataset can we get ahold of and what can we solve with it in a new way?"—perhaps innovation in this area would naturally be more ethically aligned if it came through the domain from which the data's subject derived and in partnership with the experts in its related fields of study. Would a more humanities-centered approach to research be the answer? To date, it is unfortunate that it is a minority of AI projects that have adequate consideration of the fields that their research has bearings upon. The question thus emerges: Why aren't the humanities more involved in the ethics of AI and the formation of algorithmically based research?

The answer may in part be based on the simplistic moralistic frameworks around which emerging technologies are described and promoted. To list the number of "data for good" or "AI for good" type initiatives around which research and industry have aligned would be an exhausting activity. Even though they reflect an optimism and enthusiasm for innovation and research around technology and an attitude endorsing social beneficence, they lose recognition of the inherent complexity around AI that is the key to its ethical understanding. A binary designation of AI may easily serve a marketing or fundraising agenda, but it undercuts a more critical engagement with those who in recent times are developing the tools that are allowing us to take some of the most pronounced views of our society's fabric of being. It is on the level of the philosophical complexity of the subjects that are brought to bear with the tools of AI that the humanities would find their natural entrance to collaborations in this area, yet it is precisely this space that has been occluded. As a scholar working both in data science and in the arts, I find myself a frequent apologist to my colleagues in the humanities for the philosophical positionings of the institutional structures out of which much research concerning AI is supported.

Furthermore, simplistic binary designations of AI lend to potentially problematic structures for making decisions on the future of the regulation of technology and digital marketplaces. It is likely that these approaches to emerging technologies will soon be deemed unsustainable as the inherent complexity of tech ethics and their bearings on society becomes clearer to the public. Correspondingly, the concept of having a person on the design team with domain expertise in the given subject any AI application is based upon will likely be deemed essential in the near future. It is therefore my prediction that the one-dimensional optimism encountered in simplistic and binary ethical guidelines around emerging technologies will in retrospect be characterized as blinding naiveté, and that the ethics discourse around AI will soon come to embrace the complex and nuanced implications that its design, development, and applications have on society. With these epistemological shifts, the humanities are well-positioned to be the most promising ally to the quickly growing field of AI ethics.

*(The views expressed in this article are the author's own and not necessarily those of the institutions with which she is affiliated.)*

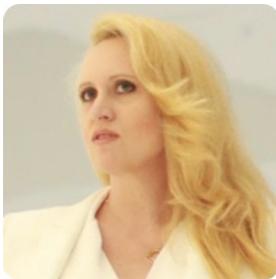

**Emily L. Spratt, Ph.D.**
Postdoctoral Research Fellow
Data Science Institute, Columbia University

Emily L. Spratt is an arts and technology academic and strategic advisor. She is currently a fellow at Columbia University in the Data Science Institute, in collaboration with the Historic Preservation Program in the Graduate School of Architecture, Planning and Preservation and the Department of Computer Science in the School of Engineering and Applied Science. She also holds advisory positions for Duke University's Ethical Tech Program, the Artificial Intelligence Finance Institute, the tech accelerator Exponential Impact, the cultural heritage foundation Iconem, The Frick Collection and Art Reference Library, and the Defense Innovator Accelerator with NSIN. It is notable that she curated the 2019 exhibition Au-delà du Terroir, Beyond AI Art for the Global Forum on AI for Humanity at the Institut de France in Paris, which was sponsored by the office of President Emmanuel Macron. Emily completed her doctorate at Princeton University. She can be reached at emilylspratt@gmail.com.

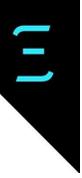

# Closing Remarks

I hope the report highlighted a few things that you missed in the past quarter, resurfaced some things that you found interesting in your readings, or provoked some critical thinking on key issues in the field of AI ethics.

Our genuine aspiration with this report is to lay out the landscape of AI ethics for you so that you can start to think about solutions and work with communities on the ground to help the entire ecosystem move towards a healthier state than its current problematic situation.

As outlined in a piece that I penned earlier this year highlighting the [importance of civic competence building in AI ethics](#) as a cornerstone to truly achieve ethical, safe, and inclusive AI, I believe that if we all come together, listen to each other, share the stage, support each other and grassroots efforts, we can get to a place where we can remember 2021 as the year where we moved from principles to practice.

MAIEI is a non-profit organization that depends on the support of individuals and organizations who, through their time and resources, help us bring all of our work in an open-source, open-access manner to all while providing fair compensation to those who work behind the scenes to make these efforts a success.

I am lucky to lean on a stellar team of individuals who help to run MAIEI and I continue to learn from them all every day. We are always open to feedback and really appreciate any insights that you can provide to help us make the *State of AI Ethics Reports* an even more valuable resource for the entire community.

Wishing you all the best in the meantime and see you next quarter!

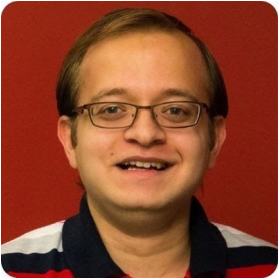

**Abhishek Gupta ([@atg_abhishek](#))**
Founder, Director, & Principal Researcher,
Montreal AI Ethics Institute

Abhishek Gupta is the founder, director, and principal researcher at the Montreal AI Ethics Institute. He is also a machine learning engineer at Microsoft, where he serves on the CSE Responsible AI Board. His book '[Actionable AI Ethics](#)' will be published by Manning in 2021.

# Support Our Work

The Montreal AI Ethics Institute is committed to democratizing AI Ethics literacy. But we can't do it alone.

Every dollar you donate helps us pay for our staff and tech stack, which make everything we do possible.

With your support, we'll be able to:

- Run more events and create more content
- Use software that respects our readers' data privacy
- Build the most engaged AI Ethics community in the world

Please make a donation today at **montrealethics.ai/donate**.

We also encourage you to sign up for our weekly newsletter *The AI Ethics Brief* at **brief.montrealethics.ai** to keep up with our latest work, including summaries of the latest research & reporting, as well as our upcoming events.

If you want to revisit previous editions of the report to catch up, head over to **montrealethics.ai/state**.

Please also reach out to **Masa Sweidan** **masa@montrealethics.ai** for providing your organizational support for upcoming quarterly editions of the *State of AI Ethics Report.*

**Note:** All donations made to the Montreal AI Ethics Institute (MAIEI) are subject to our **Contributions Policy**.